\documentclass[12pt]{iopart}
\usepackage{iopams}

\usepackage{amssymb}
\usepackage{units}
\usepackage[dvips]{graphicx}
\usepackage{multirow}

\usepackage[
            a4paper,      % A4
            ps2pdf,        % Erzeugung durch ps2pdf
            pdftitle={...},
            pdfauthor={...},
            bookmarks=true,
            bookmarksopen=true,     % Bookmarks anzeigen...
            bookmarksnumbered=true, % ...und numerieren
]{hyperref}

\newcommand{\eqref}[1]{(\ref{#1})}

\newcommand\tblspc{\rule{0pt}{2.5ex}}
\newcommand{\hlinespc}{\hline \tblspc}

\newcommand{\bfigFullPage}{\begin{figure} \begin{center} \vspace{0pt}}
\newcommand{\bfig}[1][t!]{\begin{figure}[#1] \begin{center}}
\newcommand{\efig}{\end{center} \end{figure}}
\newcommand{\btab}[1][t!]{\begin{table}[#1] \begin{center}}
\newcommand{\btabFullPage}{\begin{table} \begin{center}}
\newcommand{\etab}{\end{center} \end{table}}

\newcommand{\bq}{\begin{equation}}
\newcommand{\eq}{\end{equation}}

\newcommand{\bqq}{\begin{eqnarray}}
\newcommand{\eqq}{\end{eqnarray}}

\newcommand{\dd}{\mathrm{d}}
\newcommand{\dndeta}{\dd N_\mathrm{ch}/\dd\eta}
\newcommand{\dndy}{\dd N_\mathrm{ch}/\dd y}
\newcommand{\dndyt}{\dd N_\mathrm{ch}/\dd y_T}
\newcommand{\dndetaZero}{\dd N_\mathrm{ch}/\dd \eta|_{\eta = 0}}
\newcommand{\cms}{\sqrt{s}}

\newcommand{\cmsofT}[1]{$\sqrt{s} = \unit[#1]{TeV}$}
\newcommand{\cmsofG}[1]{$\sqrt{s} = \unit[#1]{GeV}$}
\newcommand{\expval}[1]{\langle #1 \rangle}
\newcommand{\etain}[1]{$|\eta|$~$<$~$#1$}

\newcommand{\n}{\expval{n}}
\newcommand{\N}{\expval{N}}
\newcommand{\pbar}{\bar{p}}

\newcommand{\nch}{N_\mathrm{ch}}
\newcommand{\avgnch}{\langle N_\mathrm{ch} \rangle}
\newcommand{\spps}{Sp$\bar{\mbox{p}}$S}

\newcommand{\ee}{$e^+e^-$}

\newcommand{\figref}[1]{Figure~\ref{#1}}

\mathchardef\mhyphen="2D

\begin{document}

\newcommand{\ourtitle}{Charged-Particle Multiplicity in Proton--Proton Collisions}
\review{\ourtitle}

\author{Jan Fiete Grosse-Oetringhaus$^1$, Klaus Reygers$^2$}

\address{$^1$CERN, 1211 Geneva 23, Switzerland}
%\ead{custserv@iop.org}
\address{$^2$Physikalisches Institut, Universit{\"a}t Heidelberg, Philosophenweg 12, \\69120 Heidelberg, Germany}
\address{Jan.Fiete.Grosse-Oetringhaus@cern.ch, reygers@physi.uni-heidelberg.de}
%\ead{custserv@iop.org}
\begin{abstract}
This article summarizes and critically reviews measurements of charged-particle multiplicity distributions and pseudorapidity densities in $p+p(\bar{p})$ collisions between $\sqrt{s} = \unit[23.6]{GeV}$ and $\sqrt{s} = \unit[1.8]{TeV}$. Related theoretical concepts are briefly introduced. Moments of multiplicity distributions are presented as a function of $\sqrt{s}$. Feynman scaling, KNO scaling, as well as the description of multiplicity distributions with a single negative binomial distribution and with combinations of two or more negative binomial distributions are discussed. Moreover, similarities between the energy dependence of charged-particle multiplicities in $p+p(\bar{p})$ and $e^+e^-$ collisions are studied. Finally, various predictions for pseudorapidity densities, average multiplicities in full phase space, and multiplicity distributions of charged particles in $p+p(\bar{p})$ collisions at the LHC energies of $\sqrt{s} = \unit[7]{TeV}, \unit[10]{TeV},$ and $\unit[14]{TeV}$ are summarized and compared.
\end{abstract}

\maketitle

\clearpage

\begin{footnotesize}
\tableofcontents
\end{footnotesize}

%\tableofcontents breaks headers
\markboth{\ourtitle}{\ourtitle}

\clearpage

%Style decisions:(if possible from APS styleguide)
% ,i.e.,
% space before punctuation in formulas: no!
% no space between numbers and percent 10\%

% \clearpage

\section{Introduction}

% welcome introduction
The charged-particle multiplicity is one of the simplest observables in collisions of hadrons, yet it imposes important constraints on the mechanisms of particle production. Experiments have been performed with cosmic rays, fixed target setups, and particle colliders. These measurements have been used to improve, or reject, models of particle production which are often available as Monte Carlo event generators. Considering only the number of produced charged particles is a drastic reduction of the complex information contained in the final state of a particle collision, especially if the kinematic properties are neglected. Nevertheless, the multiplicity distribution, i.e., the probability distribution of obtaining a definite number of produced particles, still contains information about particle correlations. By definition, all information about correlations is removed when the data are reduced to the average charged particle multiplicity $\langle N_\mathrm{ch}\rangle$. Distributions that still partly reflect kinematic properties are, e.g., the pseudorapidity ($\eta$) density and the transverse momentum ($p_T$) distribution which are one-dimensional projections of the kinematic properties. More sophisticated is the study of correlations of the final-state particles. Still, on a rather global level one typically studies the dependence of the average $p_T$ on the event multiplicity. More complicated correlation studies are, e.g., in the realm of Hanbury-Brown and Twiss (HBT) interferometry.

This review focuses on the charged-particle multiplicity distribution and the pseudorapidity density. Correlations will only be partly covered in the discussion of moments of the multiplicity distributions. The main topics of this review cover basic theoretical concepts and their applicability to data, experimental challenges, experimental results, as well as predictions for the Large Hadron Collider (LHC) energies. Earlier reviews can be found in \cite{DeWolf:1974ju,Giacomelli:1979nu,Carruthers:1987tv,Dremin:2000ep,Giovannini:2004yk,Kittel:2004xr,Kittel:2005fu}. The objective of this review is to give a general overview of the field, to discuss the relevant theoretical aspects, and to provide references for the reader who may want to study certain topics in more detail. From the description of collider experiments and their results the reader should obtain an understanding where the limitations of these theoretical descriptions lie and what the experimental trends as functions of centre-of-mass energy are. Furthermore, an objective is to discuss open experimental issues on which data from the LHC will provide the needed clarification. The study of the charged-particle multiplicity is an essential topic at the beginning of data-taking at the LHC. A precise characterization of the underlying event, i.e., of those particles of the event not related to the hard parton--parton scattering, is a precondition for most of the flagship research topics of the LHC.

% brief historical account
\subsection{Brief Overview of Multiplicity Measurements}
The charged-particle multiplicity is a key observable for the understanding of multi-particle production in collisions of hadrons at high energy. The probability  $P(n)$ for producing $n$ charged particles in the final state is related to the production mechanism of the particles. The multiplicity distribution follows a Poisson distribution if the final-state particles are produced independently.
In this case the dispersion $D = \sqrt{\langle n^2 \rangle - \langle n \rangle^2}$ is related to the average multiplicity as $D = \sqrt{\langle n \rangle}$. Deviations from a Poisson distribution indicate correlations. 

Measurements of multiplicity distributions provide significant constraints for particle-production models. However, the discrimination between models typically requires more differential measurements. Particle production models are typically based on Quantum Chromodynamics; however, they necessarily contain a phenomenological component as the formation of particles involves a soft scale outside the realm of perturbative techniques.

Early measurements of multiplicity distributions in $e^+e^-$ collisions at the centre-of-mass energy $\sqrt{s} = \unit[29]{GeV}$ could approximately be described with a Poisson distribution \cite{Althoff:1983ew,Derrick:1986jx}. Proton-proton collisions, on the other hand, exhibited broader multiplicity distributions. The energy dependence of the dispersion in non-single diffractive $p+p$ collisions could approximately be described as $D \propto \langle n \rangle$ up to the maximum ISR energy of $\sqrt{s} = \unit[62]{GeV}$ \cite{Breakstone:1983ns}. A simple interpretation was that the correlations in $p+p$ were related to 
an impact-parameter dependence of the charged-particle multiplicity \cite{Barshay:1973nq}.

Interest in multiplicity distributions was stimulated by the paper of Koba, Nielsen, and Olesen in 1972, in which they derived theoretically that multiplicity distributions should follow a universal scaling at high energies (KNO scaling). KNO scaling was derived based on Feynman scaling, i.e., based on the assumption that the rapidity density $\dndy$ at $y = 0$ reaches a limiting value  above a certain energy which corresponds to an asymptotic scaling of the total multiplicity  as $\langle n \rangle\propto \ln \sqrt{s}$. With $z = n/\langle n \rangle$ the function $\Psi(z) = \langle n \rangle P(n)$ was expected to asymptotically reach a universal energy-independent form.  Bubble chamber data between $\sqrt{s} \approx \unit[6]{GeV}$ and $\unit[24]{GeV}$ indicated an onset of KNO scaling already at $\sqrt{s} \approx \unit[10]{GeV}$ \cite{Slattery:1972ni}. At the ISR among other observations the relation $D \propto \langle n \rangle$ indicated that KNO scaling was satisfied \cite{Breakstone:1983ns} (although deviations were noted in \cite{Thome:1977ky}). However, it was found that the average multiplicity increased faster with energy  than $\ln \sqrt{s}$. Thus, the theoretical basis for KNO scaling was found to be empirically false.  In 1985, breaking of KNO scaling was observed by the UA5 collaboration in $p+\bar{p}$ collisions at $\sqrt{s} = \unit[540]{GeV}$ \cite{Alner:1985rj}. In a later publication UA5 concluded that KNO scaling was already violated at $\sqrt{s} = \unit[200]{GeV}$ \cite{Ansorge:1988kn}.

UA5 found that multiplicity distributions up to \cmsofG{540} can be well described by a negative binomial distribution (NBD) \cite{Alner:1985rj} which is defined by two parameters $\langle n \rangle$ and $k$. The parameter $k$ determines the width. It was found that $1/k$ increases approximately linearly with $\ln \sqrt{s}$ whereas KNO scaling corresponds to a constant, energy-independent $1/k$. However, deviations from the NBD were discovered by UA5 at \cmsofG{900} and later confirmed at the Tevatron at \cmsofG{1800} \cite{cdf_multiplicity1}. A shoulder structure appeared at $n \approx 2 \langle n \rangle$ which could not be described with a single NBD. This led to a two-component model by Giovannini and Ugoccioni in 1999 who described the measured data by a combination of two NBDs, interpreting one as a soft and one as a semi-hard component. An alternative description interpreted the results in terms of multiple-parton interactions which become more important at higher energies. The superposition of several interactions affects the multiplicity distribution and therefore potentially explains the deviation from the scaling found at lower energies.

Multiplicity measurements in $e^+e^-$ between $10 \lesssim \sqrt{s} \lesssim \unit[91.2]{GeV}$ showed that also in this system the dispersion $D$ scaled approximately linearly with $\langle n \rangle $ and that KNO scaling was approximately satisfied \cite{Abreu:1990cc}. Thus, the Poisson shape at $\sqrt{s} \approx \unit[30]{GeV}$ was merely accidental.  Also in $e^+e^-$ the NBD provided a useful description of the data. However, the Delphi experiment found that at $\sqrt{s} = \unit[91.2]{GeV}$ multiplicity distributions in restricted rapidity intervals exhibited a shoulder structure similar to the observations in $p+p(\bar{p})$ \cite{Abreu:1991yc}. This shoulder was attributed to three- and four-jet events which have larger average multiplicities than two-jet events. 

\subsection{Structure of the Review}

This review is divided into two parts. The first introduces basic theoretical concepts, the second concentrates on experimental data.

The first part discusses scaling properties of the multiplicity, i.e., Feynman and KNO scaling. We recall the definitions of various moments used in this review. Furthermore, negative binomial distributions (NBDs) and two-component models are discussed. Similarities between $p+p(\bar{p})$ and $e^+e^-$ collisions are investigated in the context of QCD.

The second part starts with the introduction of important aspects of the multiplicity analysis. It is demonstrated that measured multiplicity distributions need to be unfolded to obtain the original (true) distribution. Subsequently, measurements in the centre-of-mass energy range from $\cms = \unit[23.6]{GeV}$ to $\unit[1.8]{TeV}$ are presented. The applied analysis methods and error treatments are discussed. Selected pseudorapidity density and multiplicity distributions are shown, and their agreement with the theoretical descriptions introduced in the first part is assessed. The dependence of the multiplicity on the collision energy is analyzed. Results from hadron and lepton colliders are compared and their universalities and differences investigated. The behaviour of the moments of the multiplicity distribution as a function of $\cms$ are studied. Then we investigate how single NBDs and the combination of two NBDs describe the distributions. This part concludes with a discussion of open experimental issues.

An overview of available predictions for the LHC energy range is given in the final section of the review.

\section{Theoretical Concepts Related to the Charged-Particle Multiplicity}
\label{section_basics_multiplicity}

Before introducing various analytical descriptions of the multiplicity distribution, it is important to remark that in the case that the underlying production process can be described by uncorrelated emission the multiplicity distribution is expected to be of Poisson form. Any deviation from this indicates correlations between the produced particles. Forward--backward correlations have in fact been measured, e.g., by UA5 in $p+\pbar$ collisions \cite{Alpgard:1983xp} but are not further discussed here.

Many authors have tried to identify simple analytical forms that reproduce the multiplicity distributions at different $\cms$ requiring only a simple rescaling or changing only a few parameters as function of $\cms$. At LHC energies a regime is reached where the average collision contains multiple parton interactions whose products might undergo final-state interactions. Considering the wealth of processes that are expected at large $\cms$ it is not obvious whether approaches based on analytical forms are capable of capturing the underlying physics.

  \subsection{Feynman Scaling}

    Feynman concluded that for asymptotically large energies the mean total number of any kind of particle rises logarithmically with $\cms$ \cite{Feynman:1969ej}:
    \bq
      \N \propto \ln W \propto \ln \cms \ \ \ \mbox{with} \ \ \ W = \sqrt{s}/2.
    \eq
    His conclusions are based on phenomenological arguments about the exchange of quantum numbers between the colliding particles. He argued that the number of particles with a given mass and transverse momentum per  longitudinal momentum interval $p_z$ depends on the energy $E=E(p_z)$ as
    \bq
      \frac{\dd N}{\dd p_z} \sim \frac{1}{E}.
    \eq
    This was extended to the probability of finding a particle of kind $i$ with mass $m$ and transverse and longitudinal momentum $p_T$ and $p_z$:
    \bq
      f_i(p_T, x_F = p_z / W) \frac{\dd p_z}{E} \dd^2p_T \label{eq_feynman_d3sigma}
    \eq
    %which is the invariant cross section
    with the energy of the particle
    \bq
      E = \sqrt{m^2 + p_T^2 + p_z^2}.
    \eq
    The function $f_i(p_T, x_F)$ denotes the particle distribution. Feynman's hypothesis is that $f_i$ becomes independent of $W$ at high energies. This assumption is known as \textit{Feynman scaling} and $f_i$ is called the scaling function or Feynman function. The variable $x_F = p_z / W$, called \textit{Feynman-x}, is the ratio of the longitudinal momentum of the particle $p_z$ to the total energy of an incident particle $W$. Integration of expression~\eqref{eq_feynman_d3sigma} results in $\N \propto \ln W$. A derivation is given in Appendix~\ref{section_derivation_feynmanscaling}.

    Considering that the maximum rapidity in a collisions increases also with $\ln \cms$, it follows that:
    \bq
      \frac{\dd N}{\dd y} = \mathrm{constant},
    \eq
    i.e., the height of the rapidity distribution around mid-rapidity, the so-called plateau, is independent of $\cms$.
    Equivalently, the pseudorapidity at mid-rapidity $\dd N/\dd \eta|_{\eta = 0}$ is approximately constant if Feynman scaling holds (the pseudorapidity is defined as $\eta = (1/2) \ln [(p+p_L)/(p-p_L)] = - \ln \tan \vartheta/2$ where $p$ ($p_L$) is the total (longitudinal) momentum of the particle and $\vartheta$ the angle between the particle and the beam axis). 
    Here the transformation from $y$ to $\eta$ has to be taken into account. It depends on the average $m_T = \sqrt{m^2+p_T^2}$ of the considered particles which, however, is only weakly energy-dependent. An estimate based on the Pythia event generator shows that the ratio $(\dndy)/(\dndeta)$ changes by only 1~--~2\% from \cmsofG{100} to \unit[1]{TeV}.
    Furthermore, this transformation causes a dip in the distribution around $\eta \approx 0$ which is not present in the rapidity distribution itself
(see Section \ref{section_meas_energydependence} where measured $\dndeta$ distributions are shown).
    %a(s) = sqrt(1 - m^2/<m_T(s)^2>)
    %pion: p = 140 MeV, <p_T> 50 gev> 0.37; 900 GeV: 0.45 (anisovich)
    %a(900) - a(50) = -0.02

  \subsection{Moments}
    \label{section_theory_moments}

		To describe properties of multiplicity distributions, e.g., as a function of $\cms$, it is convenient to study their moments. All moments together contain the information of the full distribution. In practice only the first few moments can be calculated with reasonable uncertainties due to limited statistics.  The reduced $C$-moments are defined by
		\bq
		  C_q = \frac{\expval{n^q}}{\n^q} = \frac{\sum_n n^q P_n}{\left(\sum_n n P_n \right)^q} \label{eq_cmoments}
		\eq
                where $q$ is a positive integer and $P_n$ the probability for producing $n$ particles. The normalized factorial $F$-moments are defined by:
		\bq
		  F_q = \frac{\expval{n (n-1) \ldots (n-q+1)}}{\expval{n}^q}.  \label{eq_fmoments}
		\eq
		It can be shown that these moments contain in integrated form the correlations of the system of particles (see e.g. \cite{Mueller:1971ez,Dremin:2000ep}). For a Poisson distribution one obtains $F_q = 1$ for all values of $q$.
		
		The $D$-moments are defined by:
		\bq
		  D_q = \expval{(n - \n)^q}^{1/q}.  \label{eq_dmoments}
		\eq
		$D \equiv D_2$ is referred to as dispersion.
		The equations used to calculate the uncertainties of these moments are given in Appendix~\ref{section_moments_uncertainty}.

                The normalized factorial cumulants $K_q$ can be calculated recursively from the normalized factorial moments according to \cite{Mueller:1971ez,Dremin:2000ep,Giovannini:1996jz}
		\bq
		  K_q = F_q - \sum_{i=1}^{q-1} {q-1 \choose i} K_{q-1} F_i.
		\eq
                The cumulants of rank $q$ represent genuine $q$-particle correlations not reducible to the product of lower-order correlations. The $H$-moments are defined by 
		\bq
                H_q = \frac{K_q}{F_q}.
		\eq
                The $H$-moments are interesting because higher-order QCD calculations predict that these oscillate as a function of the rank $q$ \cite{Dremin:1993vq}. $K$- and $H$-moments are only briefly discussed in the following.
		For more details about the definitions of the moments see, e.g., \cite{Dremin:2000ep,DeWolf:1995pc}.

The analysis of moments helps unveil patterns and correlations in the multi-particle final state of high-energy collisions in the presence of statistical fluctuations due to the limited number of produced particles. One pattern extensively searched for experimentally was self-similar or fractal structures in multi-particle spectra \cite{DeWolf:1995pc}. The observation of fractal structures is of great interest because it imposes strong constraints on the underlying particle-production mechanism. A natural candidate for explaining self-similarity is the parton cascade \cite{Bialas:1993nw}. Bialas and Peschanski introduced the concept of intermittency to search for self-similarity in multiplicity distributions \cite{Bialas:1985jb,Bialas:1988wc}. They proposed to study the normalized factorial moments $F_q$ in decreasing pseudorapidity intervals $\delta \eta$. Self-similarity in the particle production process would then manifest itself as a power-law behaviour of the $F_q(\delta \eta)$ as a function of the bin size $\delta \eta$:
\bq
F_q(\delta \eta) \propto (\delta \eta)^{-\phi_q}.
\eq  
For more information we refer the reader to \cite{DeWolf:1995pc,Abbott:1995as,Kittel:2001gt,Kittel:2004xr}.

  \subsection{Koba--Nielsen--Olesen (KNO) Scaling}
    \label{section_theory_kno}

    \textit{KNO scaling} was suggested in 1972 by Koba, Nielsen, and Olesen \cite{Koba:1972ng}. Their main assumption is Feynman scaling.

    KNO scaling is derived by calculating
    \bqq
      \expval{n(n-1) \ldots (n-q+1)} = \nonumber \\
      \int f^{(q)} (x_1, p_{T,1}; ...; x_q, p_{T,q}) \frac{\dd p_{z,1}}{E_1} \dd p^2_{T,1} \cdots \frac{\dd p_{z,q}}{E_q} \dd p^2_{T,q}
    \eqq
    which is an extension of the expression used in the derivation of Feynman scaling (see Eq.~(\ref{eq_feynman_avgn})) that uses a function $f^{(q)}$ that describes $q$-particle correlations ($q$ particles with energy $E_q$, longitudinal momentum $p_{z,q}$, transverse momentum $p_{T,q}$, and Feynman-$x$ $x_q$). Integration by parts is performed for all $x_i$ and it is proven that the resulting function is uniquely defined by moments. This yields a polynomial in $\ln s$. With a substitution of the form $\n \propto \ln s$ the multiplicity distribution $P(n)$ is found to scale as
    \bq
      P(n) = \frac{1}{\n} \Psi(\frac{n}{\n}) + {\cal O} \left( \frac{1}{\n^2} \right), \label{eq_kno}
    \eq
    where the first term results from the leading term in $\ln s$, that is $(\ln s)^q$. The second term contains all other terms in $\ln s$, i.e., $(\ln s)^{q'}$ for $q' < q$.
    $\Psi(z := n/\n)$ is a universal, i.e., energy-independent function.
    This means that multiplicity distributions at all energies fall on one curve when plotted as a function of $z$. However, $\Psi(z)$ can be different depending on the type of reaction and the type of measured particles.

    The $C$-moments, 
    \bq
      C_q = \int_0^\infty z^q \Psi(z) \, \dd z, \label{eq_knoscaling_reducedmoments}
    \eq
    define $\Psi(z)$ uniquely \cite{Koba:1972ng}.
    Substituting $z = n/\n$ results in Eq.~\eqref{eq_cmoments}.
    % can be concluded from (3.2) in \cite{Koba:1972ng} with substitution

When the scaling hypothesis holds the moments are independent of energy. Experimentally one can determine $D^2 = \expval{n^2} - \n^2$; the relation $D/\n = const.$ follows from Eq.~(\ref{eq_kno}) (if $\Psi(z)$ is not a $\delta$ function, see \cite{Koba:1972ng}).

     It has been pointed out \cite{Zajc:1986pn} that the conclusion that the multiplicity distribution follows a universal function is only an approximation (neglecting the second term in Eq.~\eqref{eq_kno}). Therefore the exact result is that the factorial moments (see Eq.~\eqref{eq_fmoments}) are required to be constant, not the reduced moments (which follow from Eq.~\eqref{eq_knoscaling_reducedmoments}). This is addressed further in Section~\ref{section_meas_moments}.

     The description of discrete data points with a continuous function in Eq.~\eqref{eq_kno} is an approximation valid for $\langle n \rangle \gg 1$. A generalized KNO scaling which avoids this problem is described in \cite{Golokhvastov:1977tz, Szwed:1985bs}. Moreover, different scaling laws for multiplicity distributions were proposed (see e.g. \cite{Dremin:2000ep, Kittel:2004xr}), among those the so-called log-KNO scaling \cite{Hegyi:1999aa} which predicts a scaling of the form
     \bq
     P(n) = \frac{1}{\lambda(s)} \varphi \left( \frac{\ln n + c(s)}{\lambda(s)}\right) \label{eq_log_kno}
     \eq
where $\varphi$ is a universal, energy-independent function. The energy-dependent functions $\lambda(s)$ and $c(s)$ correspond to $\langle n \rangle$ and the multiplicity related to the leading particles, respectively.

  \subsection{Negative Binomial Distributions}
    \label{section_theory_nbd}

    % compared to http://en.wikipedia.org/wiki/Negative_binomial_distribution
    % p --> 1 / (1 + <n>/k)
    % k --> n
    % r --> k
    % NOTE: in binomial lower value can be either k - 1 OR n (equivalent!)

    The \textit{Negative Binomial Distribution (NBD)} is defined as
    \bq
      P^{\rm NBD}_{p, k}(n) = \left ( \begin{array}{c} n+k-1 \\ n \\ \end{array} \right ) (1-p)^n \  p^k.
    \eq
    It gives the probability for $n$ failures and $k-1$ successes in any order before the $k$'th success in a Bernoulli experiment with a success probability $p$. The NBD is a Poisson distribution for $k^{-1} \rightarrow 0$ and a geometrical distribution for $k = 1$. For negative integer $k$ and $\langle n \rangle \le -k$ the distribution is a binomial distribution where $-k$ is the number of trials and $-\langle n \rangle/k$ the success probability (see Appendix~\ref{relation_nbd_bd}). The continuation to negative integer $k$ is performed by writing the binomial in terms of the $\Gamma$ function and using the equation $\Gamma(x+1) = x \Gamma(x)$:
    \bqq
      \left ( \begin{array}{c} n+k-1 \\ n \\ \end{array} \right ) = \frac{(n+k-1)!}{n!(k-1)!} =
        \frac{\Gamma(n+k) }{\Gamma(n+1) \Gamma(k)} \nonumber \\
        = \frac{(n+k-1)\cdot(n+k-2)\cdot \ldots \cdot k}{\Gamma(n+1)}. \label{eq_nbd_gamma}
    \eqq
    The mean of the distribution $\n$ is related to $p$ by $p^{-1} = 1 + \n/k$. This leads to the form of the NBD that is commonly used to describe multiplicity distributions \cite{Alner:1985zc, Alner:1985wj}:
    \bq
      P^{\rm NBD}_{\n, k}(n) = \left ( \begin{array}{c} n+k-1 \\ n \\ \end{array} \right )
        \left ( \frac{ \n / k }{ 1 + \n/k } \right)^n \frac{1}{(1 + \n/k)^k}.
    \eq
    % from Alner:1985zc, but identical to parameterizations used in NBD Giovannini, hove
    \bfig
      \includegraphics[width=0.48\linewidth]{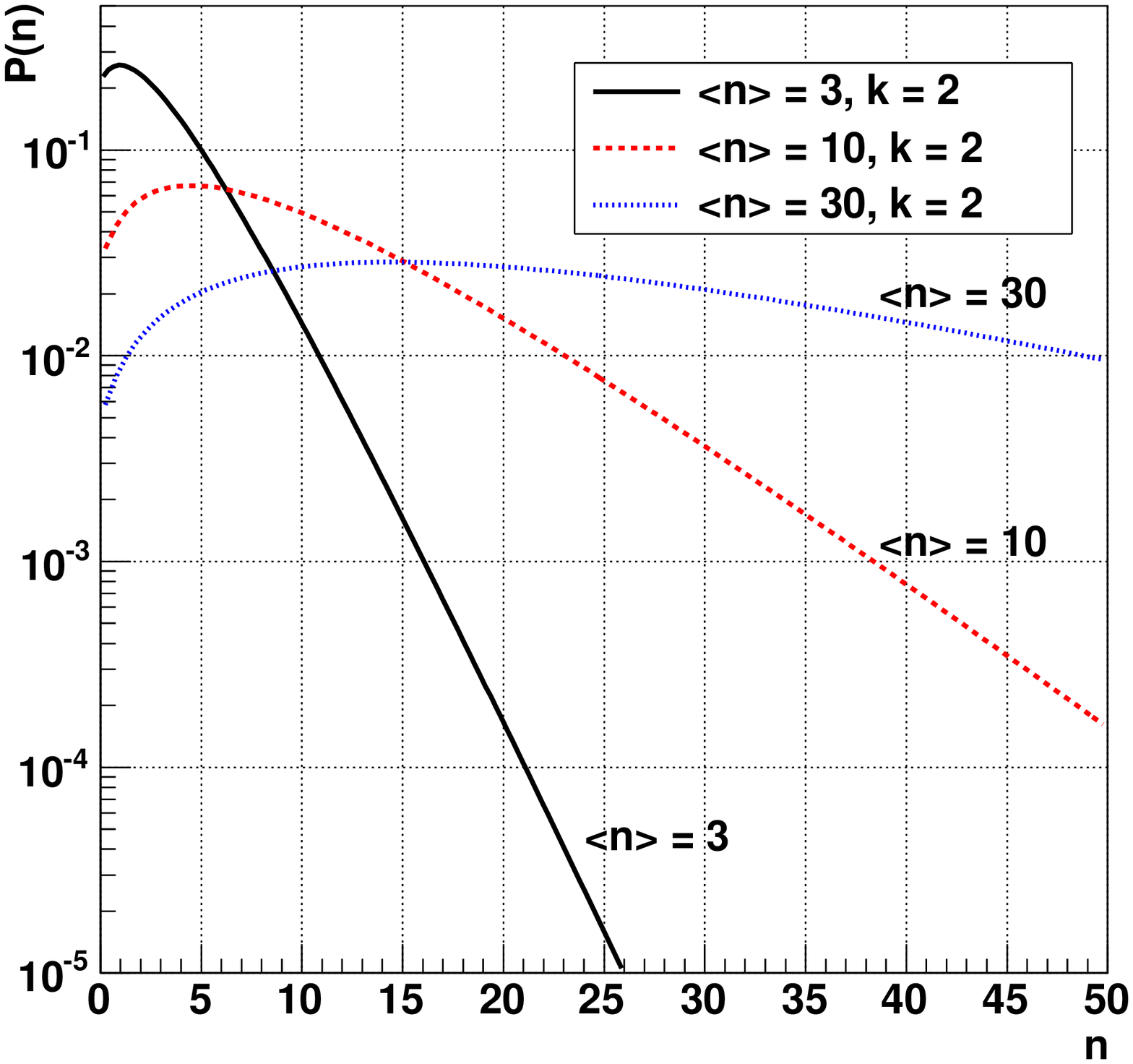}
      \hfill
      \includegraphics[width=0.48\linewidth]{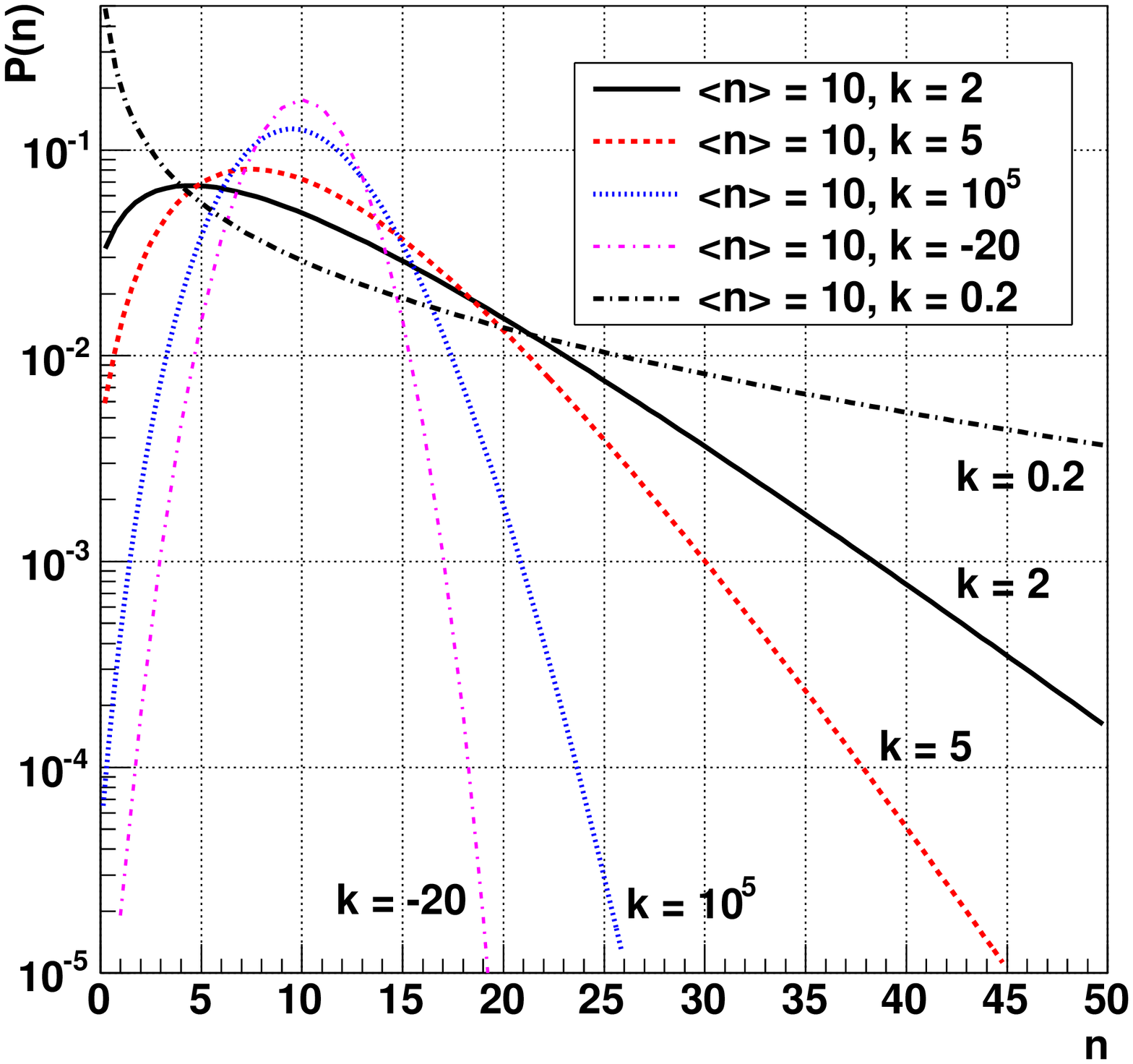}
      \caption{Examples of negative binomial distributions.}
      \label{fig_nbd_example}
    \efig
The dispersion $D$ and the second-order normalized factorial moment $F_2$ of the NBD are given by
    \bq
    D = \sqrt{\n \left(1 + \frac{\n}{k}\right)}; \quad \quad
    F_2 = 1 + \frac{1}{k}.
    \eq
    Moreover, $k$ is related to the integral of the two-particle pseudorapidity correlation function $C_2(\eta_1,\eta_2)$ as shown in \cite{Giovannini:1985mz, DeWolf:1990js}. 

    Figure~\ref{fig_nbd_example} shows normalized NBDs for different sets of parameters. NBDs for values of $k$ that lead to characteristic shapes are also shown: the case of a large $k$ where the distribution approaches a Poisson distribution is shown, the case with a negative integer $k$ where the function becomes binomial, and the case of $k$ being positive and smaller than unity.
    $P^{\rm NBD}_{\n, k}(n)$ follows KNO scaling if $k$ is constant (energy-independent).
   This can be seen from the KNO form
   \bq
   \label{eq_kno_form_nbd}
   \Psi_\mathrm{NBD}(z) = \frac{k^k}{\Gamma(k)} z^{k-1} e^{-kz}
   \eq
   which holds in the limit $\langle n \rangle/k \gg 1$ \cite{Carruthers:1987tv}.
    Therefore, studying $k$ as a function of $\cms$ for multiplicity distributions described by NBDs indicates whether KNO scaling is fulfilled. NBDs have been shown to provide a useful parameterization of multiplicity distributions in $p+p(\pbar)$ collisions as well as in various other systems including $e^+e^-$ \cite{Derrick:1986jx,Buskulic:1995xz}, $\mu+p$ \cite{Arneodo:1987qy} and central nucleus-nucleus collisions \cite{Arneodo:1987qy,Abbott:1995as}. 
    However, the NBD has been shown to underestimate particle correlations found in $e^+e^-$ data, which can be shown by studying factorial moments and cumulants \cite{Sarkisian:2000ux, Abbiendi:2001bu}.
    
    The physical origin of a multiplicity distribution following a negative binomial form has not been ultimately understood.
    However, one approach is to use the recurrence relation of collisions of multiplicities $n$ and $n+1$ \cite{Giovannini:1985mz}.
    This relation is defined such that for uncorrelated emission it is constant; any departure shows the presence of correlations.
    Evaluating
    \bq
      g(n) = \frac{(n + 1) P(n+1)}{P(n)} \label{nbd_study1}
    \eq
    for a Poisson distribution  $P(n) = \lambda^n e^{-\lambda} / n!$ (representing uncorrelated emission) yields the result that $g(n) \equiv \lambda$ is indeed constant.
    The term $n+1$ in Eq.~\eqref{nbd_study1} can be understood by considering that the particles are in principle distinguishable, e.g., by their momenta; therefore it has to be taken into account that a collision of multiplicity $n+1$ can be related to $n+1$ collisions of multiplicity $n$ (by removing any single one of the $n+1$ particles).

    For NBDs, Eq.~(\ref{nbd_study1}) can be written as
    \bqq
      g(n) = a + bn \label{nbd_study2} \ \ && \mbox{ with }\ \  k = a/b \mbox{ and } \n = a/(1-b)  \\
			&& \mbox{ or }\ \ a = \n k/(\n + k) \mbox{ and } b = \n/(\n + k). \nonumber
    \eqq
    %$a$ and $b$ can be measured experimentally.
    %[plot in the ref.]
    A model of partially stimulated emission identifies $a$ in Eq.~(\ref{nbd_study2}) with the production of particles independent of the already present particles and $bn$ with emission that is enhanced by already present particles (Bose--Einstein interference). From $g(n) = a(1+n/k)$ (which follows from Eq.~\eqref{nbd_study2}) one sees that $k^{-1}$ is the fraction of the already present particles $n$ stimulating emission of additional particles. Following these rather simple assumptions results in two conclusions that are confirmed experimentally: 1) $k$ increases when the considered $\eta$-interval is enlarged (because the range of Bose--Einstein interference is finite, the fraction of present particles stimulating further emission reduces); 2) $k$ decreases with increasing $\cms$ for a fixed $\eta$-interval (the density of particles in the same interval increases because $\n$ increases) \cite{Giovannini:1985mz}.

    The multiplicity distribution can be deduced as being of negative binomial shape within the so-called \emph{clan model} \cite{Giovannini:1985mz,Giovannini:1987tw,Giovannini:2004yk}. It describes the underlying production by a cascading mechanism.
    In the clan model a particle can emit additional particles, e.g., by decay and fragmentation. A clan (or cluster) contains all particles that stem from the same ancestor.  The ancestors themselves are produced independently.

    The production of ancestors, and thus clans, is governed by a Poisson distribution $P(N, \N)$ where $\N$ is the average number of produced clans. 
    The probability of producing $n_c$ particles in one clan $F_c(n_c)$ can be determined from the following considerations: One stipulates that without particles there is no clan:
    \bq
      F_c(0) = 0
    \eq
    and assumes that the production of an additional particle in a clan is proportional to the number of already existing particles with some probability $\tilde{p}$ (see also Eq.~\eqref{nbd_study1}):
    \bq
      \frac{(n_c+1) F_c(n_c+1)}{F_c(n_c)} = \tilde{p} n_c. \label{eq_nbd_recurrence}
    \eq
    By iteration, the following expression is obtained:
    \bq
      F_c(n_c) = F_c(1) \frac{\tilde{p}^{n_c-1}}{n_c}.
    \eq
    The multiplicity distribution that takes into account the distribution of clans and the distribution of particles among the different clans is:
    \bq
      P(n) = \sum_{N=1}^{n} P(N, \N) {\sum}^* F_c(n_1) F_c(n_2) ... F_c(n_N), \label{eq_nbd_clan}
    \eq
    where $\sum^*$ runs over all combinations $n_i$ for which $n = \sum_{i=1}^{N} n_i$ is valid.
    It can be shown that Eq.~\eqref{eq_nbd_clan} is a NBD where
    $\n = \N F_c(1)/(1-\tilde{p})$ and $k = \N F_c(1) / \tilde{p}$ \cite{Giovannini:1985mz}.
    %unknowns: F_c(1), \tilde{p}, <N>
    % <n> = <N>*<n_c> (Eq.~17) = <N>*F_c(1)/(1-\tilde{p}) (Eq.~15, 14)
    % k = a/b = <N>Fc(1)/\tilde{p} (Eq.~14)
    The average number of clans $\langle N \rangle$ and the average multiplicity $\langle n_c \rangle$ within a clan, in turn, are related to the NBD parameters $\langle n \rangle$ and $k$ via \cite{Giovannini:1985mz}
    \bq
    \langle N \rangle = \frac{\langle n\rangle}{\langle n_c\rangle} = k \ln \left(1 + \frac{\langle n\rangle}{k}\right).
    \eq

    For the case of $n=2$ it can be shown using Eqs.~\eqref{eq_nbd_recurrence} and \eqref{eq_nbd_clan} that $k$ is the relative probability of obtaining one clan with two particles with respect to obtaining two clans with one particle each.

  \subsection{Two-Component Approaches}

    \subsubsection{Combination of two NBDs}
      \label{section_twocomponent_2nbds}

      Multiplicity distributions measured by UA5 have been successfully fitted with a combination of two NBD-shaped components \cite{ua5_2nbdfits}. A systematic investigation has been performed by Giovannini and Ugoccioni who interpret the two components as a soft and a semi-hard one \cite{Giovannini:1998zb}.
      These can be understood as events with and without minijets, respectively (the authors of \cite{Giovannini:1998zb} use a definition from the UA1 collaboration: a minijet is a group of particles having a total transverse energy larger than 5 GeV): the fraction of semi-hard events found corresponds to the fraction of events with minijets seen by UA1. It is important to note that this approach combines two classes of events, not two different particle-production mechanisms in the same event. Therefore, no interference terms have to be considered and the final distribution is the sum of the two independent distributions.

      In this approach, the multiplicity distribution depends on five parameters, that may all be $\cms$ dependent:
      \bqq
          P(n) =  \alpha_\mathrm{soft} \times P^{\rm NBD}_{\n_\mathrm{soft}, k_\mathrm{soft}}(n) 
          + (1 - \alpha_\mathrm{soft}) \times P^{\rm NBD}_{\n_\mathrm{semi\mhyphen{}hard}, k_\mathrm{semi\mhyphen{}hard}}(n). \label{eq_twocomponent}
      \eqq
      The parameters and their dependence on $\cms$ are found by fitting experimental data.
      Note that $\n$ is about two times larger in the semi-hard component than in the soft component.
      Furthermore, the fits show that the soft component follows KNO scaling, whereas the semi-hard component violates KNO scaling. This is discussed in more detail in Section~\ref{section_meas_twonbd}.
      
      A modified formulation of this approach includes a third component representing events initiated by hard parton scattering. This class is also of NBD form with the parameter $k$ being smaller than 1 resulting in a substantially different shape (see \figref{fig_nbd_example}). Furthermore, the parameter $\n$ is much larger than for the other two components. For more details see \cite{Giovannini:2003ft}.

\subsubsection{Interpretation in the Framework of Multiple-Parton Interactions}
\label{section_twocomponent_doubleparton}
Above ISR energies parton-parton interactions with high momentum transfer (i.e. hard scatterings) are expected to contribute significantly to the total charged-particle multiplicity in $p+p(\bar p)$ collisions \cite{Gaisser:1985jb, Sjostrand:1987su, Wang:1990qp}. Hard parton-parton scatterings resulting in QCD jets above a transverse momentum threshold can be described by perturbative QCD. Softer interactions either require a recipe for the regularisation of the diverging QCD jet cross section for $p_T \rightarrow 0$ \cite{Sjostrand:1987su} or models for soft-particle production, see e.g. \cite{Wang:1990qp}. The transverse momentum scale $p_{T,0}$ which controls the transition from soft to hard interactions is typically around $\unit[2]{GeV}/c$. In these QCD-inspired models two or more independent hard parton-parton scatterings frequently occur within the same $p+p(\bar p)$ collision \cite{Sjostrand:1987su, Sjostrand:2004pf}. These models explain many observed features of these collisions including the increase of the total inelastic $p+p(\bar{p})$ cross section with $\sqrt{s}$, the increase of $\langle p_\mathrm{T} \rangle$ with the charged particle multiplicity $N_\mathrm{ch}$,  the increase of $\langle p_\mathrm{T} \rangle$ with $\sqrt{s}$, and the increase of $\mathrm{d}N_\mathrm{ch}/\mathrm{d}\eta$ with $\sqrt{s}$.  High-multiplicity collisions in these models are collisions with a large number of minijets. In the minijet model of ref. \cite{Wang:1990qp} up to eight independent parton-parton scatterings are expected to significantly contribute to the high-multiplicity tail of the multiplicity distribution at $\sqrt{s} = \unit[1.8]{TeV}$. The violation of KNO scaling within this model is also attributed to the onset of minijet production. Strong correlations between multiple parton interactions and the shape of the multiplicity distribution are also present in the Pythia event generator \cite{Sjostrand:2006za}. In Pythia the multiplicity distribution turns out to be strongly related to the density profile of the proton \cite{Sjostrand:1987su, Sjostrand:2004pf}. A purely analytic model based on multiple parton interactions is the  IPPI model \cite{Dremin:2004ts}. In this model the multiplicity distribution is a superposition of negative binomial distributions where each NBD represents the contribution of collisions with a given number of parton-parton scatterings.

      A  data-driven approach to define and identify double parton interactions and thus a second component in the multiplicity distribution is given in \cite{Alexopoulos:1998bi} where the multiplicity distribution is plotted in a KNO-like form and the part of the distribution for which KNO scaling holds is subtracted. This is done by comparing the distribution to a KNO fit that is valid at ISR energies. The KNO-like variable $z' = n/\expval{n_1}$, with $\expval{n_1}$ being the average multiplicity of the part of the distribution that follows KNO scaling, is used.
      Due to the large errors in the low-multiplicity bins of the specific data set at \unit[1.8]{TeV} analyzed in \cite{Alexopoulos:1998bi}, $\expval{n_1}$ cannot be satisfactorily determined. Therefore, it is found by using the empirical relation $\expval{n_1} \approx 1.25\ n_\mathrm{max}$ where $n_\mathrm{max}$ is the most probable multiplicity which is inferred from the KNO fit at ISR energies.
      %An excess at higher multiplicities is noted, proving breaking of KNO scaling.
      The authors find an interesting feature when the part that follows the KNO fit is subtracted and the remaining part is plotted (not shown here). The remaining part does not follow KNO scaling, its most probable value $z'_\mathrm{max}$ is 2, and its width is about $\sqrt{2}$ times the width of the KNO distribution. 
      % the kno variable is n/<n1> here where n1 is the mult of the part that follows kno scaling

This procedure to identify the second component is similar to the one described in the previous section.
 However, the authors of \cite{Alexopoulos:1998bi} conclude that the second part of the distribution is the result of two independent parton--parton interactions within the same collision.
      The cross sections of the two contributions ($\sigma_1, \sigma_2$) can be calculated as a function of $\cms$. 
      It is found that $\sigma_1$ is almost independent of $\cms$, whereas $\sigma_2$ increases with $\cms$. However, it remains unclear if two parton--parton interactions in the same collision evolve independently to their final multiplicity due to final-state interactions.

      The same reasoning and data are used in \cite{Matinyan:1998ja} to identify a third component, three independent parton--parton interactions. This is extended in \cite{Walker:2004tx} to predict a multiplicity distribution for the LHC design energy of \unit[14]{TeV} which is shown in Section~\ref{sec_predictions}.

\subsection{Similarities between \texorpdfstring{$p+p(\bar{p})$ and $e^+e^-$}{p+p(pbar) and e+e-} Collisions and QCD Predictions}
\label{sec_pp_ee_similarities_and_qcd}
The theoretical description of the formation of hadrons necessarily involves a soft scale so that perturbative QCD cannot be directly applied. Therefore, models for soft interactions, like those from the large class of string models, are often used to describe multiplicity distributions in collisions of hadrons (see for example the Dual Parton Model \cite{Capella:1992yb} or the Quark--Gluon String Model \cite{Kaidalov:2003au}). It is instructive to compare multi-particle production in $p+p(\bar{p})$  and $e^+e^-$  collisions. In $p+p(\bar{p})$ collisions without a hard parton-parton interaction, as well as in $e^+e^-$ collisions, particle production can be viewed as resulting from the fragmentation of colour-connected partons. In $e^+e^-$ collisions the colour field extends along the jet axis whereas in $p+p(\bar{p})$ it stretches along the beam axis. Based on this analogy it is not unreasonable to expect some similarities between particle production in $p+p(\bar{p})$ and $e^+e^-$ collisions. However, the configurations of the strings in the two cases are different. In addition, processes with different energy dependences will contribute significantly to the overall particle multiplicity at high energies \cite{Kittel:2004xr}: minijet production in hard parton-parton scattering in the case of $p+p(\bar{p})$ collisions and hard gluon radiation in $e^+e^-$ collisions. Thus, theory does not provide convincing arguments for this simple analogy, yet striking similarities were indeed observed (see Section~\ref{section_universality}). 

In $e^+e^-$ collisions moments of the multiplicity distributions are rather well described in an analytical form with perturbative QCD in the modified leading logarithmic approximation (MLLA) \cite{Khoze:1996dn}. The evolution of the parton shower is described perturbatively to rather low virtuality scales close to the hadron mass. Using the hypothesis of local parton-hadron duality (LPHD) one then assumes a direct relation between parton and hadron multiplicities. In $e^+e^-$ collisions the next-to-leading-order (NLO) prediction for the average multiplicities is given by
\bq
\label{eq_nch_ee_qcd}
\langle N_\mathrm{ch}(\sqrt{s}) \rangle = A_\mathrm{LPHD} \cdot \alpha_s^b(\sqrt{s})
\cdot \exp \left( \frac{a}{\sqrt{\alpha_s(\sqrt{s})}}\right) + A_0
\eq
with $a =  \sqrt{6 \pi} 12/23$ and $b = 407/838$ for 5 quark flavors \cite{Ellis:1991qj, Biebel:2001ka, Dissertori:2003}. Fixing the  strong coupling constant $\alpha_s$ at the $Z$ mass to $\alpha_s(M_Z^2) = 0.118$ leaves $A_0$ and $A_\mathrm{LPHD}$ as free parameters. An excellent parameterization of the experimental multiplicities can be obtained in this way (see Section \ref{section_universality}). An analytical form at next-to-next-to-next-to-leading order (3NLO) is available in \cite{Dremin:2000ep, Heister:2003aj}.

The second factorial moment for multiplicity distributions in $e^+e^-$ collisions
\bq
\label{eq_r2}
F_2 = \frac{\langle n(n-1)  \rangle}{\langle n \rangle^2} = 1 + \frac{D^2}{\langle n \rangle^2} - \frac{1}{\langle n \rangle} \;,
\eq
is given at NLO  by \cite{Malaza:1985jd}
\bq
\label{eq_r2_ee_qcd}
F_2(\sqrt{s}) = \frac{11}{8} (1 - 0.55 \sqrt{\alpha_s(\sqrt{s})}) \;.
\eq
This QCD prediction for $F_2$ is about $10\%$ above the experimental values for $\sqrt{s} = \unit[10-91.2]{GeV}$ \cite{Barate:1996fi}.
The calculation of higher moments shows that the theoretical multiplicity distributions in $e^+e^-$ collisions are well approximated by negative binomial distributions \cite{Malaza:1985jd} with
\bq
\label{eq_nbd_k_from_qcd}
1/k \approx 0.4 - 0.88 \sqrt{\alpha_s}  \;.
\eq
This implies that asymptotically ($\alpha_s \rightarrow 0$ as $s \rightarrow \infty$) the multiplicity distributions in $e^+e^-$ collisions satisfy KNO scaling. However, the KNO form of the multiplicity distributions up to the maximum LEP energy (corresponding to $\alpha_s \gtrsim 0.1$) differs significantly from the asymptotic form.

Even before QCD was known, Polyakov found that KNO scaling occurs naturally in a picture of  hadron production in a self-similar scale-invariant branching process \cite{Polyakov:1970,Hegyi:2000sp}.  For $e^+e^-$ collisions the KNO form was given as
\bq
\label{eq_kno_form_polyakov}
\psi(z) \propto a(z) \exp(- z^\mu) \quad \mathrm{with} \quad \mu > 1 \;,
\eq
where $a(z)$ is a monomial. Thus, $\psi(z)$ is a gamma distribution in $z^\mu$. In the double logarithmic approximation (DLA) of QCD, valid at asymptotic energies, the KNO form of the multiplicity distribution in jets can be calculated \cite{Bassetto:1979nt, Dokshitzer:1991wu}.  Higher-order corrections to this form were found to be large \cite{Dokshitzer:1993dc} so that the preasymptotic distributions, e.g., at LEP energies, are quite different from the asymptotic DLA form \cite{Ochs:1994gm}.

\section{Charged-Particle Multiplicity Measurements}
	\label{chapter_beforelhc}

	This part of the review presents $p+p(\pbar)$ measurements that have been performed by experiments at hadron colliders, i.e., the ISR, \spps, and Tevatron. The Intersecting Storage Rings (ISR), the very first hadron collider, was operating at CERN between 1971 and 1984. It collided $p$ on $p$, $\pbar$, and $\alpha$-particles at a maximum centre-of-mass energy of \unit[63]{GeV}. The Super Proton Synchrotron (SPS) which has operated at CERN since 1976 has accelerated in its lifetime electrons, positrons, protons, anti-protons, and ions. After modifications it operated as a collider and provided $p$ on $\pbar$ collisions with a maximum $\cms$ of \unit[900]{GeV}, at that time it was called \spps. The Tevatron at the Fermi National Accelerator Laboratory (FNAL) came into operation in 1983. It provides $p+\pbar$ collisions at energies up to \cmsofT{1.96}. In addition, results from bubble chamber experiments are included where appropriate.

   References of experimental measurements at these colliders are given, discussing their analysis methods and error treatments. A selection of measurements of experiments at these colliders is shown to assess the validity of the models that have been described above. Additionally, the experimental challenges are recalled and unresolved experimental inconsistencies are discussed.

\subsection{Analysis Techniques}
    	
    \subsubsection{Event Classes}
      \bfig
        \includegraphics[width=\linewidth]{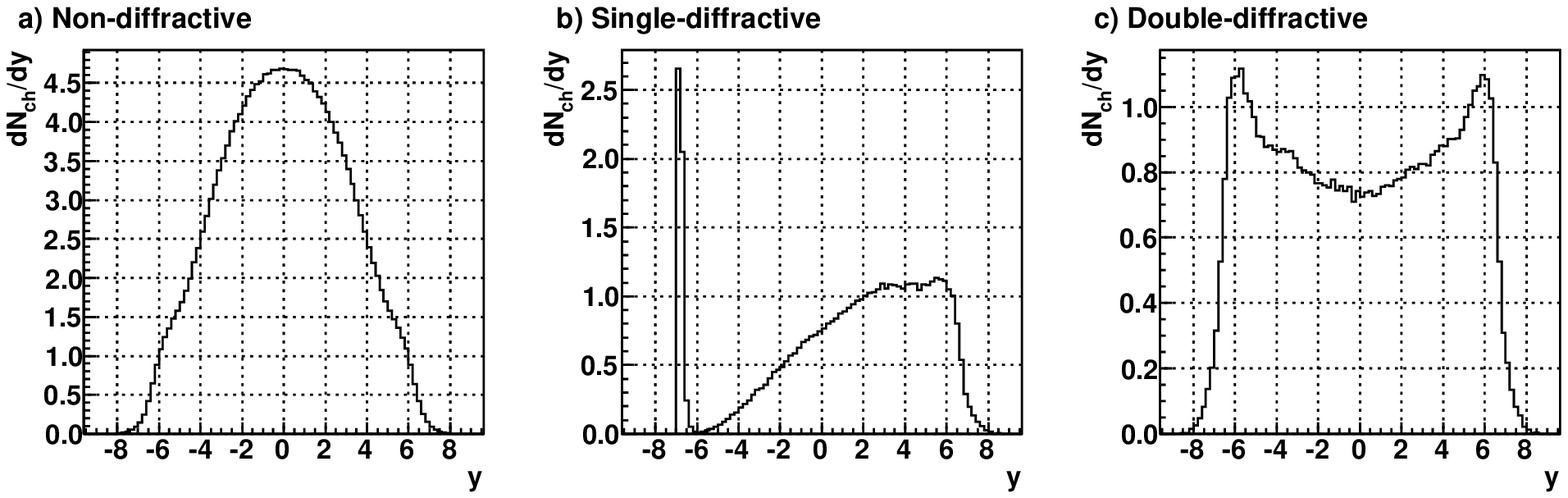}
        \caption{Rapidity distributions of charged particles per event for different processes, non-diffractive (left panel), single-diffractive (centre panel), and double-diffractive (right panel). These have been obtained with Pythia at \cmsofG{900}.}
        \label{fig_processtypes}
        %PWG0/MCTask show.C
      \efig

      Inelastic $p+p$ collisions are commonly divided into non-diffractive (ND), single-diffractive (SD), and double-diffractive (DD) events. \figref{fig_processtypes} shows rapidity distributions of those classes obtained with Pythia to illustrate their differences. Non-diffractive collisions (left panel) have many particles in the central region, with their yield steeply falling towards higher rapidities.
		  In a single-diffractive collision only one of the beam particles breaks up and produces particles at high rapidities on one side. In the centre panel only those single-diffractive collisions are shown where the particle going to positive $y$ breaks up. The other incoming particle, still intact and with only slightly altered momentum, is found near the rapidity of the beam.
		  In a double-diffractive collision (right panel) both beam particles break up and produce particles. A dip can be seen in the central region. The different scales of the three distributions should be noted. Integrating the histograms demonstrates that the average total multiplicity is about a factor of four higher in non-diffractive collisions than in diffractive collisions.
		
		  Measurements are usually presented for the sample of all inelastic collisions or non-single-diffractive (NSD) collisions, i.e., not considering the SD component. The reason for the latter choice is that trigger detectors are usually less sensitive to SD events due to their topology: few particles are found in the central region and only the incident proton is found on one side. To select a pure NSD sample for the analysis, depending on the detector geometry, SD events that pass the trigger can be rejected by their reconstructed topology, e.g., to reject events where in one hemisphere no track or only one track is found that has 80\% of the incident proton momentum. Collider detectors operating today have limited phase space acceptance at higher rapidities. Therefore they allow only a limited event-by-event decision of the occurred process and rely on Monte Carlo simulations for the subtraction of SD events. Naturally a larger systematic uncertainty is associated with this correction method.

    \subsubsection{Unfolding of Multiplicity Distributions}
      \label{section_analysis_unfolding}

			\bfig
				\includegraphics[width=\textwidth]{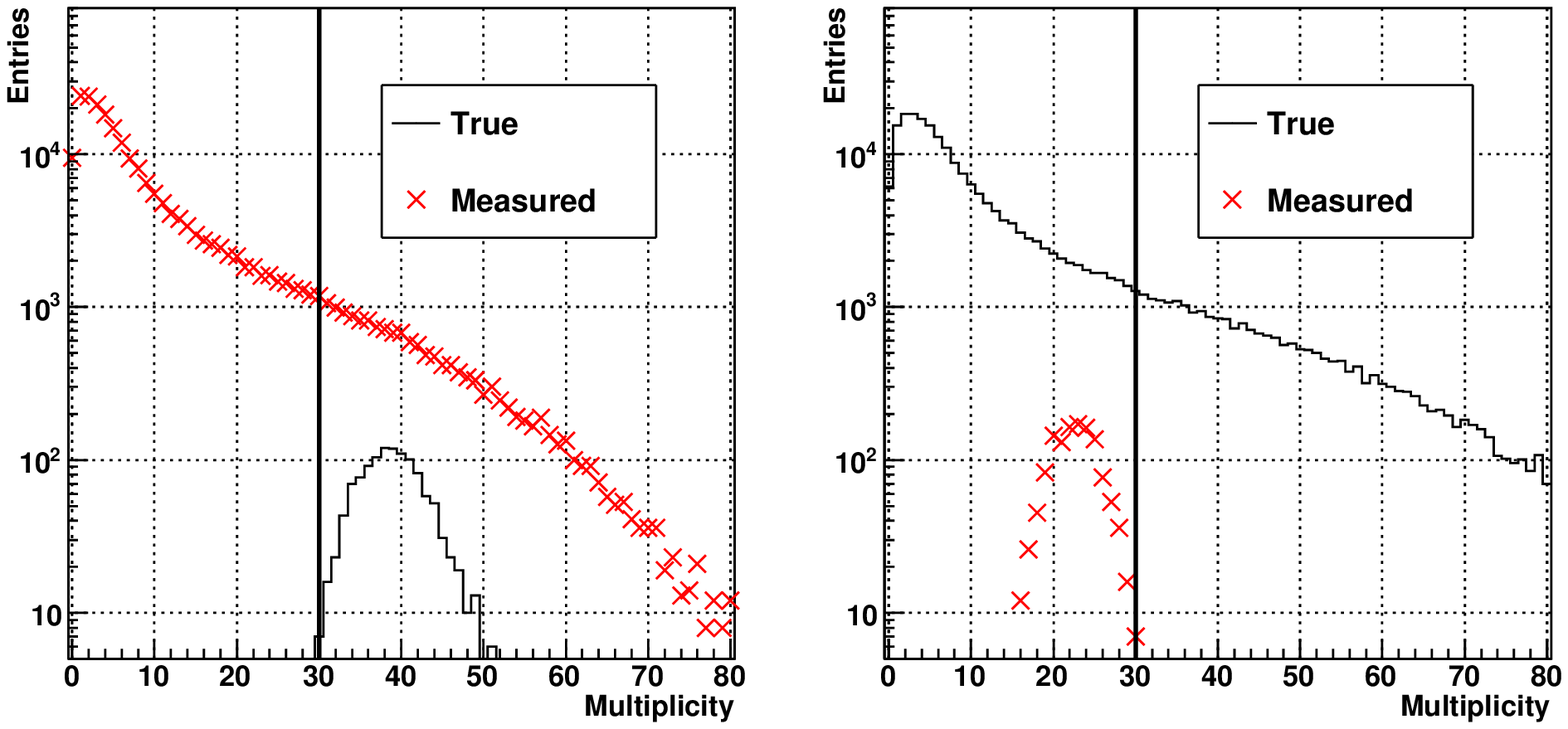}
				\caption{The need for unfolding. The left panel shows a measured spectrum in a limited region of phase space superimposed with the true distribution that caused the entries in one single measured bin (exemplarily at multiplicity 30 indicated by the line). Clearly the shape of this true distribution depends on the shape of the multiplicity distribution given by the model used (a suggestive example is if the true spectrum stopped at a multiplicity of 40: the true distribution that contributed to the measured multiplicity of 30 would clearly be different, still events at a multiplicity of 30 would be measured). Inversely, in the right panel, the true distribution is shown superimposed with the measured distribution caused by events with the true multiplicity 30 (exemplarily). The shape of this measured distribution still depends on the detector simulation, i.e., the transport code and reconstruction, but \textbf{not} on the multiplicity distribution given by the model (only events with multiplicity 30 contribute to the shown measured distribution).}
				\label{fig_mult_ModelDependencyPlot}
				% plots.C: ModelDependencyPlot()
			\efig

			Given a vector $T$ representing the true spectrum, the measured spectrum $M$ can be calculated using the detector response matrix $R$:
			\bq
				M = R T. \label{eq_mult_matrixequ}
			\eq
			The aim of the analysis is to infer $T$ from $M$. Simple weighting, i.e., assuming that a measured multiplicity $m$ is caused `mostly' by a true multiplicity $t$, would not be correct. This is illustrated in Figure~\ref{fig_mult_ModelDependencyPlot}. Analogously, adding for each measured multiplicity the corresponding row of the detector response matrix to the true distribution is also incorrect. This is model-dependent and thus may produce an incorrect result. On the other hand the measured spectrum which is the result of a given true multiplicity is only determined by the detector simulation and is independent of the assumed spectrum.
			
			Given a measured spectrum, the true spectrum is formally calculated as follows:
			\bq
				T = R^{-1} M. \label{eq_mult_inversion}
			\eq
			$R^{-1}$ cannot be calculated in all cases, because $R$ may be singular; e.g. when a poor detector resolution causes two rows of the matrix to be identical. This can in most cases be solved by choosing a more appropriate binning (combining the entries in question). Even if $R$ can be inverted, the result obtained by Eq.~(\ref{eq_mult_inversion}) contains usually severe oscillations (due to statistical fluctuation caused by the limited number of measured events and events used to create the response matrix). The effect of  the limited number of measured events can be illustrated with the following example \cite{Blobel:1984ku}: a square response matrix is assumed to describe the detector (rows: measured multiplicities; columns: true multiplicities):
			\bq
			  R = \left(
			    \begin{array}{ccccc}
			      0.75 & 0.25 & 0    &      &   \cdots \\
			      0.25 & 0.50 & 0.25 & 0    &   \\
			      0    & 0.25 & 0.50 & 0.25 &   \\
			           & 0    & 0.25 & 0.50 &   \\
			      \vdots  &      &      &      & \ddots \\
			    \end{array}
			  \right).
			\eq
			A true distribution $T$ is assumed, and the expected measured distribution $M$ is calculated using Eq.~(\ref{eq_mult_matrixequ}). The distribution $M$ is used to generate a sample of 10\,000 measurements: $\tilde{M}$. Using Eq.~(\ref{eq_mult_inversion}) the corresponding true distribution $\tilde{T}$ is calculated. Figure~\ref{fig_mult_BlobelUnfoldingExample} shows these four distributions. Although the resolution effect on the shape of the measured distribution (left histogram) is very small, the unfolded solution (right histogram) suffers from large non-physical fluctuations. Clearly, this is not the spectrum that corresponds to the true one.
			
			\bfig
			  \includegraphics[width=\textwidth]{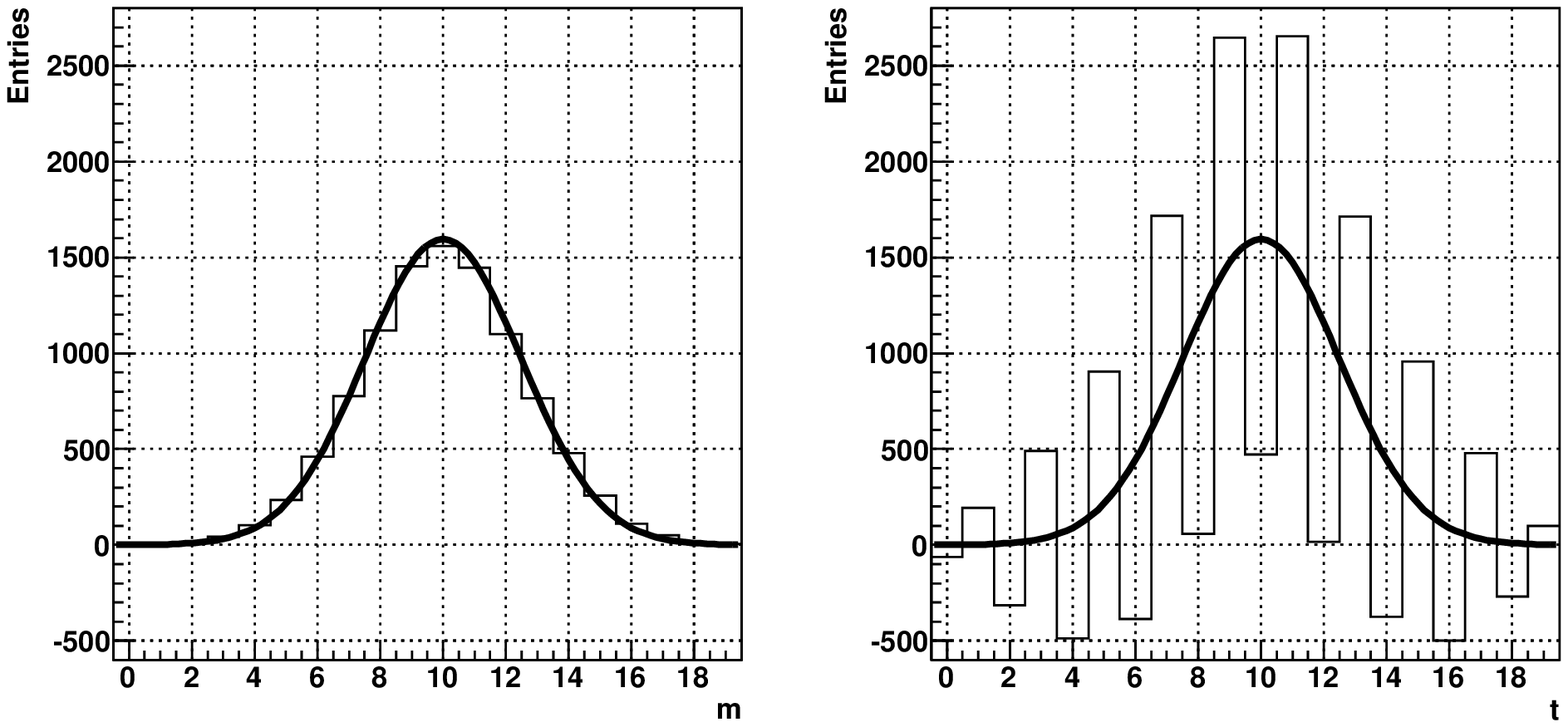}
			  \caption{Illustration of the problem with simple matrix inversion. The left panel shows a sample of the measured distribution $\tilde{M}$ with 10\,000 entries (histogram). Using Eq.~(\ref{eq_mult_inversion}) the corresponding true distribution $\tilde{T}$ is calculated, which is shown in the right panel (histogram). The overlaid function is the true shape $T$. Although the resolution effect on the shape of the measured distribution (left) is very small, the solution obtained by matrix inversion suffers from large fluctuations. The regularity is explained by the fact that the response matrix only contains entries on the diagonal and directly next to it.}
			  \label{fig_mult_BlobelUnfoldingExample}
			  %plots.C: BlobelUnfoldingExample()
			\efig
						
			The information that is lost due to the resolution cannot be recovered. To work around the consequence of non-physical fluctuations the result is usually constrained with \emph{a priori} knowledge about the smoothness of the true distribution. Methods that allow the recovery of the true distribution (with the limitation that structures in the distribution that are smaller than the resolution will not become visible) are $\chi^2$ minimization with regularization \cite{Blobel:1984ku} and Bayesian unfolding \cite{D'Agostini:1994zf, D'Agostini:1999cm}. 
			$\chi^2$ minimization allows to find the true spectrum by minimizing a $\chi^2$ function with a regularization term. These regularization schemas reduce fluctuations, e.g., by preferring solutions with small sums of the first or second derivatives, or by maximizing the entropy. Their influence has to be carefully studied to keep the bias on the unfolded solution small \cite{Cowan:2002in}.
			Bayesian unfolding is an iterative method based on Bayes' theorem which implictly regularizes the solution by limiting the number of iterations \cite{Blobel:2002pu}.
			Both methods are described and evaluated in detail in \cite{jfgo_thesis}. 
			
			Clearly, for the true distribution in full phase space only even numbers of particles occur due to charge conservation. Due to efficiency and acceptance effects in the measured spectrum also odd numbers occur. This has to be taken into account in the detector response matrix, i.e., every column corresponding to an odd number of generated particles is empty; for the number of measured particles even and odd values occur. For the correction to limited phase space this constraint does not arise.

  \subsection{Data Sample}

    \btab
      \caption{\label{table_datasample} Listed are the references of data used in this chapter. For each reference it is specified at which energy the sample was taken, to which event class it is corrected, and what kind of data ($\dndeta$ and/or multiplicity distribution) are presented.}
      \begin{tabular}{|p{2.1cm}|l|p{6.25cm}|c|c|l|}
        \hlinespc
        Experiment & Ref. & Energy & $\dndeta$ & Mult. & Remark \\
        \hlinespc
         SFM & \cite{Breakstone:1983ns} & \unit[30.4, 44.5, 52.6, 62.2]{GeV} \newline (INEL, NSD) & & X & $^{\rm a}$ \\
         \hlinespc
         Streamer \newline Chambers \newline Detector & \cite{Thome:1977ky} & \unit[23.6, 30.8, 45.2, 53.2, 62.8]{GeV} \newline (INEL) & X & X & \\
         \hlinespc
         UA1 & \cite{Albajar:1989an} & \unit[200, 500, 900]{GeV} (NSD) & & X & \\
             & \cite{Arnison:1982rm} & \unit[540]{GeV} (NSD) & X & & \\
         \hlinespc
         UA5 & \cite{Alpgard:1982zx} & \unit[53]{GeV} (INEL) & X & X & $^{\rm b}$ \\
             & \cite{Alner:1986xu} & \unit[53, 200, 546, 900]{GeV} (INEL, NSD) & X & & \\
             & \cite{Alner:1987wb} & \unit[546]{GeV} (INEL, NSD) & X & X & $^{\rm c}$ \\
             & \cite{Alner:1984is} & \unit[540]{GeV} (NSD) & & X & \\
             & \cite{Alner:1985zc} & \unit[540]{GeV} (NSD) & & X & $^{\rm d}$ \\
             & \cite{Alner:1985wj} & \unit[200, 900]{GeV} (NSD) & & X & $^{\rm e}$ \\
             & \cite{Ansorge:1988kn} & \unit[200, 900]{GeV} (NSD) & & X & \\
        \hlinespc
        P238 & \cite{Harr:1997sa} & \unit[630]{GeV} (NSD) & X & & \\
        \hlinespc
        CDF & \cite{Abe:1989td} & \unit[0.63, 1.8]{TeV} (NSD) & X & & \\
            & \cite{cdf_multiplicity1} & \unit[1.8]{TeV} (NSD) & & X & $^{\rm f}$ \\
        \hlinespc
        E735 & \cite{Alexopoulos:1998bi} & \unit[0.3, 0.5, 1.0, 1.8]{TeV} (NSD) & & X & $^{\rm g}$ \\
             & \cite{Lindsey:1991pt} & \unit[0.3, 0.5, 1.0, 1.8]{TeV} (NSD) & & X & $^{\rm h}$ \\
             & \cite{wang_thesis} & \unit[1.8]{TeV} (NSD) & & X & \\  % NSD given on p76
        \hline
        \end{tabular}
        \\
      \raggedright
      \vspace{0.2cm}
      \noindent $^{\rm a}$ Error of cross section included in multiplicity distribution. \\
      \noindent $^{\rm b}$ Comparison $p+\pbar$ vs. $p+p$; only uncorrected multiplicity. \\
      \noindent $^{\rm c}$ Comprehensive report. \\
      \noindent $^{\rm d}$ Multiplicity distribution forced to be of NBD shape. \\
      \noindent $^{\rm e}$ This data is not used here because the method has been partially revised in \cite{Ansorge:1988kn}. \\
      \noindent $^{\rm f}$ No systematic error assessment. \\
      \noindent $^{\rm g}$ Method description very limited; extrapolated from \etain{3.25} to full phase space. \\
      \noindent $^{\rm h}$ Only in KNO variables; no systematic error assessment. \\
    \etab

    The data sample considered in this review is summarized in Table~\ref{table_datasample}. Details about the different analyses are given ordered by detector and collider. Unless otherwise stated, the correction procedures described in the publications consider the effect of decays of strange and neutral particles as well as the production of secondary particles due to interactions of primary particles with the material.

    \textbf{Multiwire proportional chambers inside the Split Field Magnet detector (SFM)} \cite{Della_Negra:1977sk} at the ISR measured the multiplicity distribution for NSD and inelastic $p+p$ events at $\sqrt{s} = $~30.4, 44.5, 52.6, and \unit[62.2]{GeV} \cite{Breakstone:1983ns}.
		%The trigger required a coincidence of at least three chambers pointing to the same direction. It accepted about 95\% of the inelastic events.
		% trigger coincidence info from NP B104,365
		Between 26\,000 and 60\,000 events were collected for each of the energies.
		%and corrected for decays of strange and neutral particles.
		The SD component was removed from the sample by means of its topology: events are considered SD if in one of the hemispheres no track or only one track carrying 80\% of the incident proton's energy is found. Systematic errors have been evaluated and include the error that arises from the corrections and in the low-multiplicity region from the subtraction of elastic events.

    \textbf{A detector based on streamer chambers} \cite{Thome:1977ky} at the ISR measured pseudo\-rapidity and multiplicity distributions for inelastic events at centre-of-mass energies of 23.6, 30.8, 45.2, 53.2, and \unit[62.8]{GeV}. Between 2\,300 and 5\,900 events were measured for each energy. In the analysis corrections for the acceptance, the low-momentum cut-off (about \unit[45]{MeV/$c$}), and secondary particles are taken into account.

    \textbf{The UA1 (Underground Area 1) experiment} measured the multiplicity distribution for NSD events in the interval $|\eta| < 2.5$ at $\cms = $~200, 500, and \unit[900]{GeV}~\cite{Albajar:1989an}.
    188\,000 events were used, out of which 34\% were recorded at the highest energy. The \spps\ was operated in a pulsed mode where data were taken during the energy ramp from \unit[200]{GeV} to \unit[900]{GeV} and vice versa. Therefore the data at \unit[500]{GeV} are in fact taken in an energy range from \unit[440]{GeV} to \unit[560]{GeV}. Only tracks with a $p_T$ larger than \unit[150]{MeV/$c$} are considered for the analysis to reduce the contamination from secondaries. Although not explicitly mentioned in the publication, it is assumed for this review that the low-momentum cut-off correction is part of the acceptance correction. UA1 quotes an overall systematic error of 15\%: contributions are from strange-particle decays, photon conversions and secondary interactions (3\%), as well as the uncertainty in the acceptance (4\%). Other contributions arise from the selection criteria and uncertainties in the luminosity measurement (10\%). The luminosity measurement uncertainty only applies to the cross section measurement, not to the normalized distribution. The uncertainty due to the selection criteria is not quoted. Therefore, assuming that the systematic uncertainties were summed in quadrature, this uncertainty is 10\% and the overall systematic error without the uncertainty on the luminosity is 11\% which is the value applicable to the normalized multiplicity distribution.

    UA1 measured the $\dndeta$ distribution at \cmsofG{540} \cite{Arnison:1982rm}. The analysis used 8\,000 events that have been taken without magnetic field which reduced the amount of particles lost at low-momenta to about 1\%. The systematic error of the applied corrections is estimated by the authors to be 5\% without elaborating on the different contributions.

    \textbf{The UA5 (Underground Area 5) experiment} was running at the ISR and the \spps. A comparison of data taken in $p+p$ and $p+\pbar$ collisions at \cmsofG{53} was made \cite{Alpgard:1982zx}. 3\,600 $p+p$ events and 4\,000 $p+\pbar$ events were used. Trigger and vertex-finding efficiencies as well as acceptance effects have been evaluated with a Monte Carlo simulation tuned to reproduce ISR data. For both collision systems the comparison of the $\dndeta$ distribution was done using the uncorrected data and limited to events with at least two tracks. In this way the authors attempted to achieve lower systematic errors on the result. A ratio of $1.015 \pm 0.012$ ($p+\pbar$ over $p+p$) has been found. Furthermore, the multiplicity distributions were compared. 
    %It is found that these distributions agree within errors. 
    The authors conclude that the distributions agree within errors and that  differences between $p+p$ and $p+\pbar$ collisions are smaller than 2\%.

    UA5 measured the $\dndeta$ distribution at $\cms = 200$ and \unit[900]{GeV} for NSD events \cite{Alner:1986xu, Alner:1987wb}.  2\,100 (3\,500) events have been used for the analysis at 200 (900) GeV. It should be noted that the corrections are based on a Monte Carlo simulation that has been tuned to reproduce data measured at \cmsofG{546}. The results of the simulation were parameterized and scaled to $\cms = 200$ and \unit[900]{GeV} in order to estimate the corrections for acceptance and contamination by secondaries. The authors only mention statistical errors explicitly.

    Measurements of the multiplicity distribution have been presented in \cite{Alner:1984is, Alner:1985zc, Alner:1985wj, Ansorge:1988kn, Alner:1987wb}. The distribution is measured in different $\eta$-regions (smallest: \etain{0.2} for \unit[540]{GeV} and \etain{0.5} for \unit[200 and 900]{GeV}) up to \etain{5.0}. Furthermore, the result is presented extrapolated to full phase space.
    The analysis used 4\,000 events for \unit[200]{GeV} and 7\,000 events each for 540 and \unit[900]{GeV}. In all cases the unfolding of the measured spectrum was performed by minimizing a $\chi^2$-function. For the case of \cmsofG{540} \cite{Alner:1985zc} it was required that the resulting function is a NBD which is regarded as a strong constraint. This has to be taken into account when interpreting the result at \unit[540]{GeV}. The distributions at 200 and \unit[900]{GeV} were unfolded using the maximum-entropy method \cite{Ansorge:1988kn} which is considered to be a less restrictive assumption. The assessment of the systematic errors is not very comprehensive and an uncertainty of about 2\% is quoted.

    \textbf{A Forward Silicon Micro-Vertex detector} that was tested in the context of a proposed hadronic B-physics experiment (P238) measured the $\dndeta$ distribution at forward rapidity at \cmsofG{630} \cite{Harr:1997sa}. A sample of 5 million events is corrected for tracks from secondaries (2\%) and SD events (0.5\%). Acceptance and resolution effects are corrected using Monte Carlo simulations tuned to UA5 data. Their magnitude as well as the magnitude of the trigger- and vertex-efficiency correction are not detailed. A normalization error of 5\% dominates the systematic error. It is attributed to inconsistent results when only the $x$ or $y$ tracking information is used compared to the case where both of them are used. Other effects such as detector efficiency, misalignment, and the SD cross section are considered by the authors not to significantly contribute to the systematic uncertainty.

    \textbf{The CDF (Collider Detector at Fermilab) experiment} \cite{Abe:1988me}, a detector at the Tevatron collider, measured the $\dndeta$ distribution at \cmsofG{630} and \unit[1.8]{TeV} with their so-called Vertex Time-Projection Chambers (VTPCs) \cite{Abe:1989td}. These VTPCs have been replaced after years of operation by a silicon detector. The authors do not mention whether the corrections correspond to NSD or inelastic events. However, the trigger configuration requires a hit on both sides. This points to the fact that the trigger is insensitive to the majority of SD events. Furthermore, the authors compare their measurement to NSD data from UA5 which confirms that CDF obtained their result in NSD events. 2\,800 (21\,000) events have been used for the analysis at 630 \unit[(1\,800)]{GeV}. Only events with at least 4 tracks are considered to reduce beam-gas background. The authors state that they ``do not correct for events missed by the trigger or selection procedure'' and estimated that the selection procedure misses $(13 \pm 6)\%$ of the events. This is surprising because the normalization for $\dndeta$ would be significantly wrong if this correction was not applied. This is not the case shown in the comparison to UA5 data. Tracks with $p_T < \unit[50]{MeV/\emph{c}}$ are not found due to the magnetic field and a correction of $(3 \pm 2)\%$ is applied to account for this loss. A systematic error assessment is made; the error is dominated by uncertainties in the tracking efficiency and ranges from 3\% (at $\eta = 0$) to 15\% (at $|\eta| = 3.25$).
 
    CDF measured the multiplicity distribution in various $\eta$-intervals for NSD events at \cmsofT{1.8} \cite{cdf_multiplicity1}. The publication does not mention the number of events used in the analysis. A systematic-error assessment is reported to be ongoing, but has not yet been published. It is unclear if an unfolding method was used.
    
    A further multiplicity distribution measurement based on a large event sample is in preparation by CDF \cite{CDFMult}. This study considers only tracks with a $p_T$ larger than \unit[0.4]{GeV/$c$}. 

    \textbf{The E735 experiment} \cite{Lindsey:1991pt} at the Tevatron collider measured the multiplicity distribution of NSD events at energies of $\cms =$ 0.3, 0.5, 1.0, and \unit[1.8]{TeV} \cite{Alexopoulos:1998bi}. The extrapolation to full phase space has been done by the authors based on Pythia simulations. They provide no further information about the statistics used, the corrections, and in particular the question as to whether an unfolding procedure was used. This has to be taken into account when the result is interpreted.

    In \cite{Lindsey:1991pt} multiplicity distributions of NSD events are presented in intervals of \etain{1.62} and \etain{3.25} as well as extrapolated to full phase space for the four aforementioned energies. A total number of 25 million events is mentioned, however only a subset is used for the multiplicity analysis whose size is not mentioned. The results are only presented in KNO variables. The data were unfolded using the maximum-entropy method. A systematic error assessment has not been performed.

    Reference \cite{wang_thesis} presents multiplicity distributions in \etain{1.57}, \etain{3.25}, and extrapolated to full phase space at \cmsofT{1.8} of NSD events. About 2.8 million events have been used and unfolded with an iterative method similar to the mentioned Bayesian unfolding. Systematic uncertainties have been evaluated concentrating on the effect of the cuts to reduce contamination by single-diffractive and beam-gas events. In \cite{Lindsey:1991pt} and \cite{wang_thesis} a correction for strange-particle decays is not explicitly mentioned, but it can be assumed to have been part of the Monte Carlo simulation used to obtain the correction factors.

\subsection{Multiplicity Distributions from \texorpdfstring{$\cms$}{sqrt(s)} = 20 to 1800 GeV}

    In the following sections the theoretical and phenomenological concepts introduced in the first part of the review are applied to selected multiplicity distributions. An example for KNO scaling as well as the fit with a single NBD and a combination of two NBDs is shown. The available distributions in full phase space are shown together in multiplicity and KNO variables to assess the validity of KNO scaling, which is further discussed in Section~\ref{section_meas_moments}.
		
		\bfig
			\includegraphics[width=0.48\linewidth]{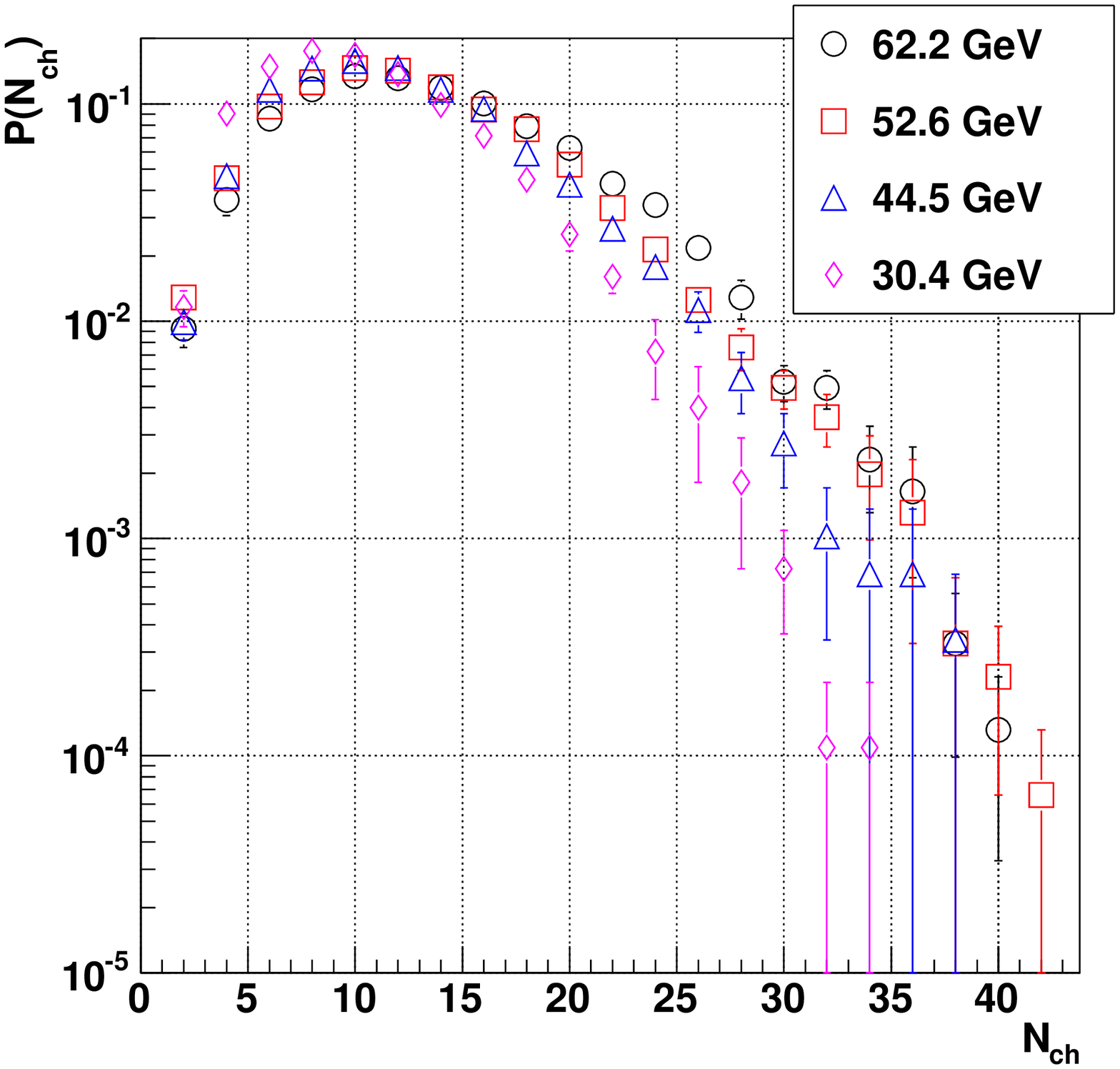}
			\hspace{0.02\linewidth}
			\includegraphics[width=0.48\linewidth]{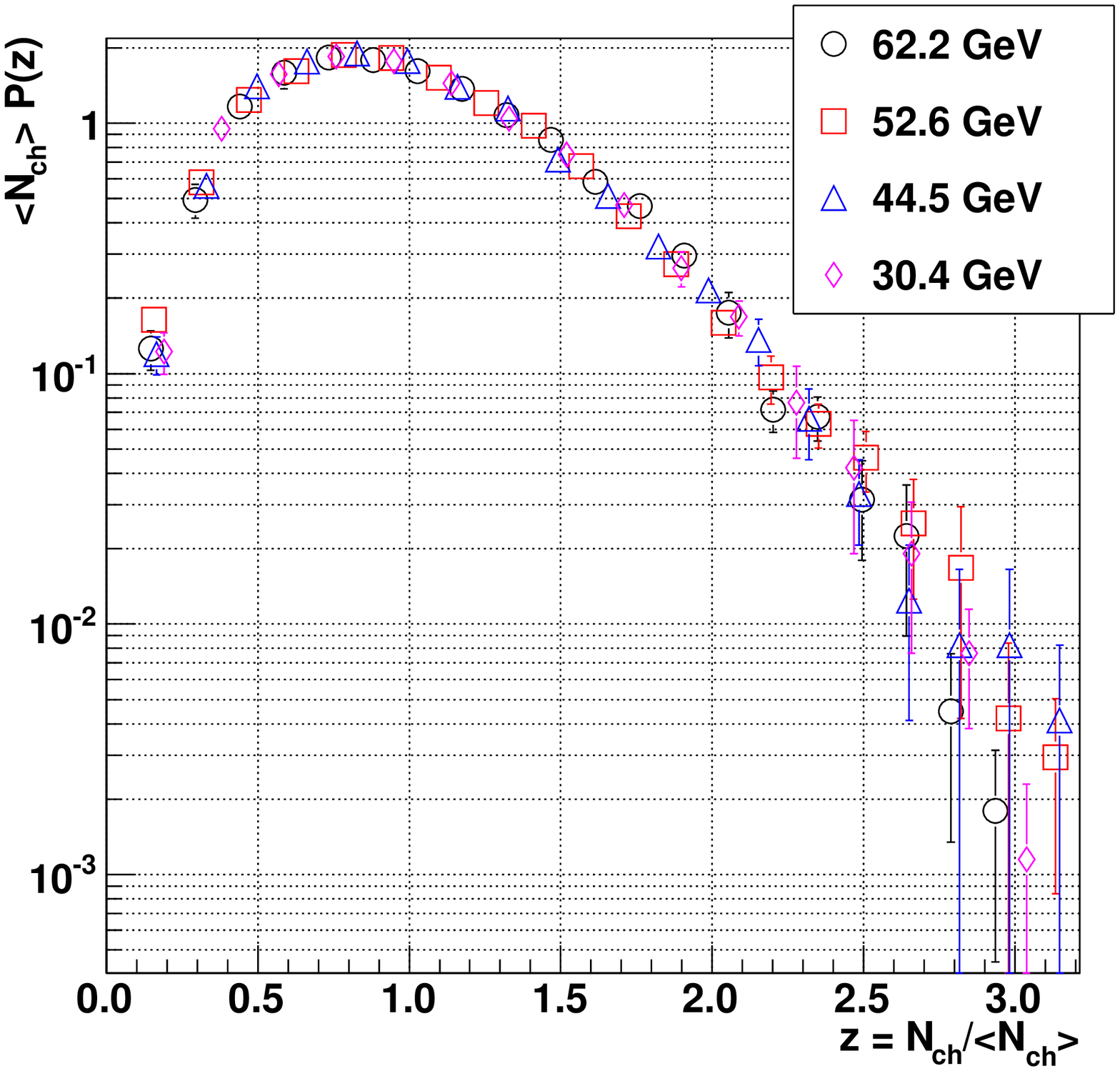}
			\caption{KNO scaling at ISR energies. The figure shows normalized multiplicity distributions for NSD events in full phase space vs. multiplicity (left panel) and using KNO variables (right panel). The data were measured by the SFM \cite{Breakstone:1983ns}.}
			\label{fig_isr_mult}
			%hepdata: http://durpdg.dur.ac.uk/cgi-bin/hepdata/testreac/10406/1/q
			%NSD data taken from paper in table (numbers), but not on hepdata
		\efig

		Figure~\ref{fig_isr_mult} shows multiplicity distributions in full phase space for NSD events taken at the ISR. The distribution is shown in multiplicity and KNO variables. The latter indicates that KNO scaling is fulfilled at ISR energies (the moments of these distributions are analyzed further below).

		\bfig
			\includegraphics[width=0.48\linewidth]{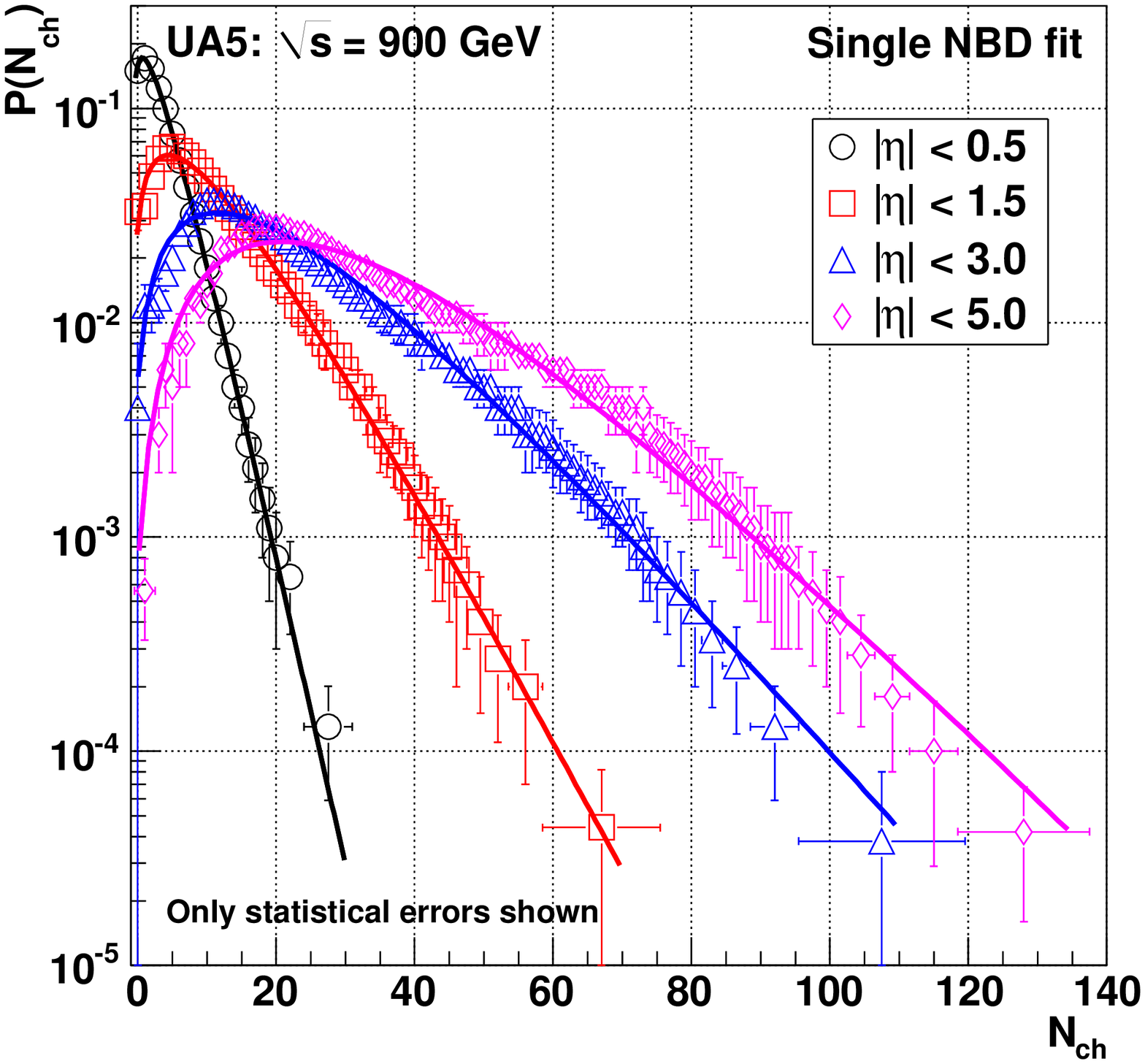}
			\hfill
			\includegraphics[width=0.48\linewidth]{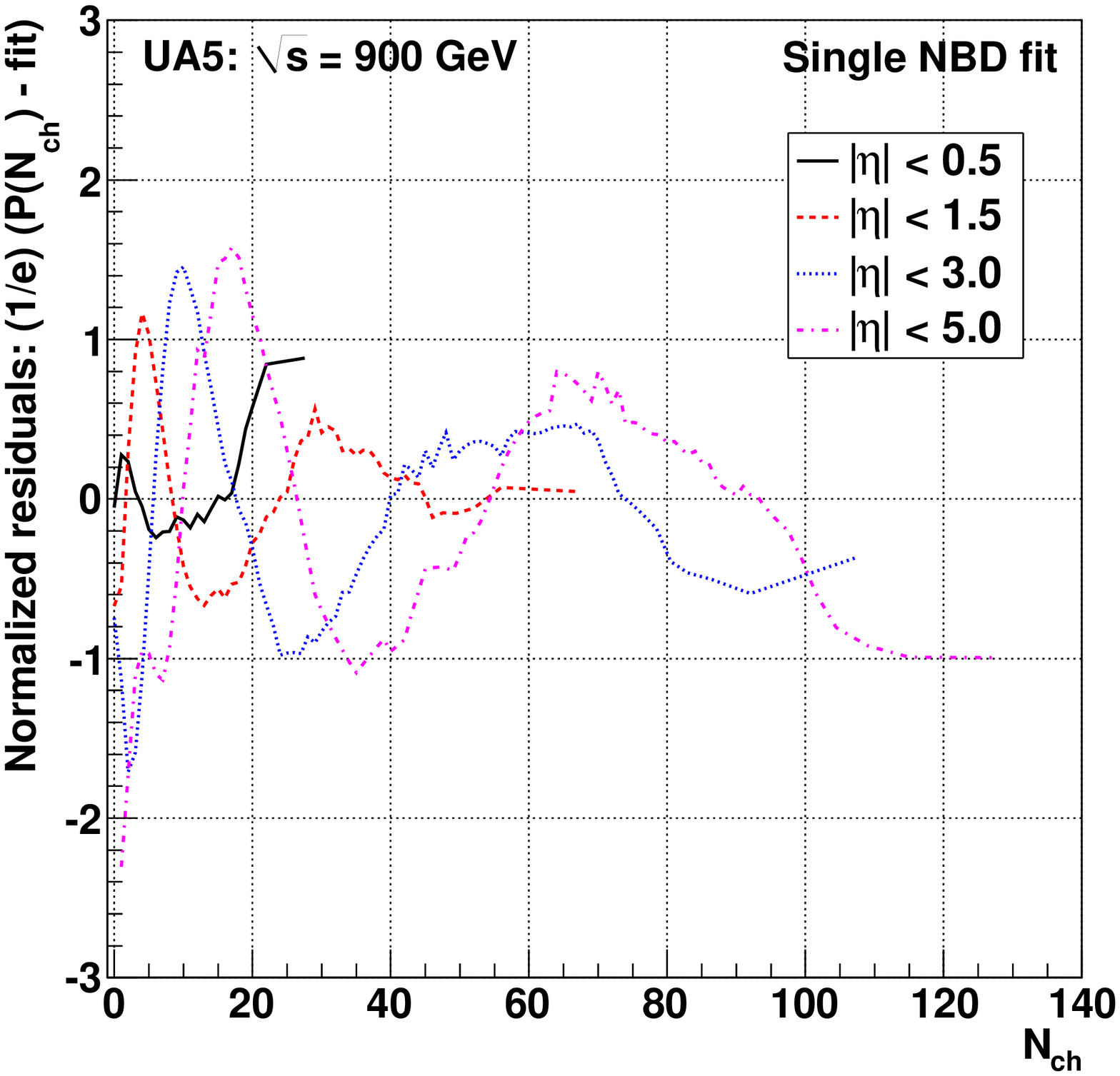}
			\includegraphics[width=0.48\linewidth]{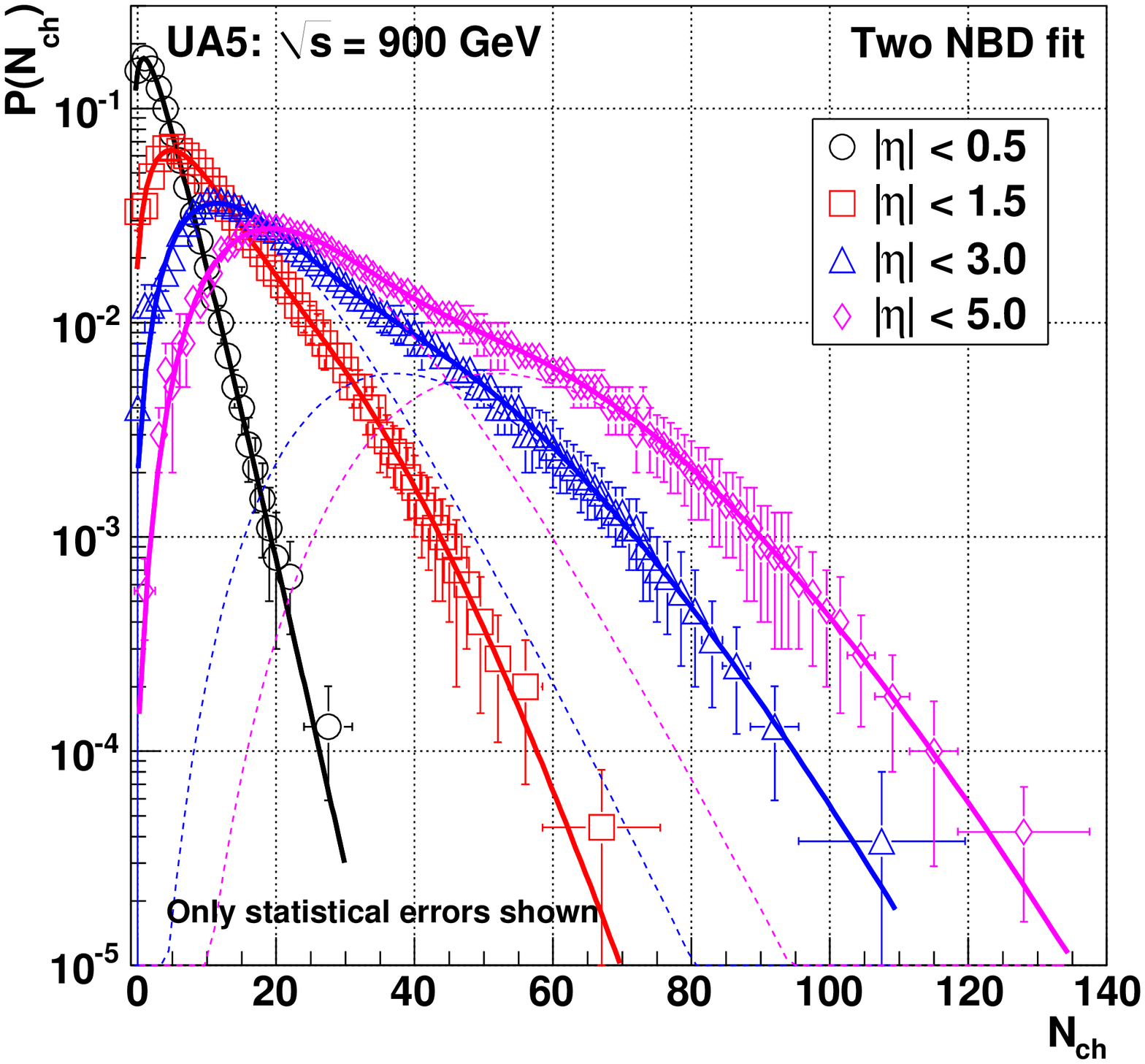}
			\hfill
			\includegraphics[width=0.48\linewidth]{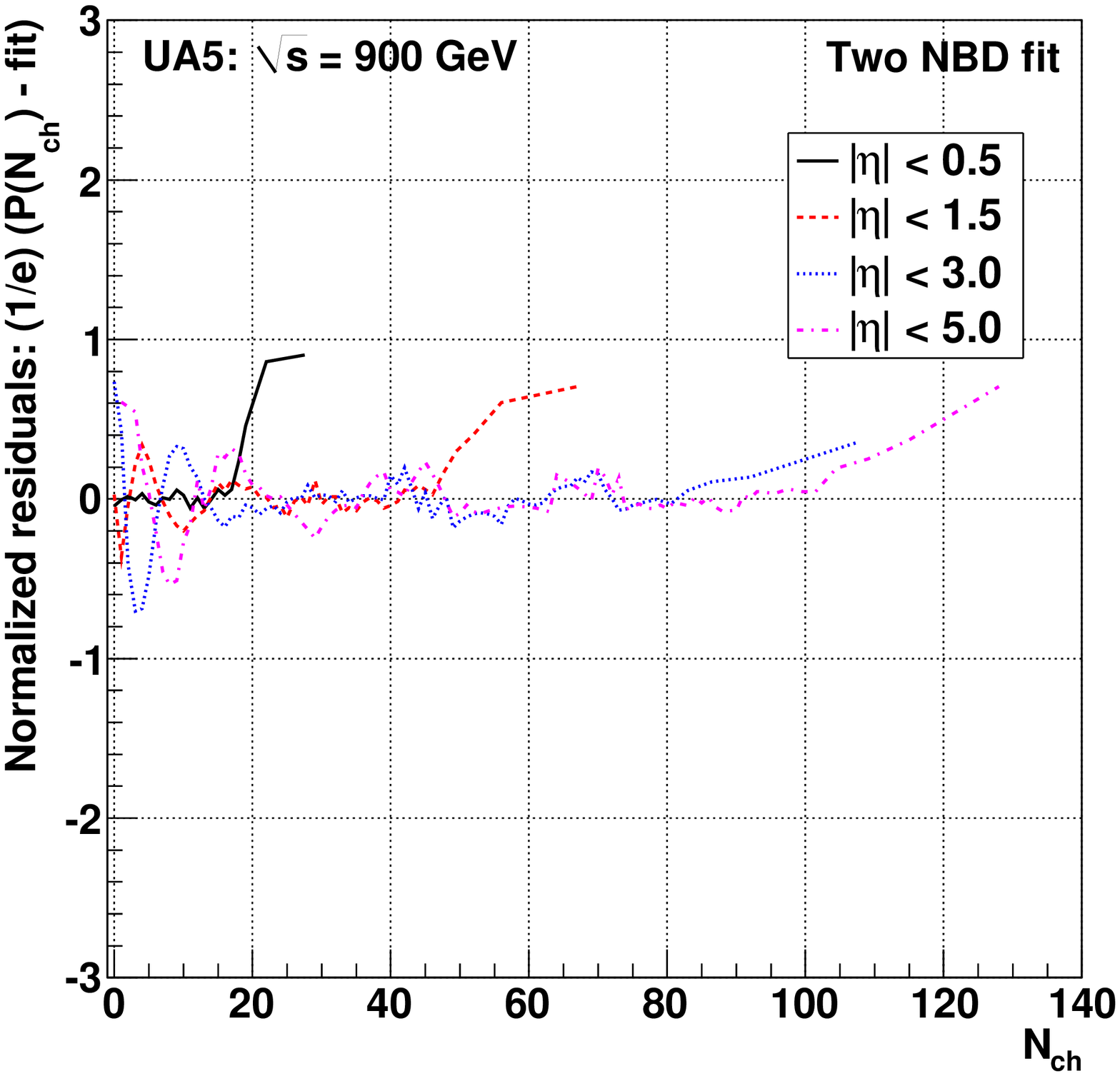}
			\caption{Normalized multiplicity distributions of NSD events at \cmsofG{900} in various rapidity intervals are shown fitted with single NBDs (top left panel) or a combination of two NBDs (bottom left panel). The two contributing NBDs (dashed lines) are shown exemplarily for $|\eta| < 3.0 \mbox{ and } 5.0$. The right panels show the normalized residuals with respect to the corresponding fits defined by $(1/e) (P(N_\mathrm{ch}) - \textrm{fit})$ with $e$ being the error on $P(N_\mathrm{ch})$. These are smoothed over four data points to reduce fluctuations. The data were measured by UA5 \cite{Ansorge:1988kn}.}
			\label{fig_ua5_mult_eta}
			%http://durpdg.dur.ac.uk/cgi-hepdata/hepreac/1926373
			%hepdata/ua5/mult_eta.C (root v5-22-00)
		\efig
				
		Multiplicity distributions are described by NBDs up to \cmsofG{540} in full phase space as well as in different $\eta$-ranges. This behaviour does not continue for \cmsofG{900}. Figure~\ref{fig_ua5_mult_eta} shows multiplicity distributions together with NBD fits in increasing pseudorapidity ranges at \unit[900]{GeV} (top left panel). The respective normalized residuals are also shown (top right panel). The NBD fit works very well for the interval $|\eta| < 0.5$, but it becomes more and more obvious with increasing $\eta$-range that the region around the most probable multiplicity is not reproduced. The structure found around the peak gave rise to the two-component approach, discussed previously, in which the data are fitted with a combination of two NBDs. The bottom left panel of Figure~\ref{fig_ua5_mult_eta} shows these fits with Eq.~(\ref{eq_twocomponent}), and normalized residuals (bottom right panel) to the same data which yields good fit results for all pseudorapidity ranges.

      \bfig
        \includegraphics[width=0.48\linewidth]{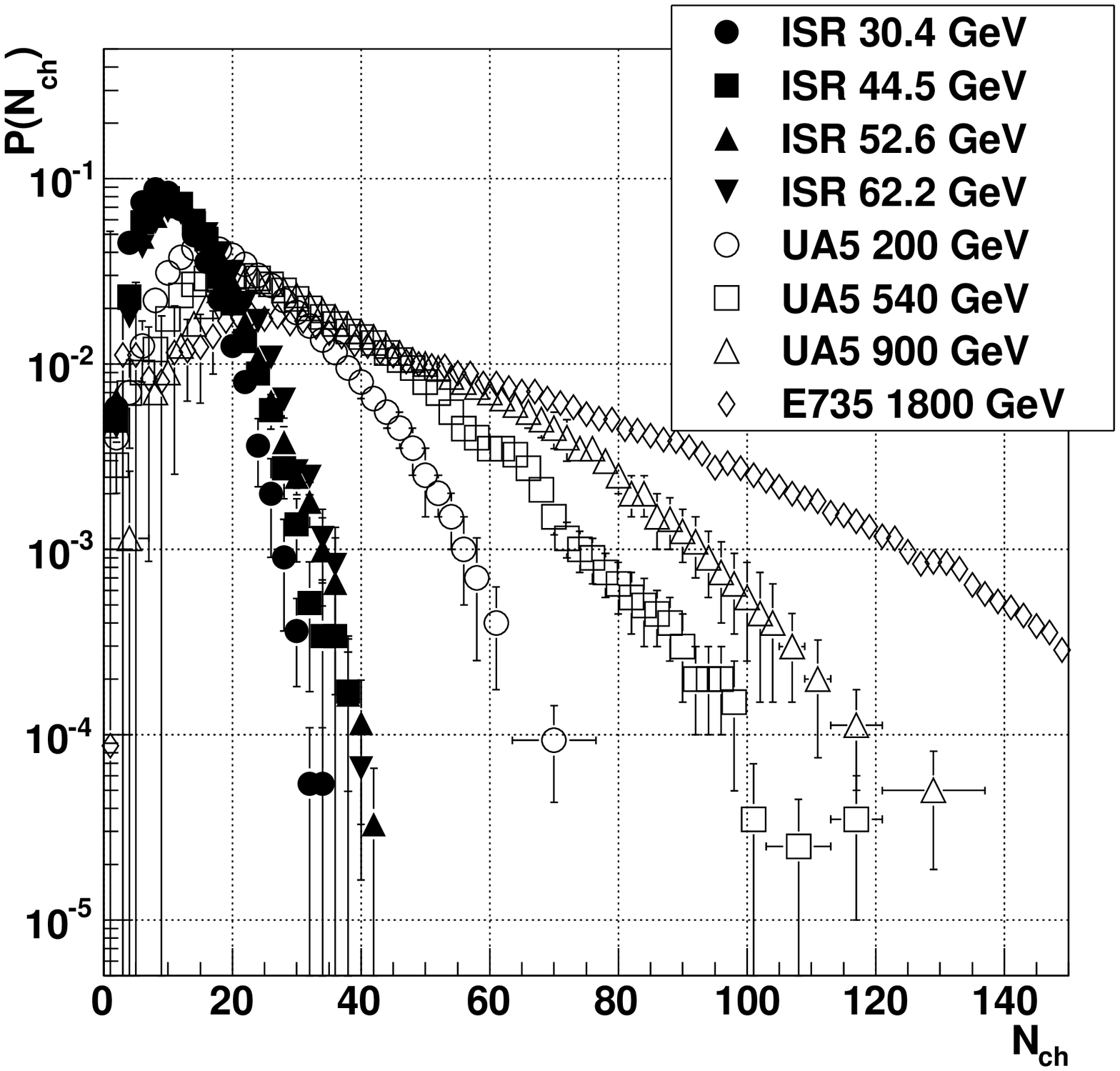}
        \hfill
        \includegraphics[width=0.48\linewidth]{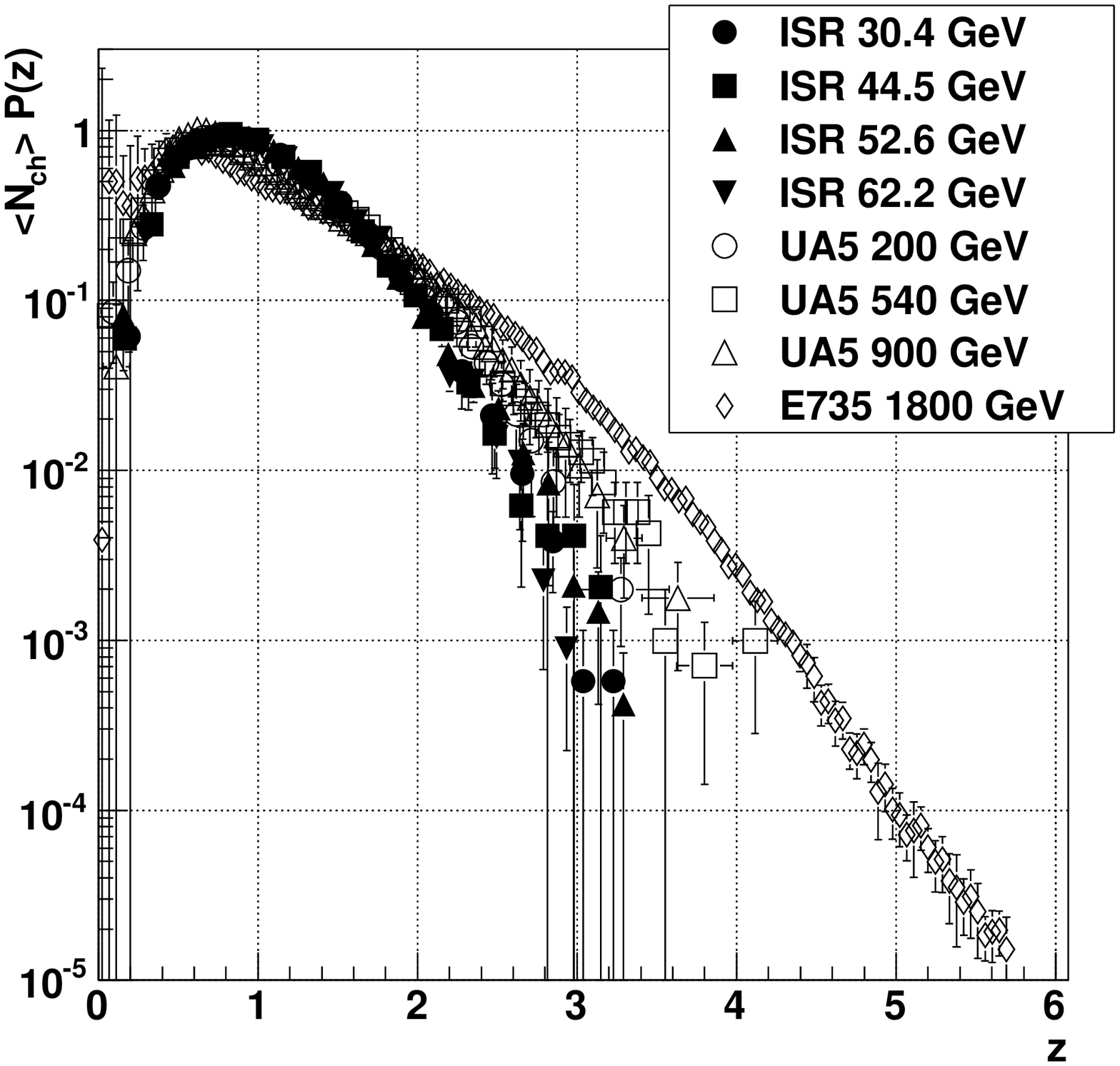}
        \caption{\label{fig_summary_knoviolation} Multiplicity distributions of NSD events in full phase space in multiplicity variables (left panel) and in KNO variables (right panel). Data points from \cite{Breakstone:1983ns, Ansorge:1988kn, Alner:1985rj, Alexopoulos:1998bi}.}
        %KNOViolation.C
      \efig

     To assess the validity of KNO scaling all available multiplicity distributions are drawn as function of the KNO variable $z = \nch/\expval\nch$. This is shown in Figure~\ref{fig_summary_knoviolation} for NSD events in full phase space from 30 to \unit[1800]{GeV}. Although it is evident that the high-multiplicity tail does not agree between the lowest and highest energy dataset, no detailed conclusion is possible for the data in the intermediate energy region. Further conclusions are derived from  
the study of the moments of these distributions, see Section~\ref{section_meas_moments}.
  	
\subsection{\texorpdfstring{$\dndeta$ and $\expval{N_\mathrm{ch}}$ vs. $\cms$}{dNch/deta and <Nch> vs. sqrt(s)}}
\label{section_meas_energydependence}

\bfig
\includegraphics[width=0.8\linewidth]{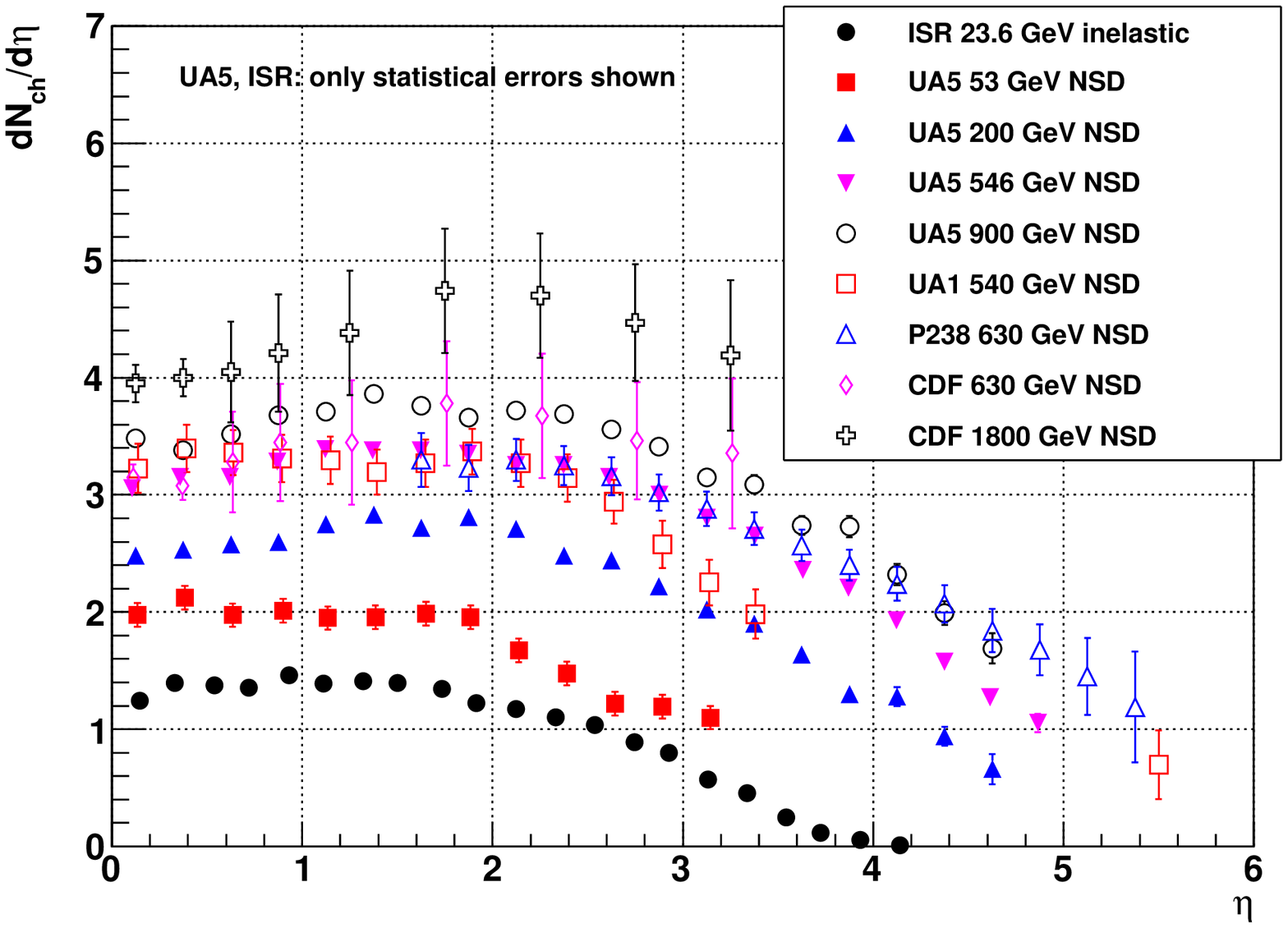}
\caption{$\dndeta$ at different $\cms$. Data points from \cite{Thome:1977ky, Arnison:1982rm, Alner:1985zc, Ansorge:1988kn, Abe:1989td, Harr:1997sa}.}
\label{fig_summary_dndeta}
% hepdata: various; see .C file
\efig
		
Figure~\ref{fig_summary_dndeta} shows $\dndeta$ at energies ranging over about two orders of magnitudes, from the ISR (\cmsofG{23.6}) to the Tevatron (CDF data, \cmsofT{1.8}). Increasing the energy results in an increase in multiplicity. The multiplicity of the central plateau increases together with the variance of the distribution. Note that the data points at the lowest energy are for inelastic events, the other data points refer to NSD events. We recall that the dip around $\eta \approx 0$ is due to the transformation from rapidity $y$ to pseudorapidity $\eta$.

		\bfig
			\includegraphics[width=0.48\linewidth]{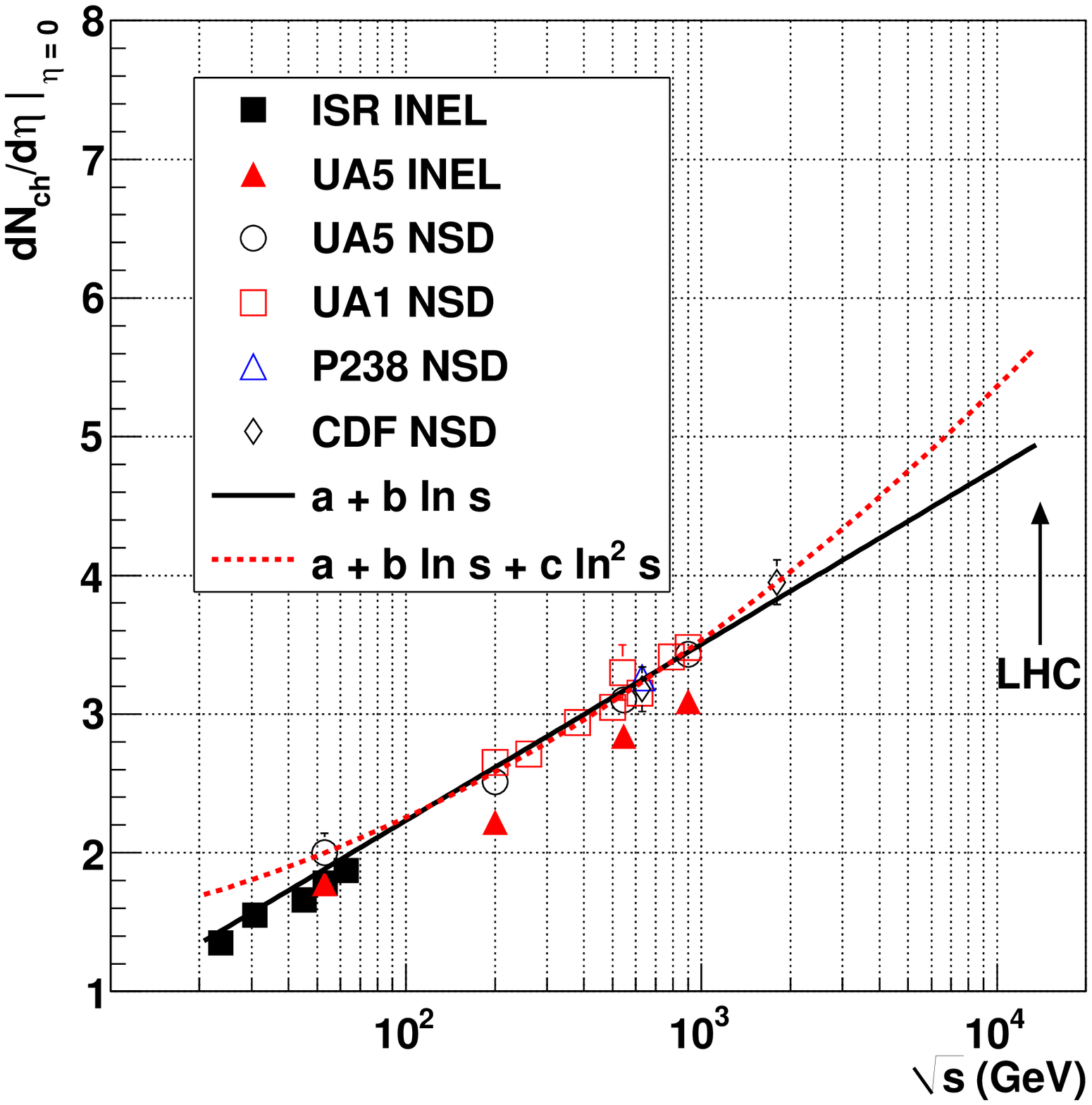}
			\hfill
			\includegraphics[width=0.48\linewidth]{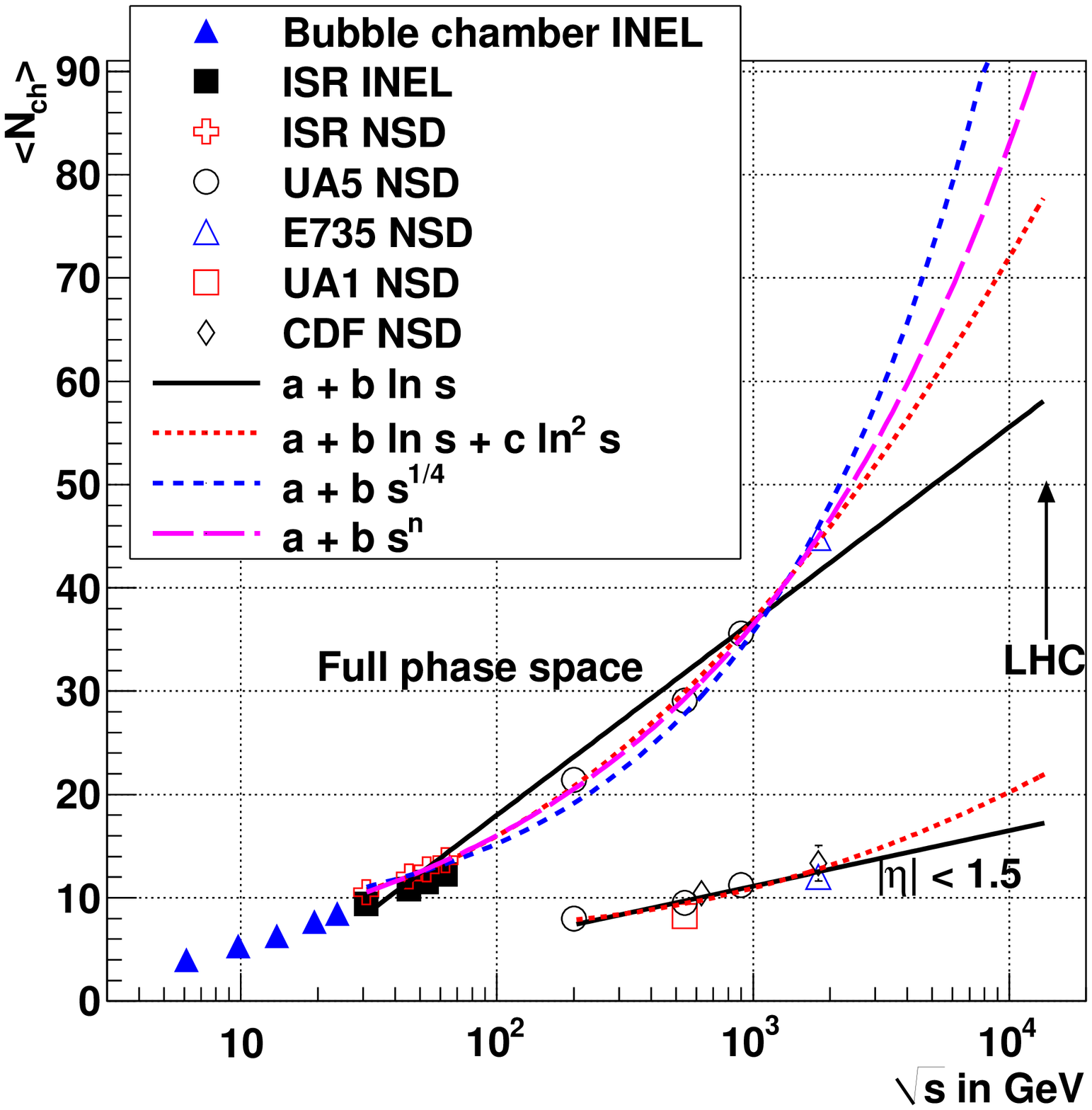}
			%hepdata: various; see .C file
			%AvgMult.C
			\caption{\label{fig_summary_mult_vs_sqrts} $\dndeta |_{\eta=0}$ (left panel) and $\expval{N_\mathrm{ch}}$ vs. $\cms$ in full phase space and \etain{1.5} (right panel) as a function of $\cms$. Data points from \cite{Slattery:1972ni, Whitmore:1973ri, Thome:1977ky, Arnison:1982rm, Alner:1984is, Breakstone:1983ns, Alner:1985zc, Ansorge:1988kn, Alexopoulos:1998bi, Abe:1989td, Albajar:1989an, cdf_multiplicity1, wang_thesis}.}
		\efig
		
		The left panel of Figure~\ref{fig_summary_mult_vs_sqrts} shows $\dndetaZero$ as a function of $\cms$. Closed symbols are data for inelastic events; open symbols for NSD events. $\dndetaZero$ increases with increasing $\cms$ violating Feynman scaling. Two fits are shown for the NSD data: a fit with $a + b\ln s$ ($a=-0.308, b=0.276$, solid black line) and $a + b \ln s + c\ln^2 s$ ($a=1.347, b=-0.0021, c=0.0013$, dashed red line). Due to the fact that different published values include different errors, e.g., no systematic errors for the UA5 data, the errors are not used for the fit. The $\ln s$ dependence was used to describe the data at centre-of-mass energies below \unit[1]{TeV}. Data at a higher energy from CDF showed deviations from this fit \cite{Abe:1989td}. The additional $\ln^2 s$ term yields a better result ($\chi^2/ndf$ reduced by about a factor of two, although the $\chi^2$ definition is not valid without using the errors); this fit suggests that $\dndetaZero$ increases faster than $\ln s$. The fits are extrapolated up to the nominal LHC energy of \cmsofT{14}.
		
    The right panel of Figure~\ref{fig_summary_mult_vs_sqrts} shows the average multiplicity $\expval\nch$ as a function of $\cms$. Data are shown for full phase space and for a limited rapidity range of \etain{1.5}. In publications two different approaches are found to obtain average values in a limited $\eta$-range. The first uses a normalization to all events having at least one track in the considered phase space. The second approach uses a normalization to the total considered cross section (inelastic or NSD) including events without any particle in the considered range (data shown here). While the latter is the more evident physical observable, the former does not depend on the efficiency to measure the total cross section which renders it less dependent on model assumptions used in the evaluation of the trigger efficiency. Data from bubble chambers at low $\cms$ are included in \figref{fig_summary_mult_vs_sqrts}: from the Mirabelle chamber at Serpukhov, Russia \cite{Slattery:1972ni} and from several bubble chambers at FNAL \cite{Whitmore:1973ri}. 

Both sets of NSD data in full phase space are fitted with four different functional forms. For full phase space the logarithmic dependence does not reproduce the data and is only shown to demonstrate the violation of Feynman scaling; the form $a+b \ln s +c \ln^2 s$ fits the data well ($a=16.65, b=-3.147, c=0.334$). The form $a+b s^{1/4}$ inspired by the Fermi-Landau model \cite{Fermi:1950jd,Wong:2008ex} provides a reasonable fit with $a=5.774$ and $b=0.948$. However, since the parameter $a$ in this form is related to contributions to the multiplicity from the leading particles it is not expected to be much larger than two. Hence, one can conclude that the Fermi-Landau form $\langle N_\mathrm{ch} \rangle \sim s^{1/4}$ fails to describe the $p+p$ data. The form $a+b s^n$ (\cite{Albini:1975iu}) provides a good description of the data with $a=0, b=3.102, n=0.178$. At $\cms=\unit[14]{TeV}$ the different fits differ significantly so that measurements at the LHC will easily reject inadequate parameterizations.
				
\subsection{Universality of Multiplicities in \texorpdfstring{$p+p(\bar{p})$ and $e^+e^-$}{p+p(pbar) and e+e-}}
\label{section_universality}
Charged-particle multiplicities in  $e^+e^-$ collisions are found to be larger than the multiplicity in $p+p(\bar{p})$ collisions at the same centre-of-mass energy (see Figure~\ref{fig_universality}). The multiplicities in $e^+e^-$  and $p+p(\bar{p})$ collisions become strikingly similar when the $p+p(\bar{p})$ points are plotted  at half their collision energy \cite{Back:2006yw, Akindinov:2007rr}. This leads to the concept of an effective energy $E_\mathrm{eff}$ in $p+p(\bar{p})$ collisions available for particle production \cite{Albini:1975iu, Basile:1980ap, Basile:1980by, Basile:1981jz}. This concept emerged already in the 1970s in the study of high-energy cosmic rays \cite{Feinberg:1972ph}.  In this picture the remaining energy is associated with the two leading baryons which emerge at small angles with respect to the beam direction:
\bq
 E_\mathrm{eff}  = \sqrt{s} - (E_\mathrm{lead,1}+E_\mathrm{lead,2}); \quad
\langle E_\mathrm{eff} \rangle = \sqrt{s} - 2 \langle E_\mathrm{leading} \rangle \;.
\label{eq_eff_energy}
\eq
It has been speculated that $E_\mathrm{eff}$ or correspondingly the inelasticity $K = E_\mathrm{eff} / \sqrt{s}$ are related to the 3-quark structure of the nucleon \cite{Chliapnikov:1990bc, Hoang:1994qz, Sarkisyan:2004vq}. In this simple picture the interaction of one of the three valence quarks in each nucleon would correspond to an average inelasticity $\langle K \rangle \approx 1/3$.

Here we estimate the coefficient of inelasticity $K$ of a $p+p(\bar{p})$ collisions by comparing $p+p(\bar{p})$ with $e^+e^-$ collisions. Given a parameterization $f_{ee}(\sqrt{s})$ of the $\sqrt{s}$ dependence of $\langle N_\mathrm{ch} \rangle$ in $e^+e^-$ collisions one can fit the $p+p(\bar{p})$ data with \cite{Chliapnikov:1990bc}
\bq
f_{pp}(\sqrt{s}) = f_{ee}(K \cdot \sqrt{s}) + n_0 \;.
\label{eq_fit_pp}
\eq
The parameter $n_0$ corresponds to the contribution from the two leading protons to the total multiplicity and is expected to be close to $n_0 = 2$.

To parameterize the multiplicity data in $e^+e^-$ collisions we use the analytic QCD expressions of Eq.~(\ref{eq_nch_ee_qcd}). The strong coupling constant $\alpha_s$ was fixed at the $Z$ mass to $\alpha_s(M_Z^2) = 0.118$ leaving $A_0$ and $A_\mathrm{LPHD}$ as fit parameters. The second form is from a 3NLO calculation \cite{Dremin:2000ep, Heister:2003aj} where the normalization and the $\Lambda$ parameter in the expression for  $\alpha_s$ were taken as fit parameters. Both forms yield excellent fits of the $e^+e^-$ data and essentially provide the same extrapolation for $\sqrt{s} > \unit[206]{GeV}$ where no data are available.

A fit with Eq.~(\ref{eq_fit_pp})  describes  the $p+p(\bar{p})$ well and yields $K = 0.35 \pm 0.01$ and $n_0 = 2.2 \pm 0.19$. The fraction of the effective energy, the inelasticity, is studied in more detail in the right panel of Figure~\ref{fig_universality}. The inelasticity $K$ is determined for each $p+p(\bar{p})$ point by solving
\bq
f_{ee}(K \sqrt{s_{pp}} - \Delta m) = \langle N_\mathrm{ch} \rangle_{pp} - n_0 \;.
\eq
In the simple quark-scattering picture the offset $\Delta m$ takes the contribution of the masses of the two participating constituent quarks to the centre-of-mass energy into account. Depending on the values for $\Delta m$ and the leading particle multiplicity $n_0$ different inelasticities can be defined.  In Figure~\ref{fig_universality} the  three cases $K_1$ ($n_0 = 0$, $\Delta m = 0$), $K_2$ ($n_0 = 2.2$, $\Delta m = 0$), and $K_3$
($n_0 = 2.2$, $\Delta m = 2/3 \, m_\mathrm{proton}$) are shown.  The inelasticity $K_1$ decreases from $\sim 0.55 - 0.6$ at ISR energies to $0.4$ at $\sqrt{s} = \unit[1.8]{TeV}$. The inelasticities $K_2$ and $K_3$ appear to be energy independent at $\sim 0.35$, in remarkable agreement with the expectation of $1/3$ in the simple quark-scattering picture.
\bfig
\includegraphics[width=0.48\linewidth]{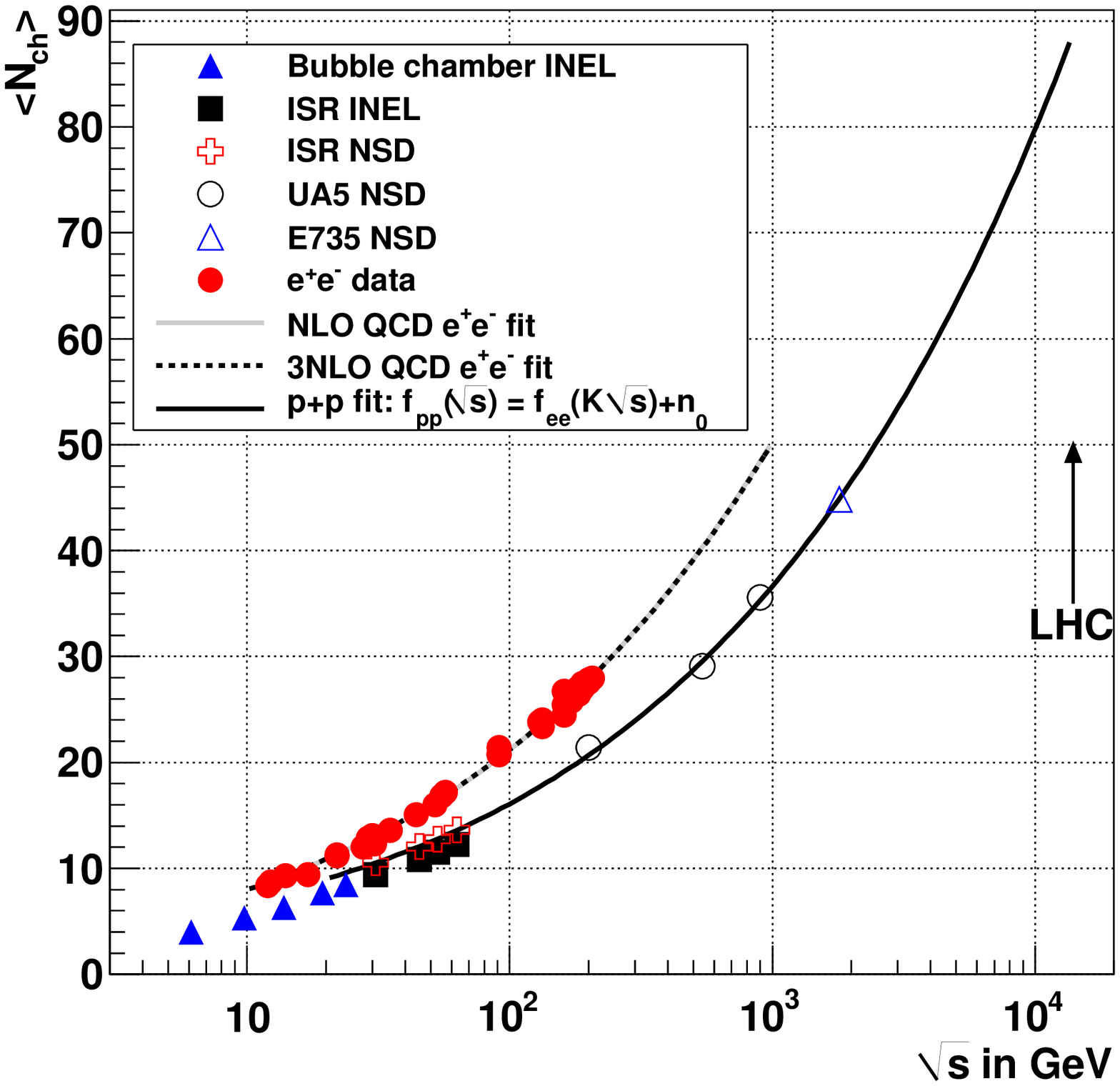}
\hfill
\includegraphics[width=0.48\linewidth]{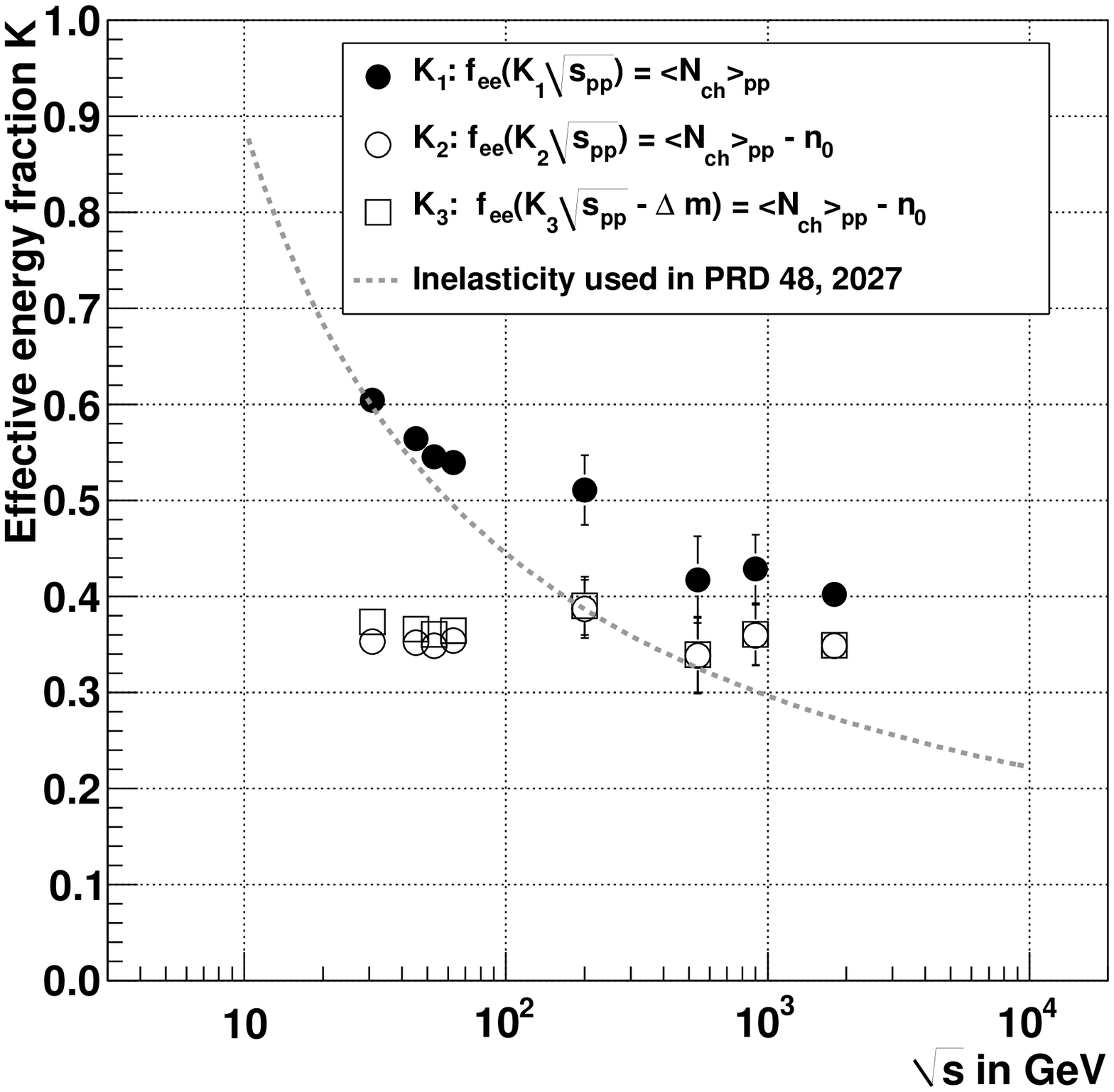}
\caption{Left panel: Comparison of charged particle multiplicities in $p+p(\bar{p})$ and $e^+e^-$ collisions ($e^+e^-$ data taken from the compilation in \cite{Dissertori:2003}). 
Note that NLO QCD fit (solid gray line) and 3NLO QCD fit (dashed line) of the $e^+e^-$ data are almost identical and lie on top of each other. Right panel: The inelasticity in $p+p(\bar{p})$ calculated for three different assumptions. The $\sqrt{s}$ dependence of the inelasticity assumed in the theoretical study \cite{Kadija:1993ie} is shown for comparison.}
\label{fig_universality}
\efig  	

The similarity between $\avgnch$ in $e^+e^-$ and $p+p(\bar{p})$ collisions when the effective energy is taken into account raises the question as to whether these similarities still persist in more differential observables like rapidity distributions. Note that a remarkable similarity was observed between $\dndeta$ per participating nucleon pair in central Au+Au collisions at $\sqrt{s_{NN}} = \unit[200]{GeV}$ and in $e^+e^-$ collisions at $\sqrt{s} = \unit[200]{GeV}$ \cite{Back:2006yw}. This suggests that the effective energy in central Au+Au collisions is close to $100\%$ of the beam energy, most likely due to the multiple interactions of the nucleons. In the left panel of Figure~\ref{fig_rapidity_ee_pp} $\dndeta$ distributions from $p+p(\bar{p})$ collisions are compared with rapidity distributions $\dndyt$ with respect to the thrust axis from $e^+e^-$ collisions. Datasets are compared for which $\sqrt{s_{pp}} \approx (2 \div 3) \sqrt{s_{ee}}$. For the shown cases the $\dndeta$ distribution in $p+p(\bar{p})$ are broader than the $\dndyt$ distributions. This might indicate the contribution from beam-particle fragmentation in $p+p(\bar{p})$. Note, however, that based on the Landau hydrodynamic picture a simple relation between $\dd N_\mathrm{ch}/\dd \eta|_{\eta=0}^{p+p,\sqrt{s}}$ and $\dd N_\mathrm{ch}/\dd y_T|_{y_T=0}^{e^+e^-,\sqrt{s}/3}$ was suggested in \cite{Sarkisyan:2004vq,Sarkisyan:2005rt}.
The width $\lambda$ of the distribution defined as $\lambda = \avgnch / \dndeta|_{\eta=0}$ and $\lambda = \avgnch / \dndyt|_{y_T=0}$, respectively,  is shown in the right panel of Figure~\ref{fig_rapidity_ee_pp}. Based on the QCD calculation in \cite{Tesima:1989ca} $\lambda$ is expected to scale linearly with $\sqrt{\ln s}$. As shown in Figure~\ref{fig_rapidity_ee_pp} this form does not describe the $p+p(\bar{p})$ data which are well parameterized with $\lambda = a + b \ln s$. The Landau hydrodynamic model also predicts a linear $\sqrt{\ln s}$ dependence of $\lambda$ \cite{Carruthers:1973ws, Steinberg:2004vy, Steinberg:2007iv} and hence also fails to describe the $p+p(\bar{p})$ data.

\bfig
\includegraphics[width=0.48\linewidth]{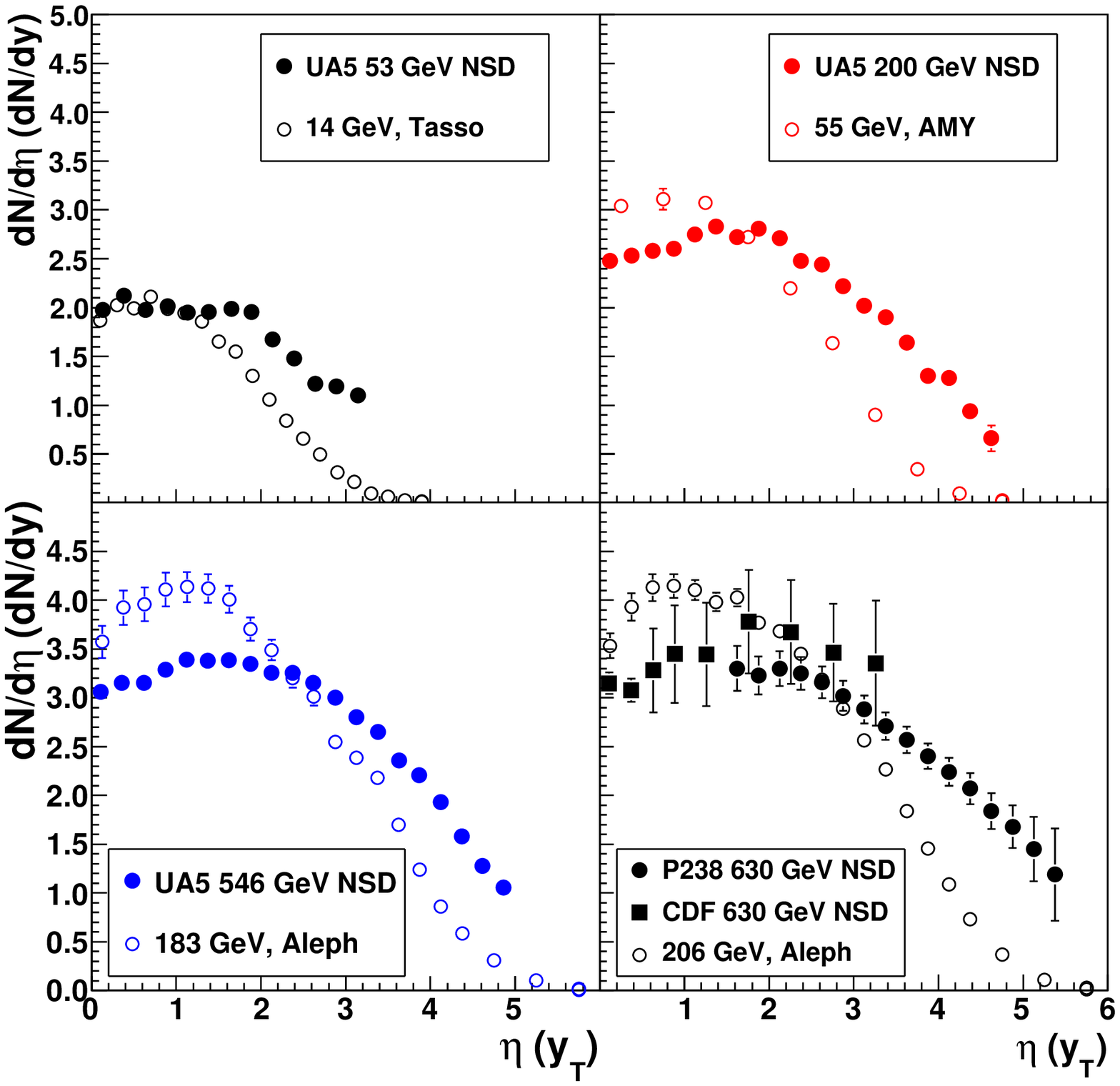}
\hfill
\includegraphics[width=0.48\linewidth]{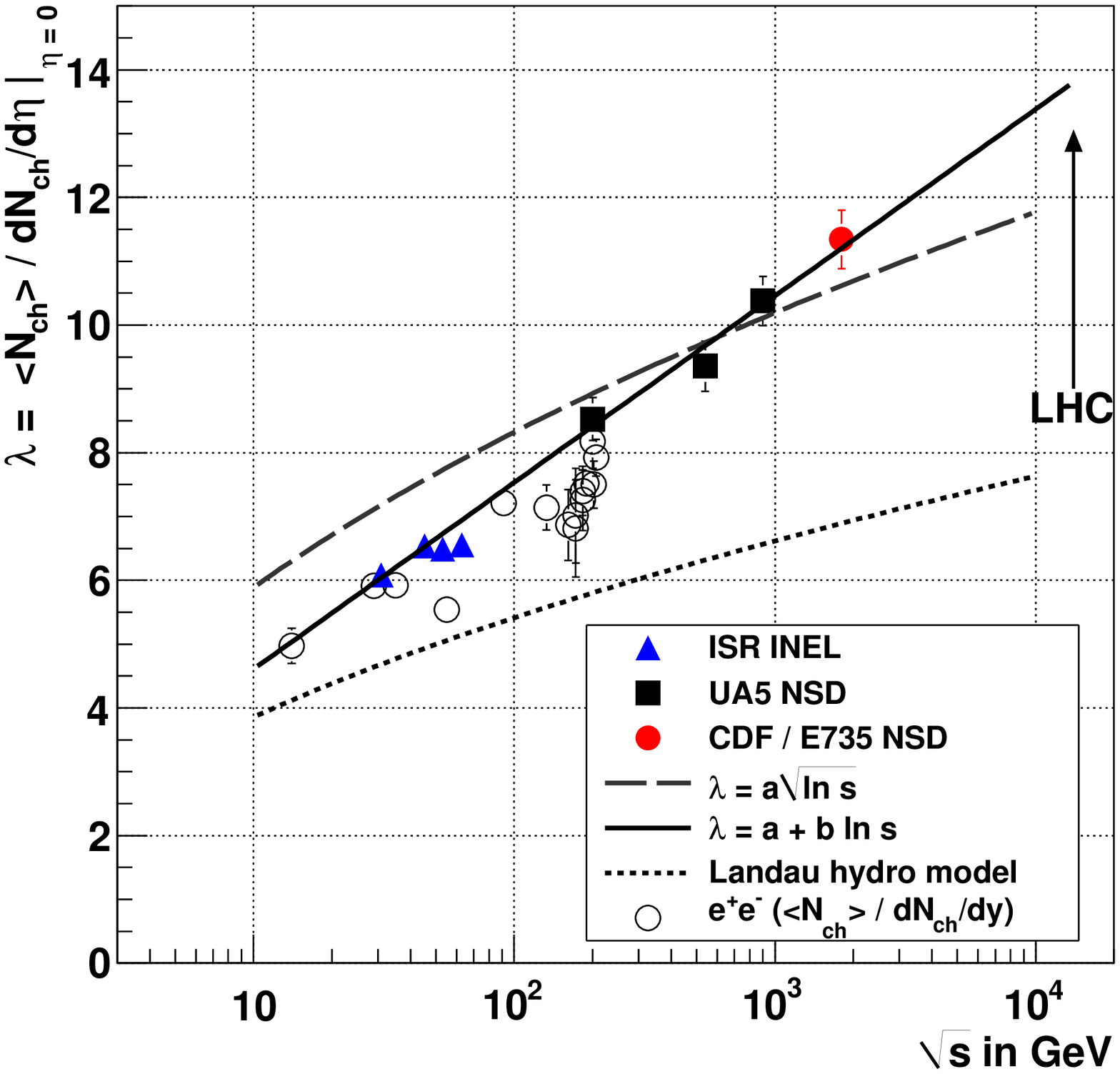}
\caption{Left panel: Comparison of $\eta$ ($p+p(\bar{p})$) and $y_T$ distributions ($e^+e^-$) at different energies. The variable $y_T$ is the rapidity with respect to the thrust axis of the $e^+e^-$ collision. Right panel: The width $\lambda$ of the $\eta$ distributions ($p+p(\bar{p})$) and $y_T$ distributions ($e^+e^-$) as a function of  $\sqrt{s}$.  Note that the difference between inelastic and non-single diffractive collisions is neglected by fitting the combined  $p+p(\pbar)$ data with $\lambda = a + b \ln \sqrt{s}$. In case of the Landau model $\langle N_\mathrm{ch} \rangle/(\dndy|_{y=0}) = \sqrt{2 \pi L}$ where $L = \ln (\sqrt{s}/(2m_p))$ is shown. Data points for $e^+e^-$ from \cite{Heister:2003aj,Abbiendi:2002mj,Buskulic:1992hq,Li:1989sn,Braunschweig:1988qm,Althoff:1983ew,Petersen:1987bq}.}
\label{fig_rapidity_ee_pp}
\efig  	

It will be interesting to see whether this universality of multiplicities in $e^+e^-$ and $p+p(\bar{p})$ collisions also holds at LHC energies.  This universality appears to be valid at least up to Tevatron energies despite its rather weak theoretical foundation 
(see Section~\ref{sec_pp_ee_similarities_and_qcd}). Under the assumptions that $K_2$ remains constant at about 0.35 also at LHC energies and that the extrapolation of the $e^+e^-$ data with the 3NLO QCD form is still reliable at $\sqrt{s} \approx \unit[5]{TeV}$ one can use the fit of $p+p(\bar{p})$ data to predict the multiplicities at the LHC. This yields $\avgnch \approx 70.9$ at $\unit[7]{TeV}$, $\avgnch \approx 79.7$ at $\unit[10]{TeV}$ and  $\avgnch \approx 88.9$ at $\unit[14]{TeV}$. Extrapolating the ratio $\lambda = \avgnch / (\mathrm{d}N_\mathrm{ch}/\mathrm{d} \eta )_{\eta=0}$ with the form $\lambda = a + b \ln \sqrt{s}$ (see Figure~\ref{fig_rapidity_ee_pp}) these multiplicities correspond to $\mathrm{d}N_\mathrm{ch}/\mathrm{d} \eta |_{\eta=0} \approx 5.5$ at $\unit[7]{TeV}$, $\mathrm{d}N_\mathrm{ch}/\mathrm{d} \eta |_{\eta=0} \approx 5.9$ at $\unit[10]{TeV}$ and $\mathrm{d}N_\mathrm{ch}/\mathrm{d} \eta |_{\eta=0} \approx 6.4$ at $\unit[14]{TeV}$.

  \subsection{Moments}
    \label{section_meas_moments}

    \bfig
      \includegraphics[width=0.48\linewidth]{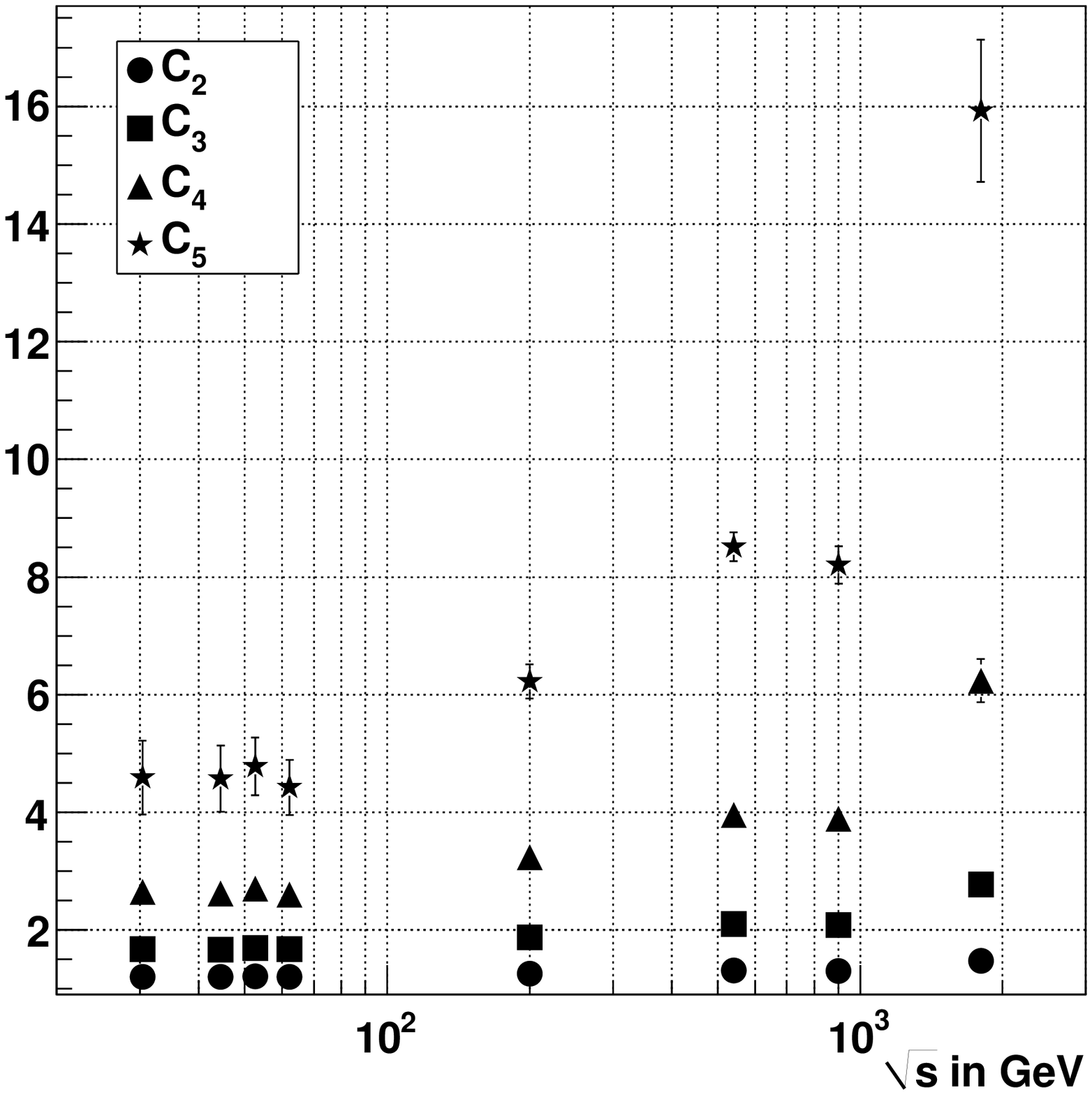}
      \hfill
      \includegraphics[width=0.48\linewidth]{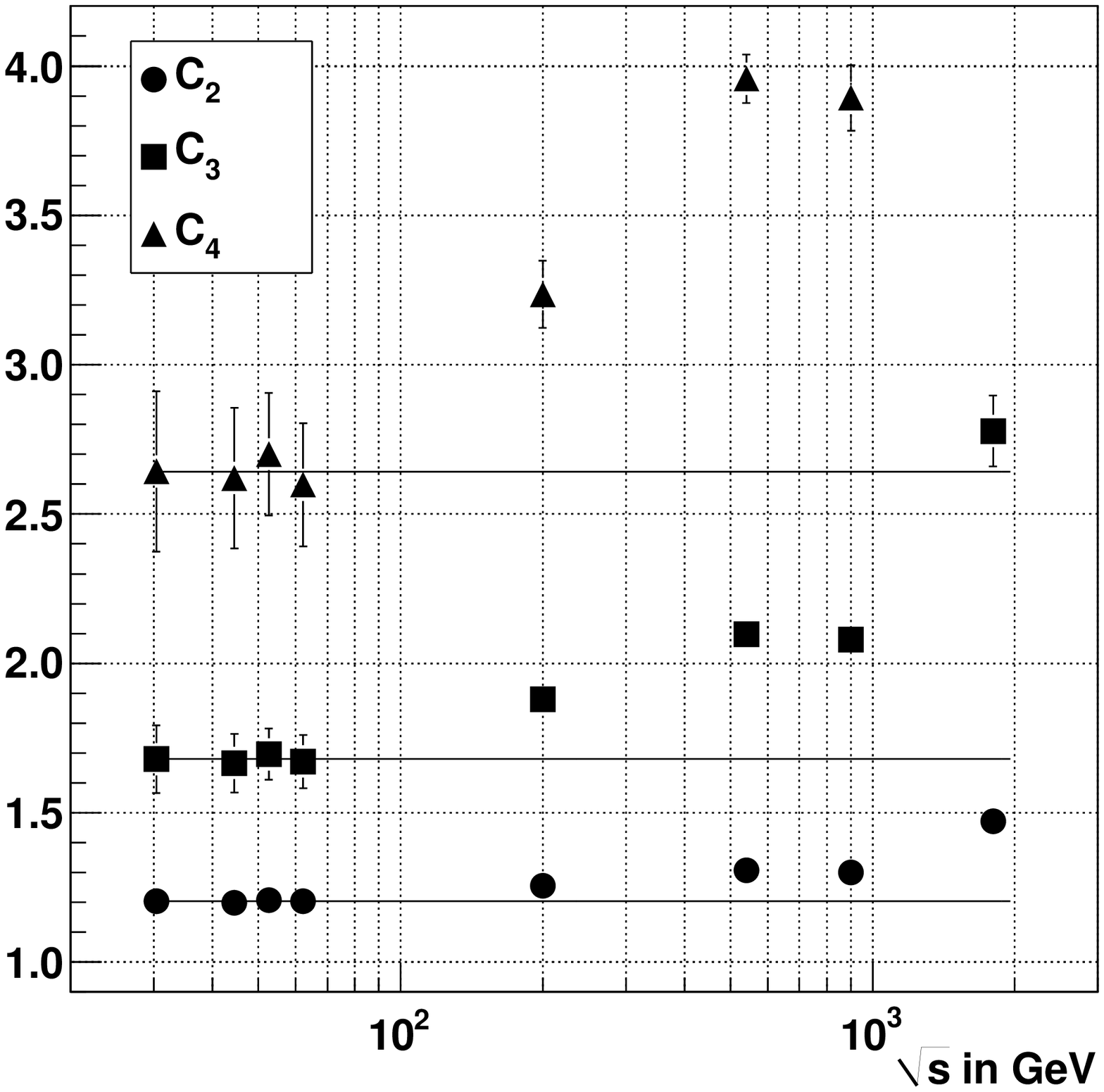}
      \caption{The reduced $C$-moments $C_2$ to $C_5$ are shown at different $\cms$ for distributions of NSD events in full phase space. The right panel shows a zoom. The lines are constant functions fitted to the low-energy data points from the ISR. The data are from \cite{Breakstone:1983ns, Ansorge:1988kn, Alner:1985zc, Alexopoulos:1998bi}.}
      \label{fig_summary_moments_c}
      %MomentsVsCMS.C
    \efig

The moments of the multiplicity distributions as defined in Section~\ref{section_theory_moments} will now be used to identify general trends as function of $\cms$ and to study the validity of KNO scaling. First the reduced $C$-moments, Eq.~\eqref{eq_cmoments}, are studied. The left panel of Figure~\ref{fig_summary_moments_c} shows $C_2$ to $C_5$ from $\cms = 30$ to \unit[1800]{GeV}. These have been calculated from the available multiplicity distributions and are consistent with published values where available. 
However, for the ISR the uncertainties are overestimated due to the fact that the uncertainties on the normalized distributions include the uncertainty on the cross section. 
At lower energies data from bubble-chamber experiments show that the moments are constant (see e.g. \cite{Alner:1985wj} for a compilation). In the right panel a constant is fitted to the data points from the ISR. This emphasizes that for $\cms$ larger than at ISR, the moments increase significantly with energy.

    \bfig
      \includegraphics[width=0.48\linewidth]{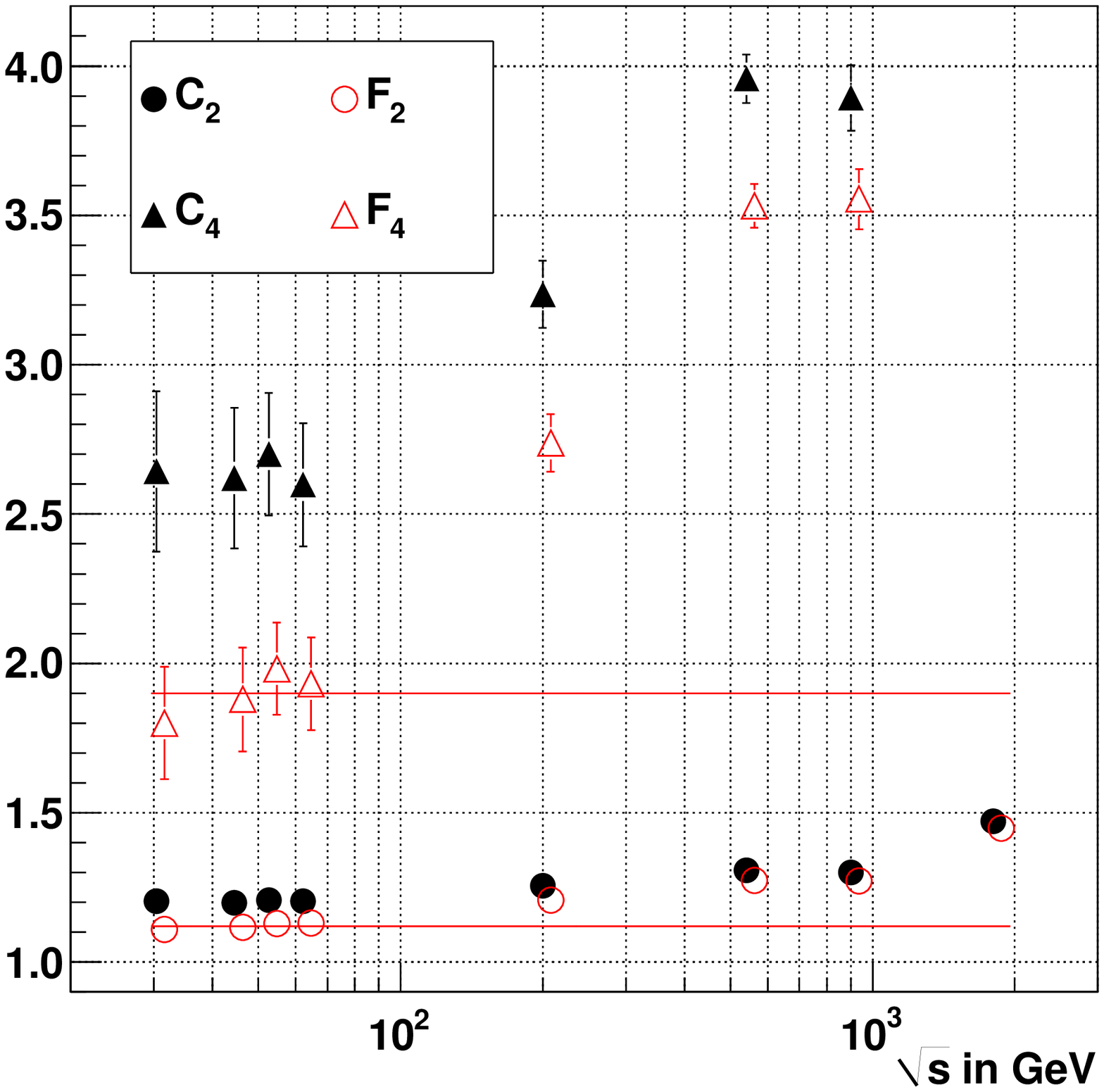}
      \hfill
      \includegraphics[width=0.48\linewidth]{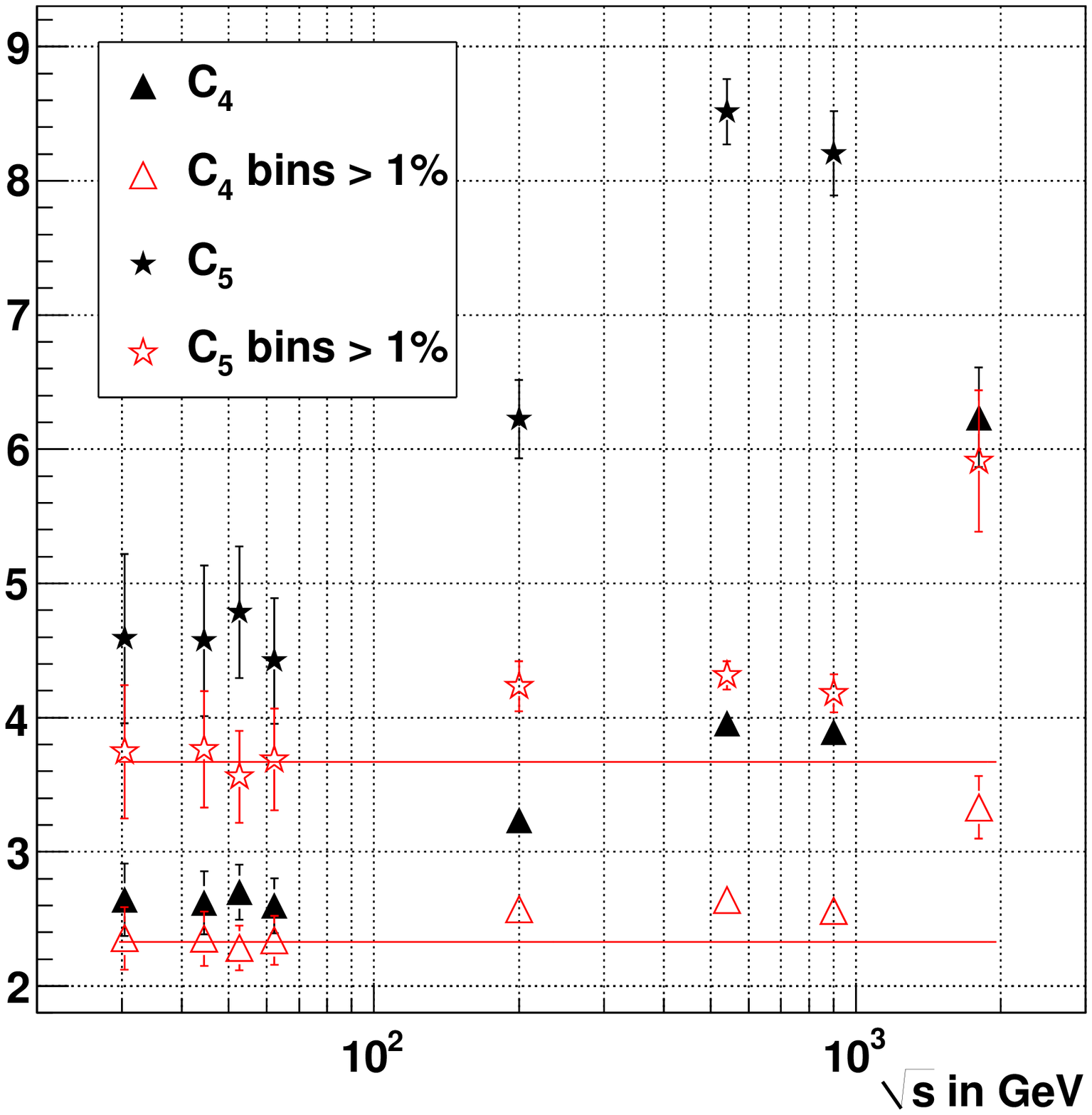}
      \caption{\label{fig_summary_moments_f_and_c_influence} Left panel: $C$- and $F$-moments at different $\cms$. The lines are constant functions fitted to the low-energy data points from the ISR.
      Right panel: Influence of high-multiplicity bins on $C$-moments. The moments $C_4$ and $C_5$ are shown once using all bins for the calculation and once only bins that contain at least 1\% of the topological cross section. The lines are constant functions fitted to the low-energy data points from the ISR. Data from \cite{Breakstone:1983ns, Ansorge:1988kn, Alner:1985zc, Alexopoulos:1998bi}.}
    \efig

    However, as argued in Section~\ref{section_theory_kno}, the conclusion about constant $C$-moments follows from KNO scaling only in an approximation. Therefore the behaviour of factorial moments is analyzed. Exemplarily $F_2$ and $F_4$ are shown in the left panel of Figure~\ref{fig_summary_moments_f_and_c_influence} compared to their $C$-moments counterparts. Also these increase with increasing $\cms$. Both, $C$- and $F$-moments, show an increase with $\cms$ and allow the same conclusion about the validity of KNO scaling.

    It is important to note the influence of the tail of the distribution, i.e., of bins at high multiplicity, on the moments; especially on the higher ones. The right panel of Figure~\ref{fig_summary_moments_f_and_c_influence} compares $C_4$ and $C_5$ calculated from a subset of bins, excluding the ones that are below 0.01 (i.e. less than 1\% of events occur at this multiplicity) which mainly excludes high-multiplicity bins, with the values calculated with all bins. The value of 0.01 is approximately the smallest bin content in the data from ISR. The difference is significant which shows that the moments will change if more events are collected at a given energy. 
    Nevertheless, we see the increase of the moments with $\cms$ although less pronounced. One may ask of course why the moment calculated with all bins and the moment calculated from the subset do not agree within uncertainties. This is due to the fact that for all bins without entries an uncertainty of 0 is assumed which is incorrect. Assuming a Poisson distribution in each bin (with an unknown mean), a bin with no measured entries has an upper limit of 2.3 at 90\% confidence level (see e.g. \cite{Amsler:2008zzb}). However, following this strictly would mean to assign this error for all bins without entries up to infinity. Consequently, also the uncertainty on the moments goes to infinity.

    \bfig
      \includegraphics[width=0.5\linewidth]{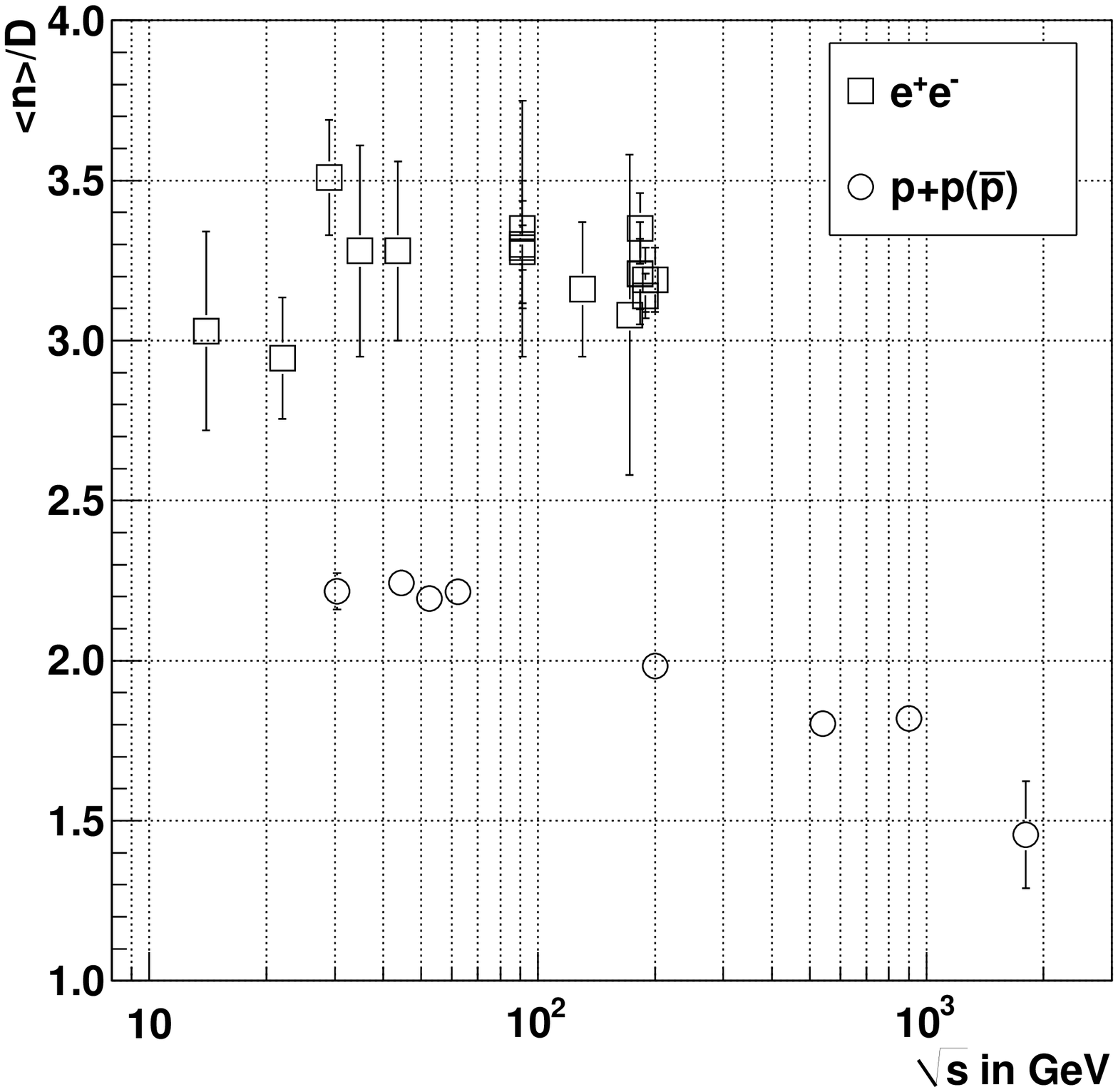}
      \caption{Ratio of average multiplicity  and  dispersion as a function of $\cms$ for $p+p(\bar{p})$ and $e^+e^-$ data. Data from \cite{Breakstone:1983ns, Ansorge:1988kn, Alner:1985zc, Alexopoulos:1998bi} ($p+p(\bar{p})$) and \cite{Braunschweig:1989bp,Derrick:1986jx,Abreu:1990eh,Decamp:1991uz,Abreu:1990cc,Achard:2001ut,Abreu:1996va,Acciarri:1998gz,Abbiendi:1999sx,Abreu:2000gw} ($e^+e^-$). More $p+p$ data points at lower energies are shown in \cite{Giacomelli:1979nu}.}
      \label{fig_summary_moments_dispersion}
      %Dispersion.C
    \efig

    In \figref{fig_summary_moments_dispersion} the ratio of the average multiplicity $\n$ and the dispersion $D$ is shown.
It is constant when KNO scaling holds \cite{Koba:1972ng}. Results for $e^+e^-$ are shown in addition to the $p+p(\bar{p})$ data.
For $p+p(\bar{p})$ the ratio is clearly not constant, while it is approximately constant for $e^+e^-$ albeit with significantly larger errors. At the same $\cms$ the multiplicity distribution in $p+p(\bar{p})$ is significantly broader than in $e^+e^-$.

    In summary, the $C$- and $F$-moments increase with $\cms$, even considering the influence of high-multiplicity bins. Furthermore, $\n/D$ is not constant. These facts clearly demonstrate that KNO scaling is broken.
    
    CDF has addressed the question as to whether the violation of KNO scaling is related to a special class of events \cite{Acosta:2001rm}. They use events at \unit[1.8]{TeV} and only tracks with a $p_T$ above \unit[0.4]{GeV/$c$}. Here, a weak KNO scaling violation is reported in \etain{1.0}. Furthermore, when they divide their data sample into two parts, they can confirm KNO scaling for the soft part of their events and at the same time rule it out for the hard part. In \cite{Acosta:2001rm} soft events are defined as events without clusters of tracks with a total transverse energy above \unit[1.1]{GeV}, regarded as jets.
    
    Two further interesting features are observed together with the onset of KNO scaling violations \cite{Albajar:1989an}: the average transverse momentum that was about \unit[360]{MeV/$c$} at ISR energies starts to increase. Furthermore, a $\cms$ dependent correlation between the average-$p_T$ and the multiplicity is measured. Both observations point to the fact that the influence of hard scattering becomes important at these energies.

    As mentioned earlier, higher-order QCD calculations predict oscillations of the $H$-moments as function of the rank. The uncertainties of moments increase with the rank (see e.g. \figref{fig_summary_moments_c} for $C$-moments); this fact applies also to the $H$-moments. The search for oscillations requires the calculation of moments up to ranks of 10 -- 20. The data studied in this review allow to calculate these moments only with large uncertainties. There are indications of oscillations but definite conclusions require a deeper study and distributions with large statistics that can hopefully be obtained at the LHC.

\subsection{NBD Parameters \texorpdfstring{$\expval\nch$ and $k^{-1}$}{<Nch> and k-1}}

\bfig
\includegraphics[width=0.48\linewidth]{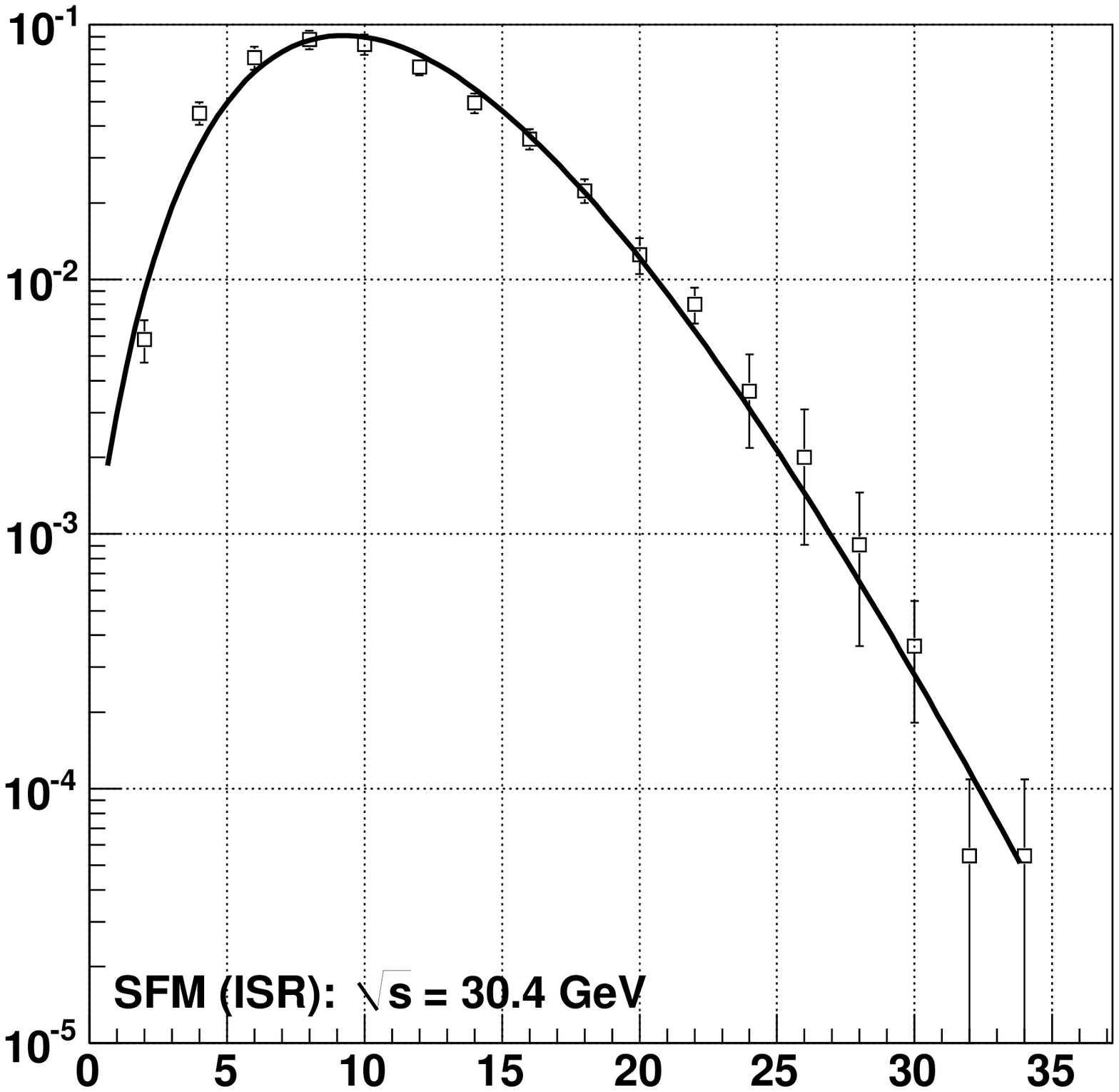}
\hfill
\includegraphics[width=0.48\linewidth]{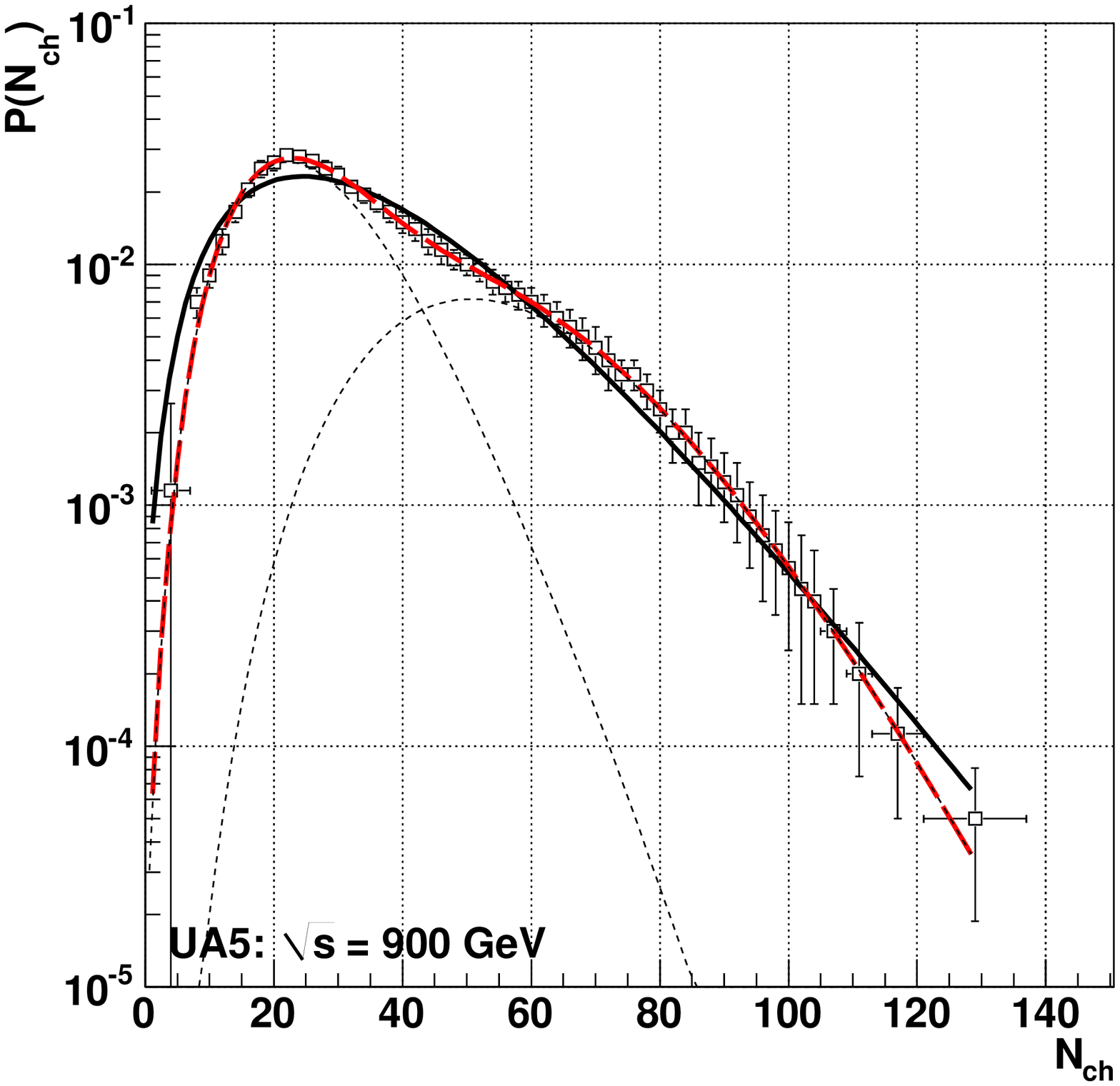}
\caption{\label{fig_summary_nbd_example} Multiplicity distribution at \cmsofG{30.4} measured at the ISR \cite{Breakstone:1983ns} and \cmsofG{900} measured by UA5 \cite{Ansorge:1988kn} fitted with a single NBD (both panels) and a combination of two NBDs (right panel).}
% NBDVsCMS.C
\efig

    Fitting the multiplicity distribution with a single NBD is satisfactory up to about \unit[540]{GeV}; at \unit[900]{GeV} deviations become clearly visible. Distributions at larger $\cms$ can be successfully fitted with a combination of two NBDs.

    \figref{fig_summary_nbd_example} shows exemplarily multiplicity distributions from ISR and UA5 fitted with a NBD. While in the former the NBD reproduces the shape very well, in the latter structures (especially around the peak) are visible that are not reproduced by the fit. Interestingly the $\chi^2/ndf$ of the fit at \cmsofG{900} is still good (see the left panel of \figref{fig_nbd_pars_and_inv_k}).

\bfig
\includegraphics[width=0.48\linewidth]{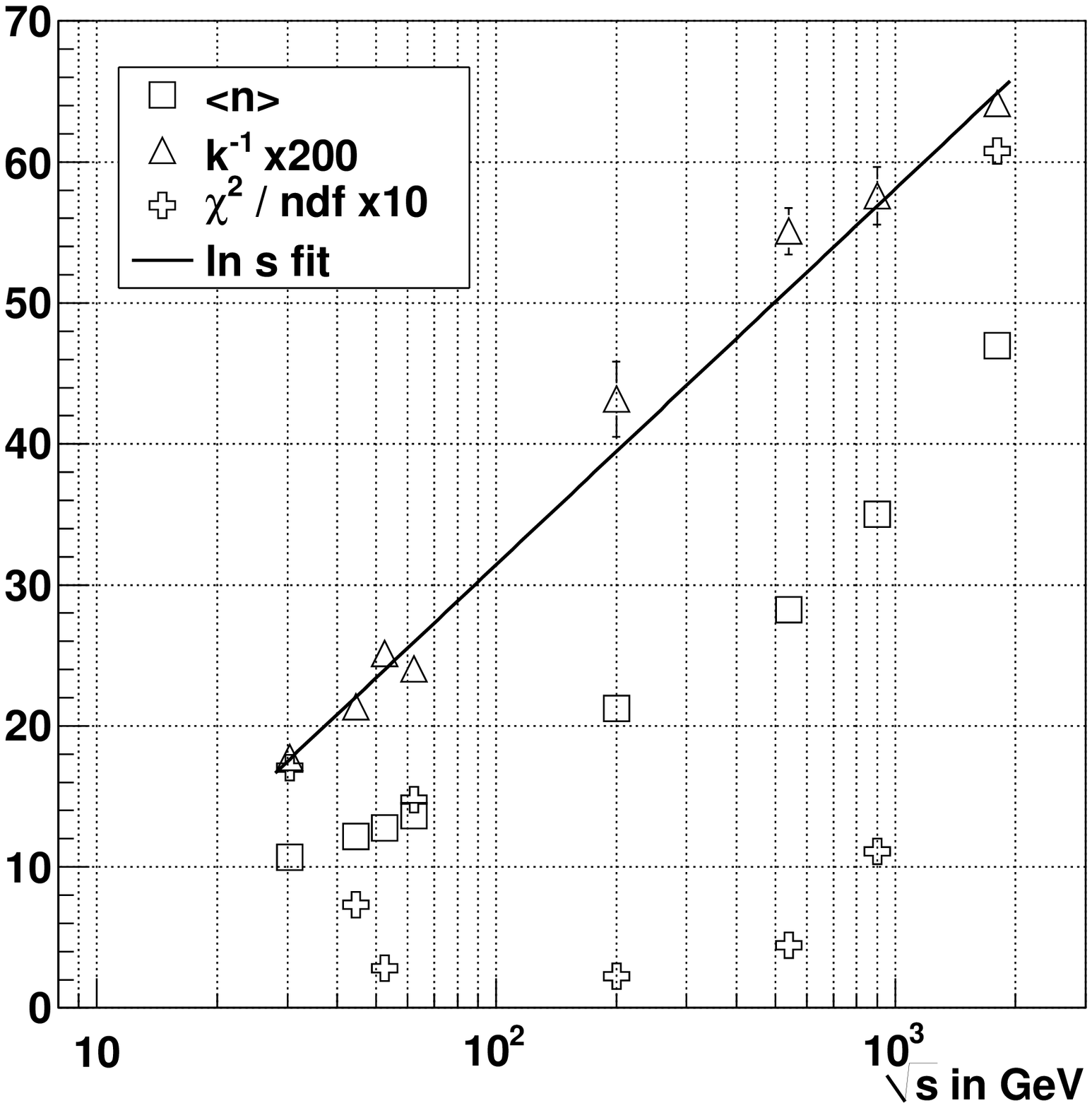}
\hfill
\includegraphics[width=0.48\linewidth]{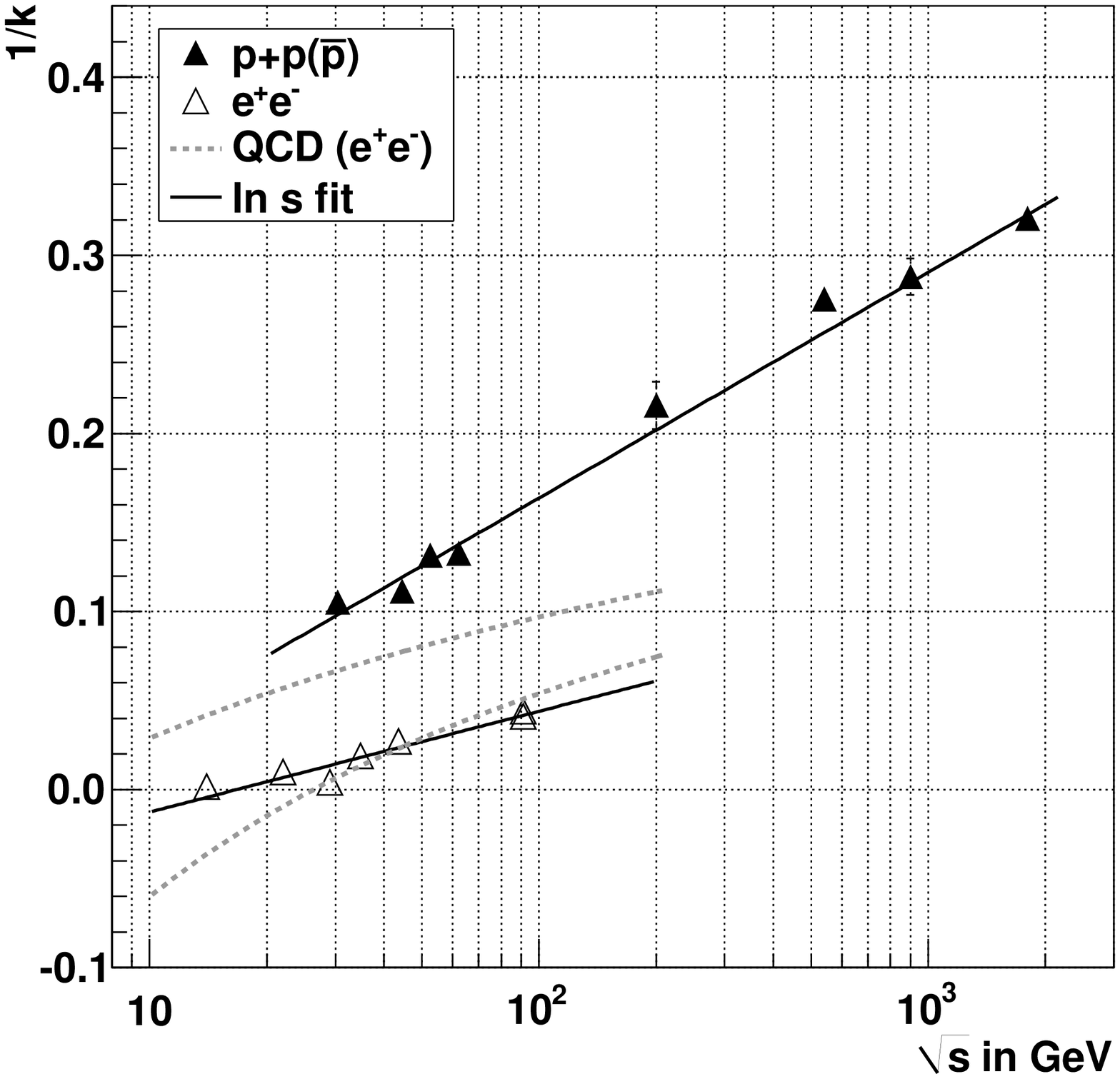}
\caption{\label{fig_nbd_pars_and_inv_k} Left panel: Parameters of a single NBD fit and corresponding $\chi^2/ndf$. $k^{-1}$ and $\chi^2/ndf$ are scaled for visibility. Right panel: Parameters of a single NBD fit compared between $p+p$ (fits performed here) and \ee\ (from \cite{Braunschweig:1989bp, Derrick:1986jx, Decamp:1991uz, Abreu:1990cc}). The area between the dashed lines corresponds to the predictions for $1/k$ from \cite{Malaza:1985jd}.}
% NBDVsCMS.C
\efig

    In \figref{fig_nbd_pars_and_inv_k} (left panel) the obtained fit parameters $\n$ and $k^{-1}$ are shown for datasets in full phase space at different $\cms$ as well as the $\chi^2/ndf$ of the fits. The shown $\chi^2/ndf$ for the data from the ISR is underestimated because as previously mentioned the uncertainties on the normalized distributions include the uncertainty on the cross section. This uncertainty has two components, one applicable to the measurement at a given $\cms$ and one global scale uncertainty \cite{Amaldi:1979kd}. Adding these linearly and removing them from the uncertainty of the normalized distribution leads to an increase of the $\chi^2/ndf$ of about 25\%.
    % with uncertainties from the publication
    % x[0]=30.4, y[0]=17.0405, ex[0]=0, ey[0]=0
    % x[1]=44.5, y[1]=7.30162, ex[1]=0, ey[1]=0
    % x[2]=52.6, y[2]=2.79686, ex[2]=0, ey[2]=0
    % x[3]=62.2, y[3]=14.7806, ex[3]=0, ey[3]=0
    % reducing by linearly added uncertainty on the x-section
    % x[0]=30.4, y[0]=20.2662, ex[0]=0, ey[0]=0
    % x[1]=44.5, y[1]=8.73599, ex[1]=0, ey[1]=0
    % x[2]=52.6, y[2]=3.49546, ex[2]=0, ey[2]=0
    % x[3]=62.2, y[3]=18.0641, ex[3]=0, ey[3]=0
    The average multiplicity $\n$ increases linearly with $\ln \cms$ like it was already discussed in Section~\ref{section_meas_energydependence}. $k^{-1}$ increases with $\cms$ and can be fitted with a function of the form $a + b \ln \cms$. KNO scaling corresponds to a constant, energy-independent $k$. \figref{fig_nbd_pars_and_inv_k} (right panel) compares $k^{-1}$ from $p+p$ and \ee\ data. Both can be fitted with the same functional form, but the values for \ee\ are generally lower, indicating a narrower distribution.
    An extensive compilation of $k^{-1}$ in $p+p$ and $e^+e^-$ collisions can be found in \cite[Figure 2.5]{Kittel:2004xr}. For $e^+e^-$, this compilation includes $k^{-1}$ at lower energies. It is argued in \cite{Kittel:2004xr}  that for LEP energies $k^{-1}$ tends to flatten, i.e., that the KNO scaling regime is reached.
    A discussion about the parameters of NBDs fitted to $p+p$ data can also be found in \cite{Dash:2009iz}.

  \subsection{Two NBD Fits}
    \label{section_meas_twonbd}

      Deviations between the multiplicity distribution and the fit with a single NBD are found at highest \spps\ energies. The combination of two NBDs (Eq.~\eqref{eq_twocomponent}) yields better agreement with the data. Both fit attempts are shown in the right panel of \figref{fig_summary_nbd_example} for \cmsofG{900}. Fits with two NBDs can be performed unconstrained or following an approach that constrains the parameters as, e.g., suggested in \cite{Giovannini:1998zb}.

      In \cite{Giovannini:1998zb} first the average multiplicity of the soft component $\n_\mathrm{soft}$ using only data below \cmsofG{60} and the total average multiplicity $\n_\mathrm{total}$ using available data up to $\cms$ of \unit[900]{GeV} are fitted. A logarithmic dependence is assumed for $\n_\mathrm{soft}$, while additionally for $\n_\mathrm{total}$ a $\ln^2$-term is added. 
      
      Following the assumption based on a minijet-analysis by UA1 \cite{Giovannini:1998zb} that the semi-hard component has about twice the average multiplicity than the soft component, $\alpha$ can be calculated from $\n_\mathrm{soft}$ and $\n_\mathrm{total}$. Two variants are considered, variant A in which $\n_\mathrm{semi\mhyphen{}hard} = 2\n_\mathrm{soft}$, and variant B with $\n_\mathrm{semi\mhyphen{}hard} = 2\n_\mathrm{soft} + 0.1 \ln^2 \cms$. 
      
      The parameter $k_\mathrm{soft}$ is found to be rather constant between \unit[200 and 900]{GeV} and thus set to $k_\mathrm{soft} = 7$. Three scenarios are then presented in \cite{Giovannini:1998zb} for the extrapolation to higher energies. The first assumes that KNO scaling is valid above \unit[900]{GeV} ($k_\mathrm{semi\mhyphen{}hard} \approx 13$).
      Scenario 2 fits $k_\mathrm{total}$ with:
      \bq
        k_\mathrm{total}^{-1} = a + b \ln \cms.
      \eq
      Scenario 3 fits a next-to-leading order QCD prediction to $k_\mathrm{semi\mhyphen{}hard}$:
      \bq
        k_\mathrm{semi\mhyphen{}hard}^{-1} \approx a - \sqrt{b/\ln (\cms/Q_0)}. \label{eq_2nbd_ksemihard}
      \eq
      Note that in the original publication \cite[Eq.~(12)]{Giovannini:1998zb} Eq.~\eqref{eq_2nbd_ksemihard} is incorrectly printed but used correctly in the calculations and figures. The correct formula can be found in \cite{Giovannini:2004yk}.
      The free parameters $a$, $b$, $Q_0$ are then found by fitting the data. These three scenarios (1--3) can be combined with the aforementioned variants A and B, resulting in a total of six possibilities. Here we restrict ourselves to only three of them (1--3 combined with A).
      
      \figref{2nbd_fit_comp_withref} shows the functional forms found in \cite{Giovannini:1998zb}. These are compared to the unconstrained results obtained from fitting the distributions with Eq.~\eqref{eq_twocomponent}. Note that only the data from \cmsofG{200} to \unit[900]{GeV} were used to fit the functional forms in \cite{Giovannini:1998zb}. There are clear differences, e.g., at \unit[200]{GeV} for $\n_\mathrm{semi\mhyphen{}hard}$ and at \unit[540]{GeV} for $k_\mathrm{semi\mhyphen{}hard}^{-1}$. One also observes large errors for certain fits showing that several solutions with similarly small $\chi^2/ndf$ exist. The $\chi^2/ndf$ is, as expected, generally better for unconstrained fitting. In several cases the $\chi^2/ndf$ is significantly lower than 1 which is unexpected and might be attributed to the requirement of smoothness in the unfolding procedure. 
      
      At \unit[1.8]{TeV} the fraction of soft events is much larger in the unconstrained fit; also the other fit parameters do not follow the extrapolations. Consequently, for the constrained fit, the $\chi^2/ndf$ is very large at this energy, the fit is not very good. 
      \figref{2nbd_comp_e735} shows the multiplicity distribution in full phase space at \cmsofT{1.8} compared to the predictions of this model. Only scenario~3 follows the general trend of the distribution. However, none of the curves reproduces the distribution in detail.
      
      \bfig
        \includegraphics[width=0.48\linewidth]{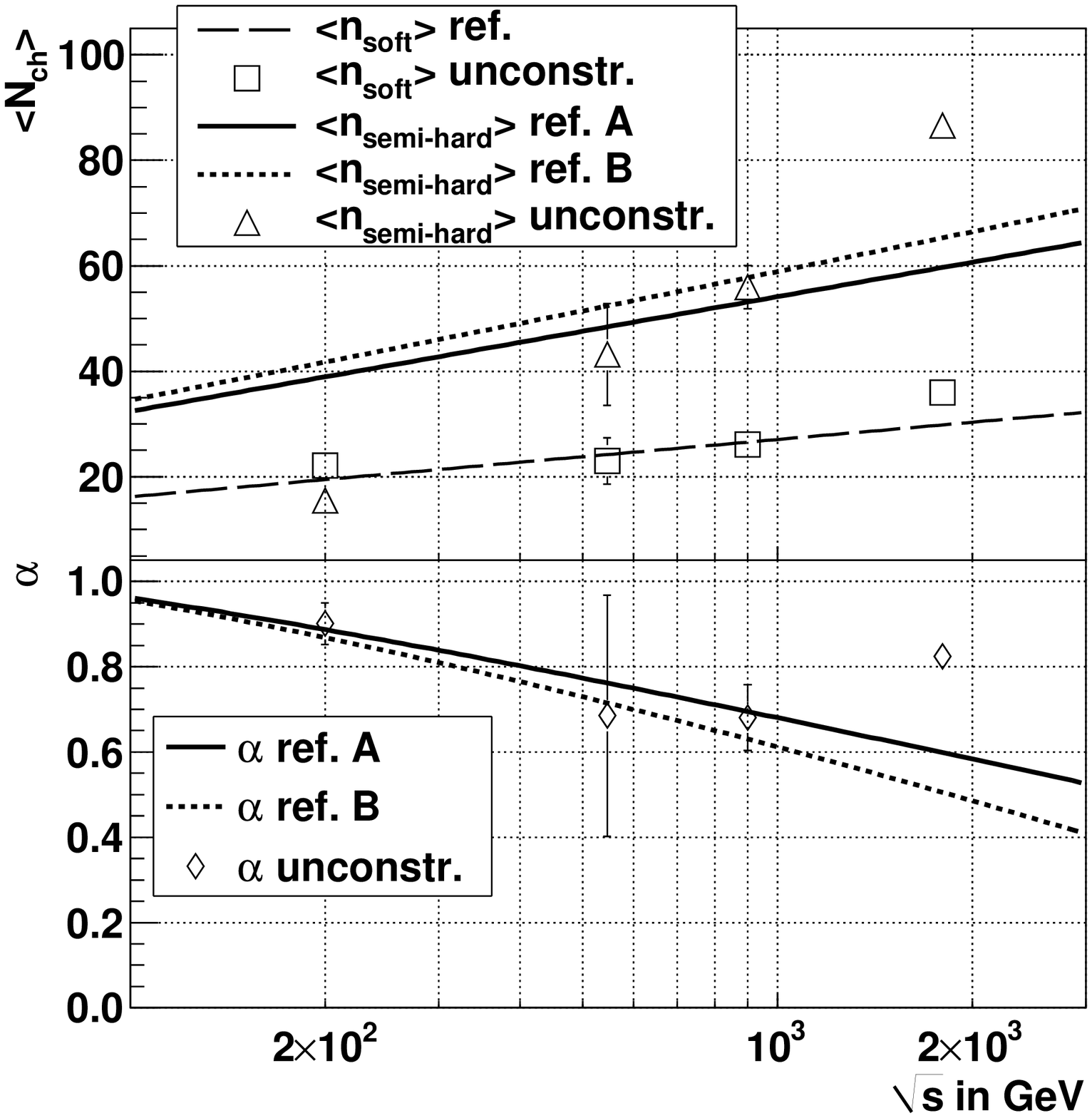}
        \hfill
        \includegraphics[width=0.48\linewidth]{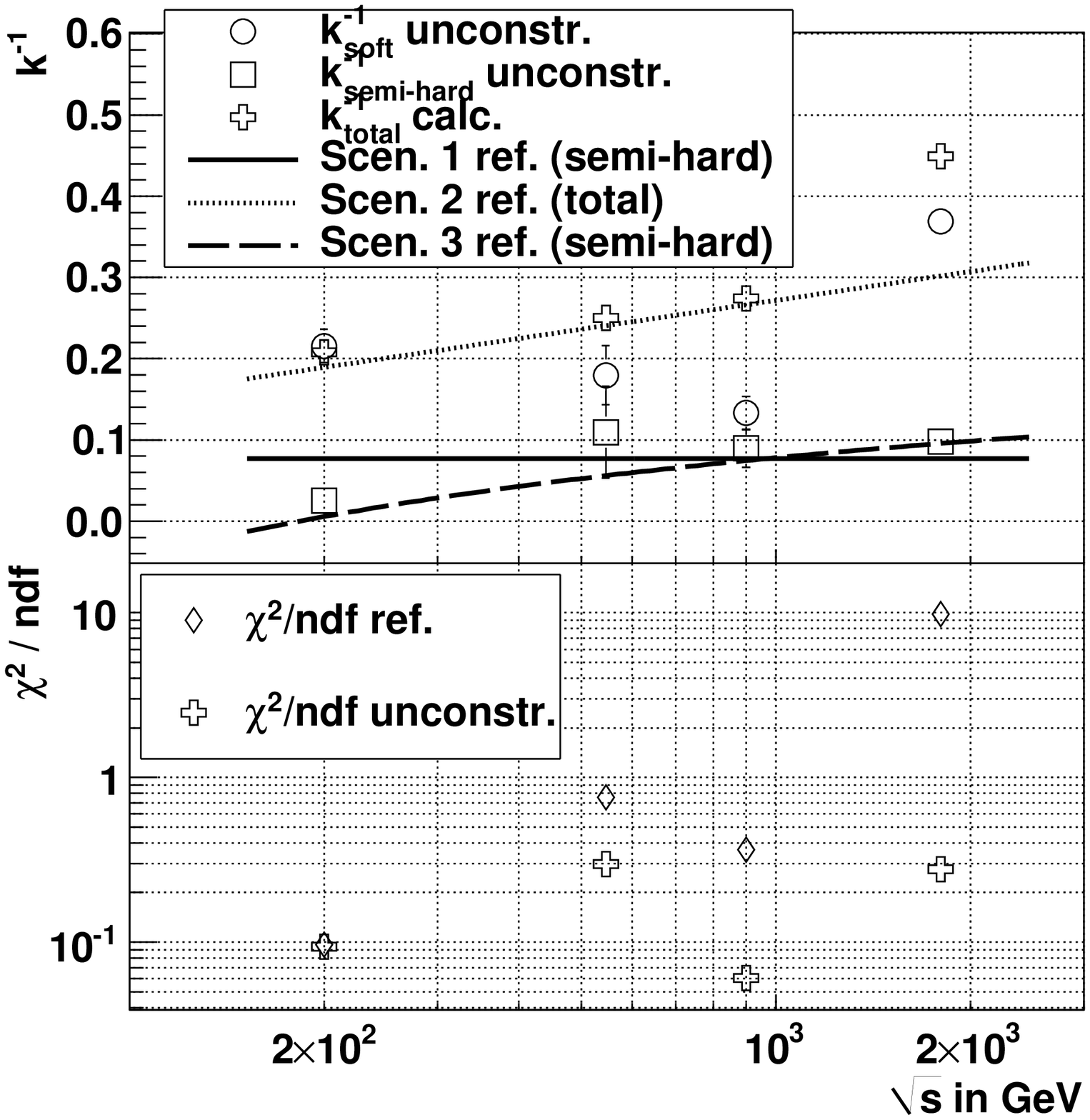}
        \caption{\label{2nbd_fit_comp_withref} The functional forms (lines) found in \cite{Giovannini:1998zb} are compared to unconstrained fits of all five parameters (points). In addition the $\chi^2/ndf$ is shown. The data at largest $\cms$ are from E735 \cite{Alexopoulos:1998bi}; the others are from UA5 \cite{Ansorge:1988kn, Alner:1987wb}. A and B in the legends refer to variants A and B in \cite{Giovannini:1998zb} (see text).}
      \efig
            
        \bfig
          \includegraphics[width=0.48\linewidth]{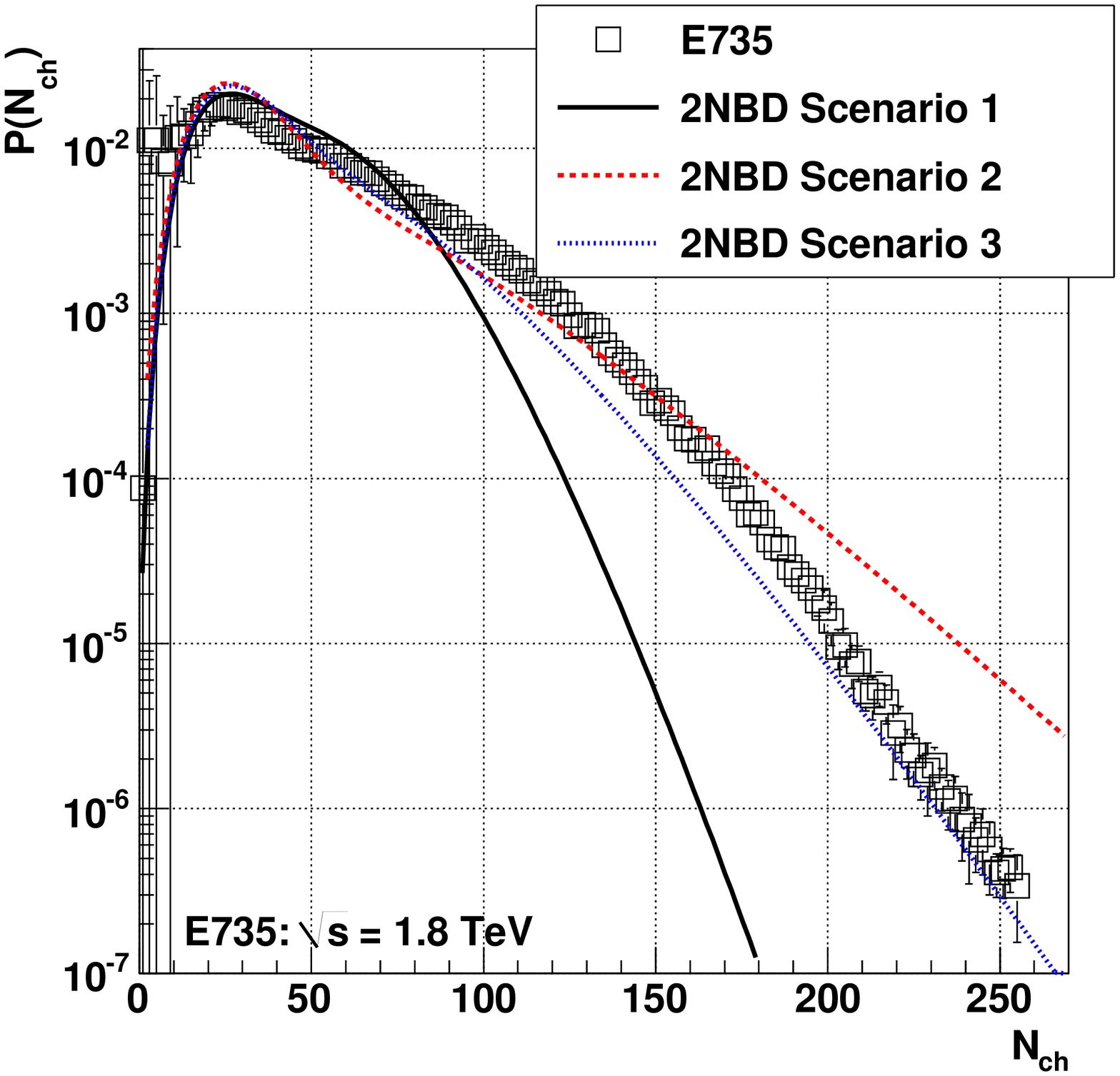}
          \hfill
          \includegraphics[width=0.48\linewidth]{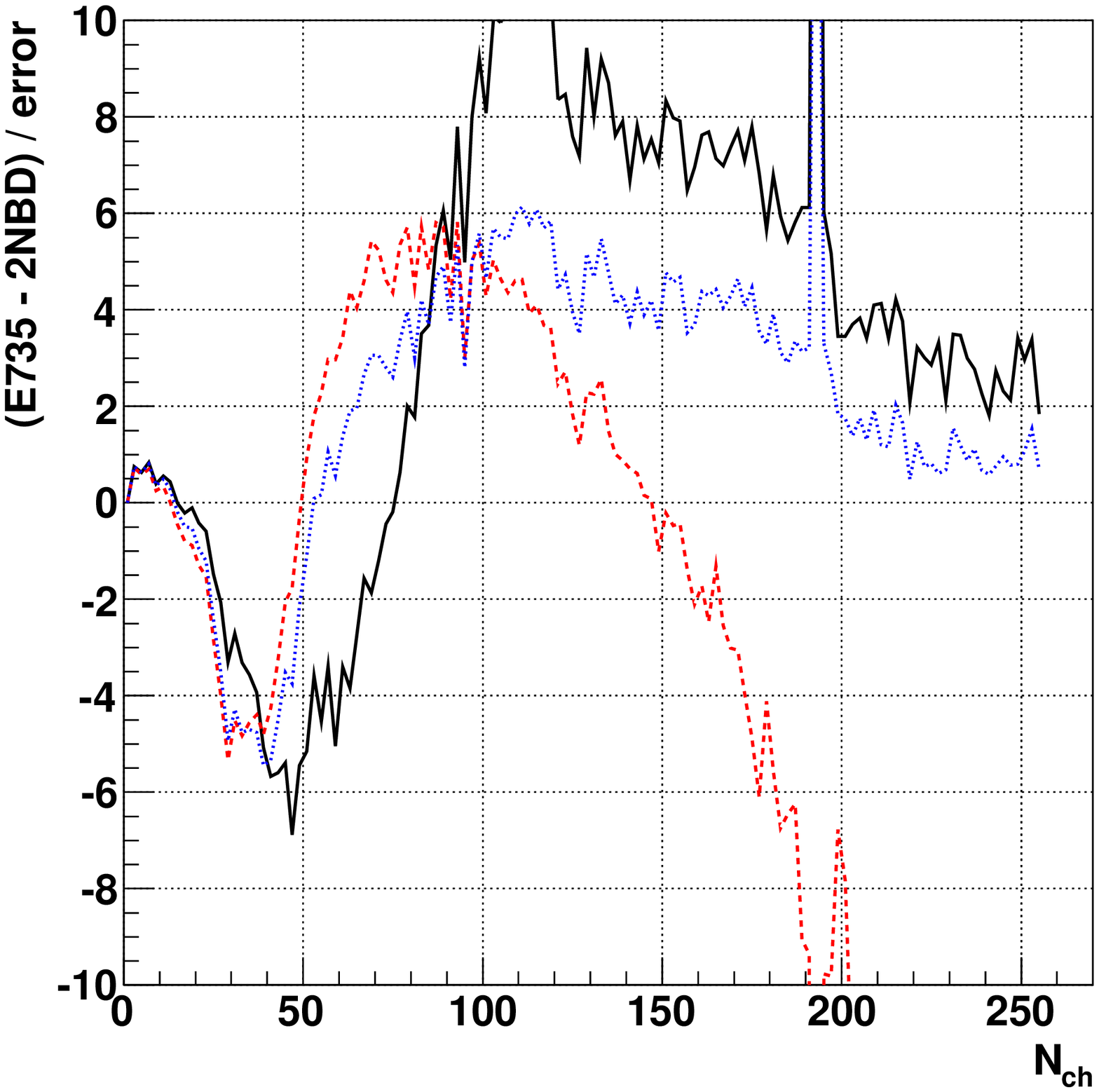}
          \caption{\label{2nbd_comp_e735} Comparison between the predictions of the two-component model \cite{Giovannini:1998zb} with the E735 measurement in full phase space at \cmsofT{1.8} \cite{Alexopoulos:1998bi}. The right panel shows normalized residuals between data and the predictions.}
        \efig
        
        \bfig
          \includegraphics[width=0.48\linewidth]{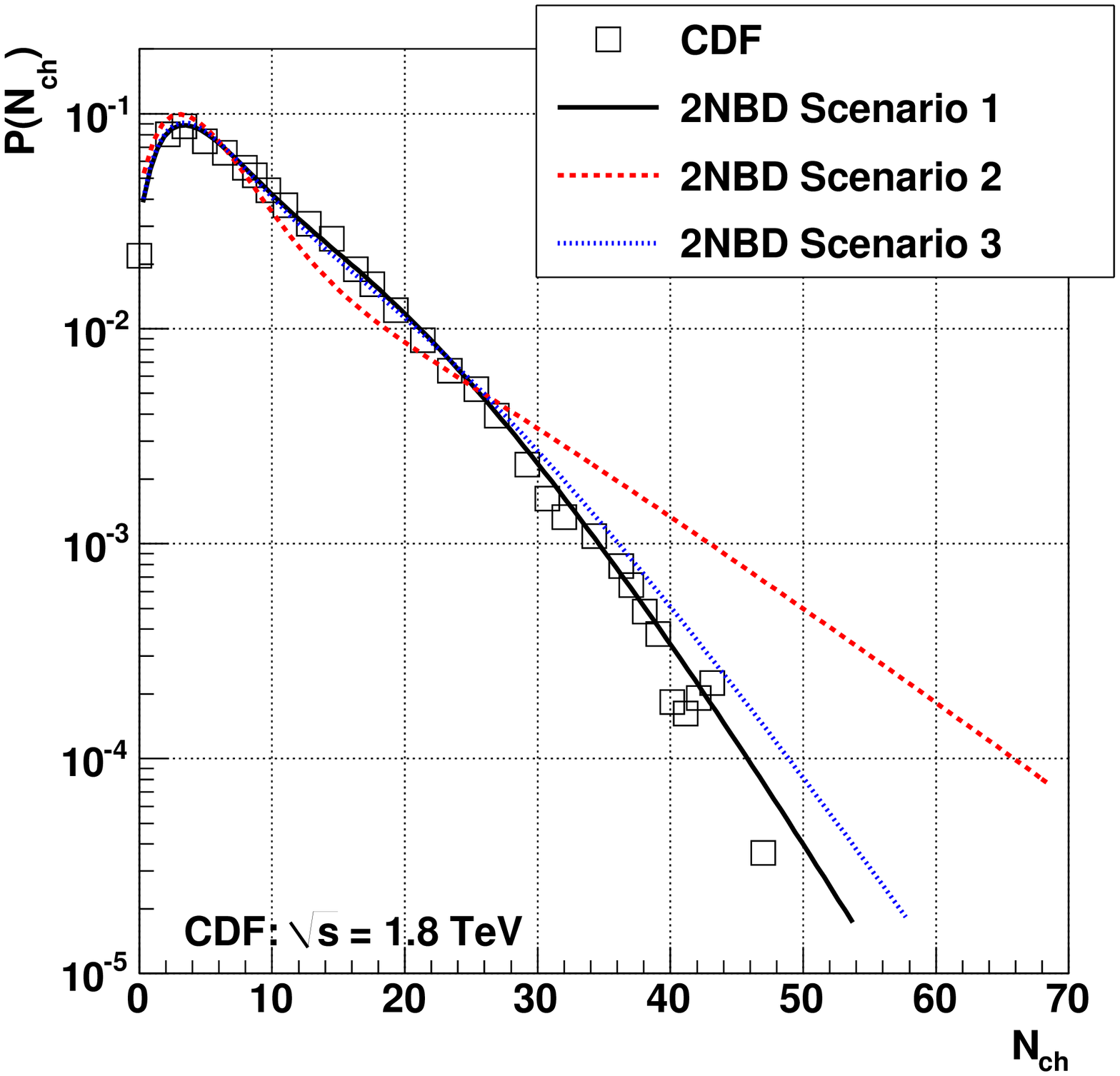}
          \hfill
          \includegraphics[width=0.48\linewidth]{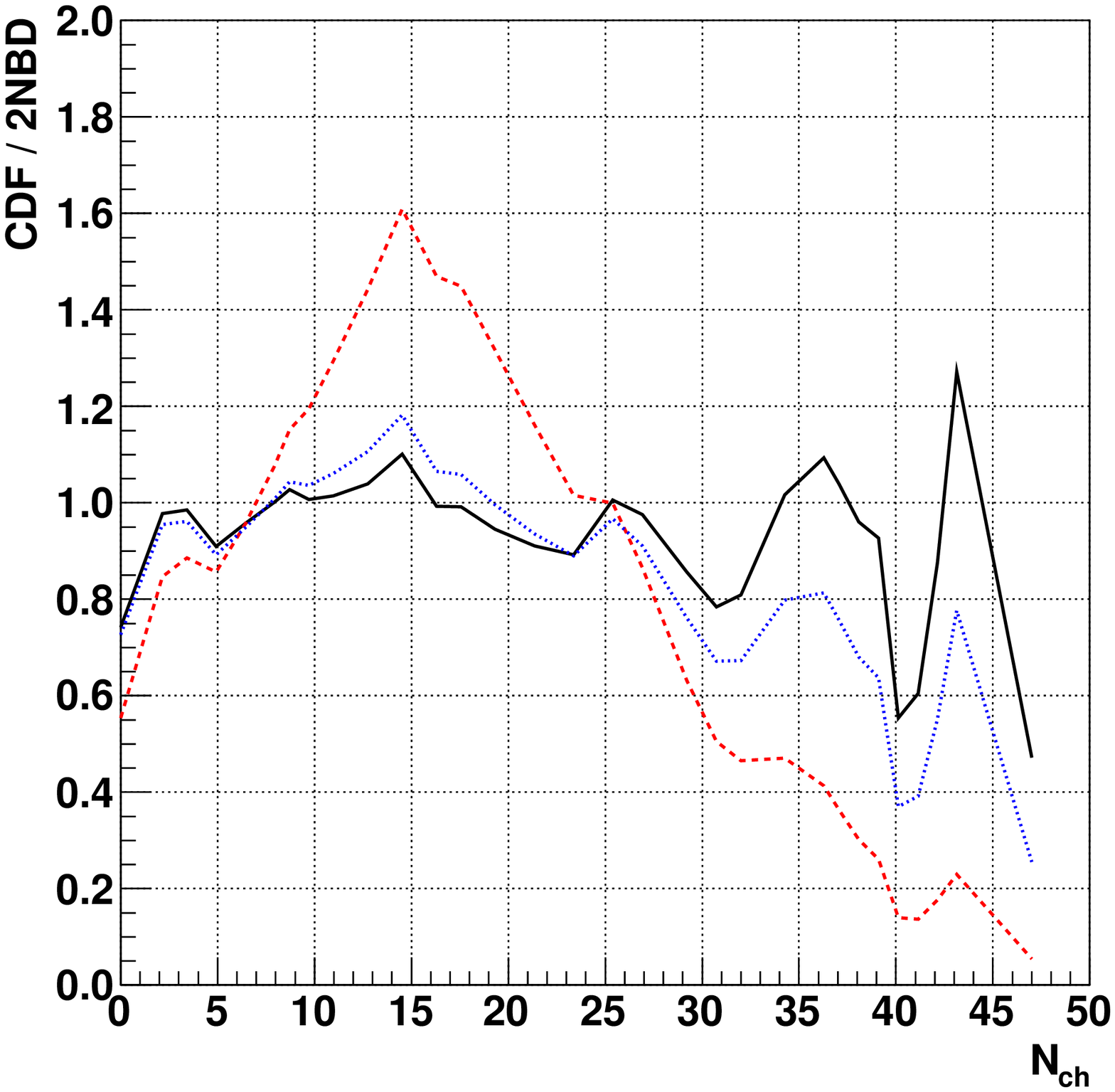}
          \caption{\label{2nbd_comp_cdf} Comparison between the predictions of the two-component model  \cite{Giovannini:1999tw} with the CDF measurement in \etain{1} at \cmsofT{1.8} \cite{cdf_multiplicity1}. The right panel shows the ratio between data and the predictions.}
        \efig

        Figure~\ref{2nbd_comp_cdf} shows a comparison of predictions of this model (using values from the authors derived for limited phase space in \cite{Giovannini:1999tw}) with data from CDF in \etain{1} at \cmsofT{1.8}. Scenario 1 and 3 reproduce the spectrum reasonably well.

      We conclude that unconstrained fits with two NBDs work successfully with a reasonable low $\chi^2/ndf$ for all distributions considered here. However, general trends as a function of $\cms$ cannot readily be identified. As an alternative in \cite{Giovannini:1998zb} parameters are fixed following certain assumptions resulting in more systematic fit results. However, the results are in some cases significantly different from the parameters obtained by the unconstrained fits.

  \subsection{Open Experimental Issues}
    \label{section_experimentalissues}

    This section addresses open experimental issues and presents some comparison plots between data of disagreeing experiments.

		\bfig
			\includegraphics[width=0.48\linewidth]{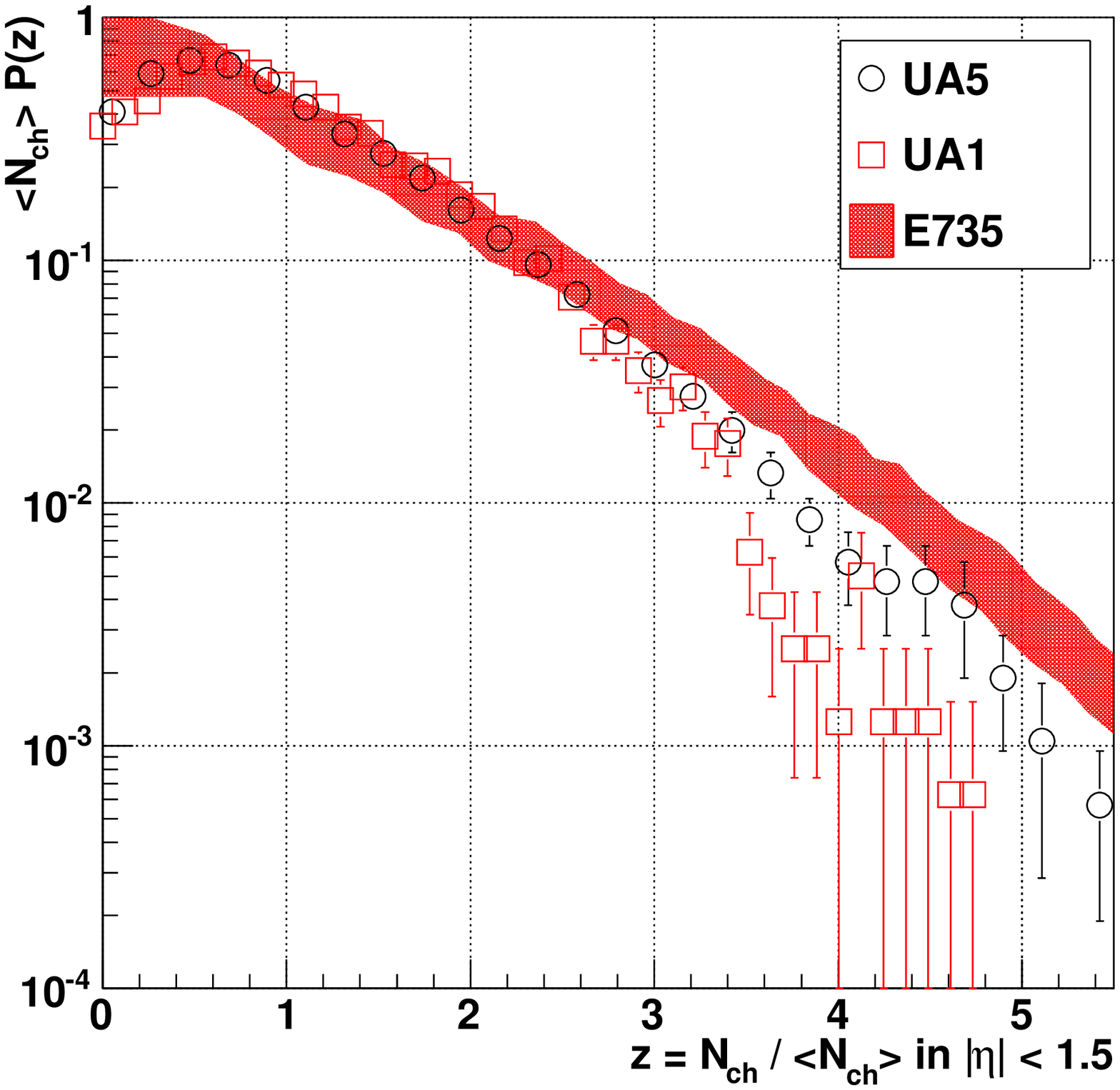}
			\caption{\label{fig_inconsistencies_ua5_ua1} Multiplicity distribution of NSD events measured by UA1 and UA5 at \cmsofG{540} in $|\eta| < 1.5$ shown in KNO variables \cite{Arnison:1982rm, Alner:1985zc}. Furthermore, data from E735 at \cmsofT{1.8} in \etain{1.62} is shown \cite{Lindsey:1991pt}. See the text for an explanation of the E735 band.}
			%hepdata: UA5: http://durpdg.dur.ac.uk/cgi-hepdata/hepreac/1353390
			%         UA1:
			%KNOViolation.C
    \efig
		
		A direct comparison between UA1 and UA5 at \cmsofG{540} in limited $\eta$-regions and in KNO variables shows that the two experiments agree in their confirmation of KNO scaling in the interval $|\eta| < 0.5$. However, they disagree in the interval $|\eta| < 1.5$, but the violation of KNO scaling in the UA5 data is only due to an excess of events with $z > 3.5$, i.e., events that have more then 3.5 times the average multiplicity. This comparison has been performed in \cite{Albajar:1989an} and is shown for $|\eta| < 1.5$ in Figure~\ref{fig_inconsistencies_ua5_ua1}. It also includes E735 data in \etain{1.62} which shows better agreement with the UA5 data in the tail. The band corresponds to the region where data points are taken from the original figure which is of poor quality \cite[Figure 2]{Lindsey:1991pt}. It corresponds to data points from \cmsofG{300} to \unit[1.8]{TeV}. The band most likely overestimates the error bars of the single points. Nevertheless, E735 confirmed KNO scaling in \etain{1.62} based on their data; this was done only by comparing the distributions in KNO variables and not by studying the moments \cite{Lindsey:1991pt}. A final conclusion about the slight KNO violation in \etain{1.5} cannot be made at present.

		\bfig
			\includegraphics[width=0.48\linewidth]{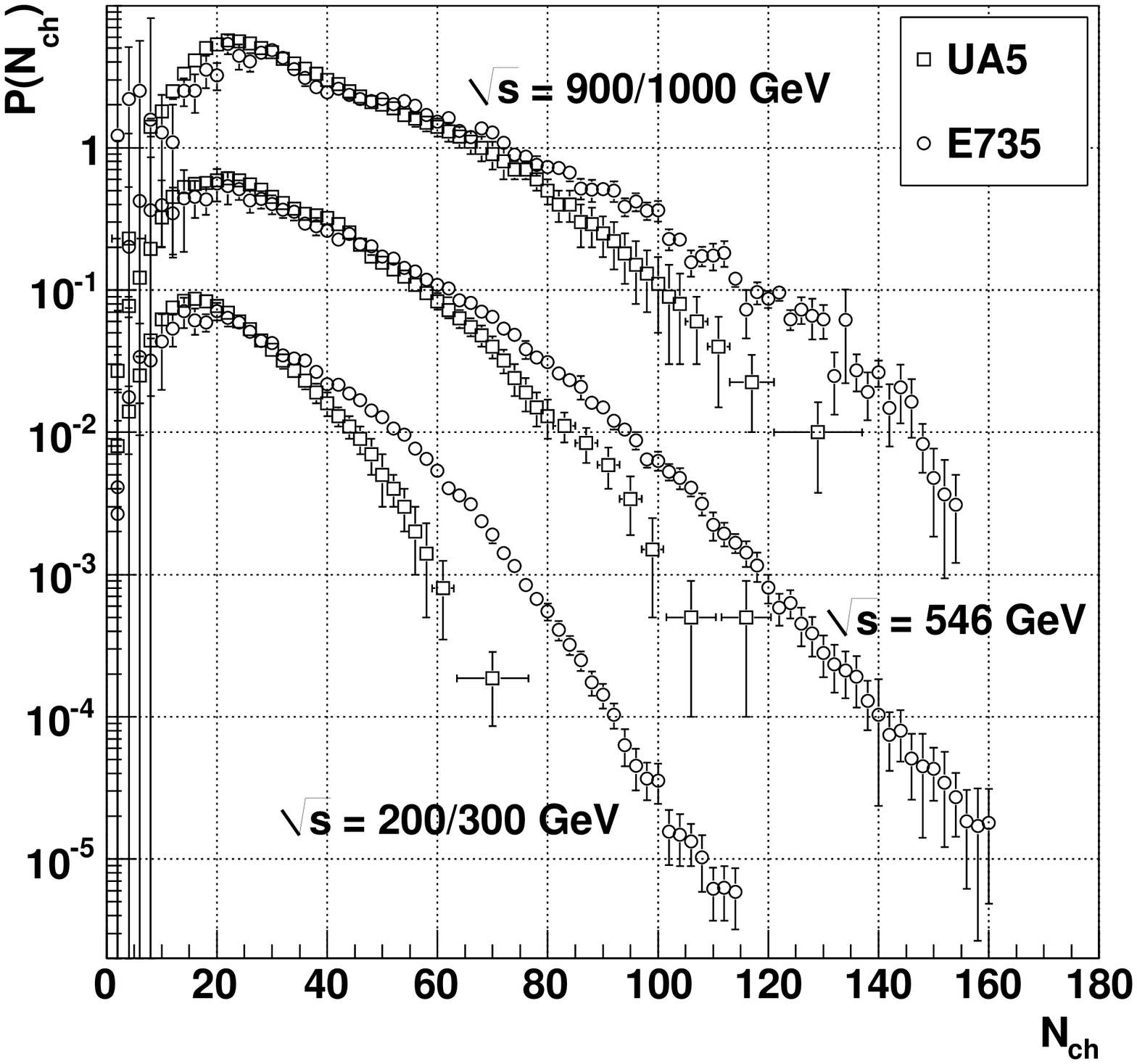}
			\hfill
			\includegraphics[width=0.48\linewidth]{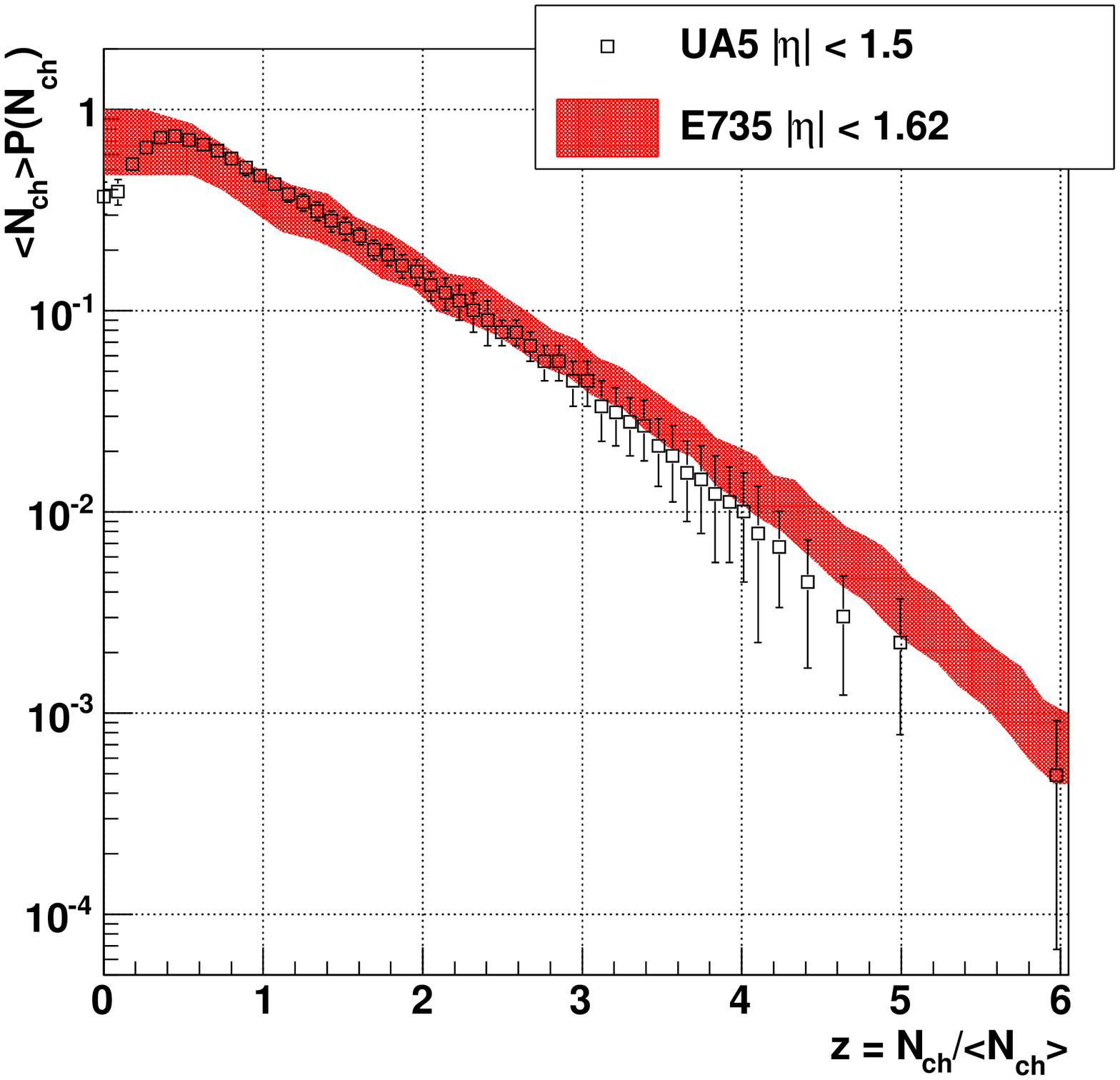}
			\includegraphics[width=0.48\linewidth]{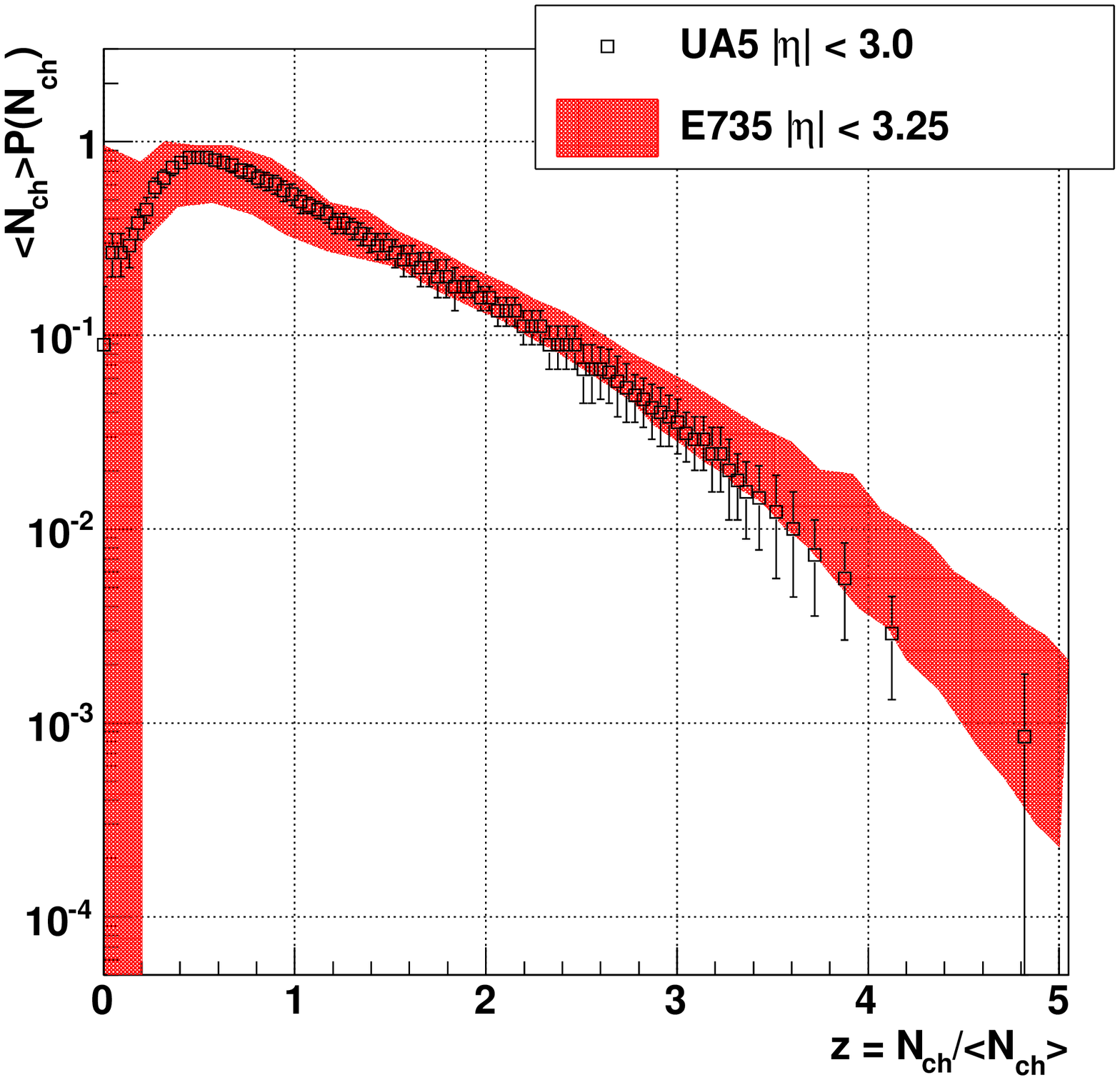}
			\hfill
			\includegraphics[width=0.48\linewidth]{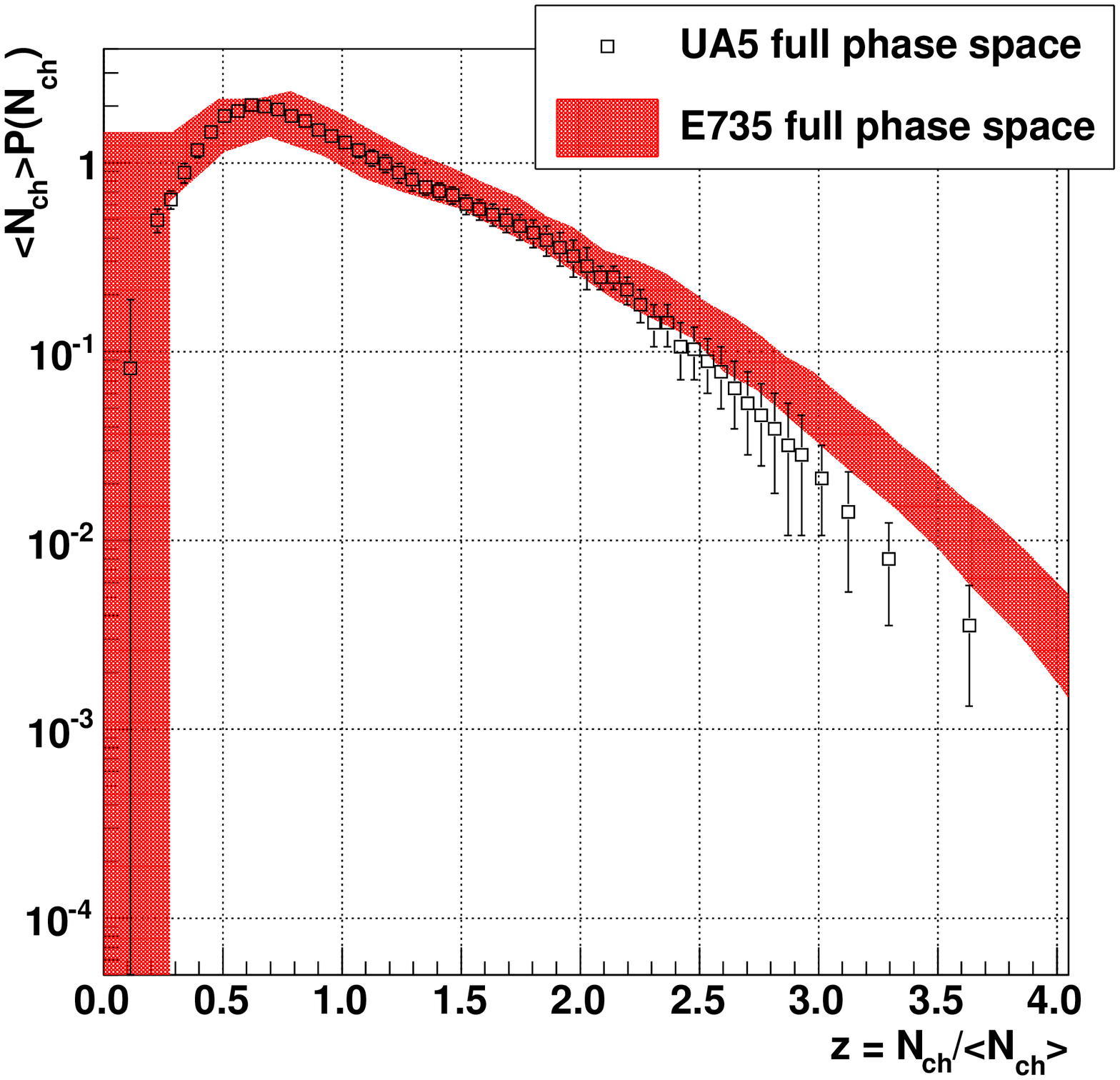}
		  \caption{Top left panel: \label{fig_inconsistencies_ua5_e735} Comparison of UA5 and E735 data in full phase space at three different energies. Data from \cite{Ansorge:1988kn, Alner:1987wb, Alexopoulos:1998bi}.
		  Other panels: Comparison between UA5 \cite{Ansorge:1988kn} at \cmsofG{900} and E735 \cite{Lindsey:1991pt} in approximately equivalent $\eta$-regions (top right and bottom left panels) and in full phase space (bottom right panel).}
		  %UA5VsE735.C
		  %UA5VsE735.C --> UA5VsE735_KNO
		\efig

    Reference~\cite{Alexopoulos:1998bi} compares multiplicity distributions in full phase space from E735 and UA5 at three different energies (see the top left panel of \figref{fig_inconsistencies_ua5_e735}). The distributions disagree especially in their tails. This inconsistency has been frequently quoted \cite{Walker:2004tx, Kittel:2004xr}. However, it is important to note that the data from E735 are extrapolated from \etain{3.25} to full phase space which may imply a significant systematic uncertainty. Also the data from UA5 are extrapolated, in this case starting from \etain{5}. This is less of a problem since an estimation based on Pythia at \cmsofG{900} shows that 86\% (64\%) of the particles are emitted in \etain{5} (3.25). A direct comparison of data from E735 and UA5 in an $\eta$-region where both detectors are sensitive is therefore very interesting. The top right and bottom left panel of Figure~\ref{fig_inconsistencies_ua5_e735} show such a comparison for similar $\eta$-regions: \etain{1.5}~(1.62) and \etain{3.0}~(3.25) for UA5 (E735). The bottom right panel of \figref{fig_inconsistencies_ua5_e735} shows the comparison in full phase space. Due to the fact that the E735 data are only available in KNO variables, the UA5 data are shown superimposed in KNO variables, too. No scaling correction has to be applied due to the different $\eta$-regions because the data are shown in KNO variables and thus already scaled with the average multiplicity. It can be seen that in both $\eta$-intervals the distributions agree within errors (apart from very low multiplicities in the smallest $\eta$-region). The discrepancy appears going to full phase space. Hence, a systematic effect in the extrapolation procedure may be suspected as the cause of the discrepancy.

    Furthermore, it should be noted that results from CDF and E735 deviate from each other in similar phase space regions, see \figref{fig_inconsistencies_cdf_e735}. Studying the multiplicity distribution in KNO variables shows that CDF results at \unit[1.8]{TeV} are closer to the UA5 results at \unit[900]{GeV} than to the E735 results at \unit[1.8]{TeV} (plot not shown).

		\bfig
			\includegraphics[width=0.48\linewidth]{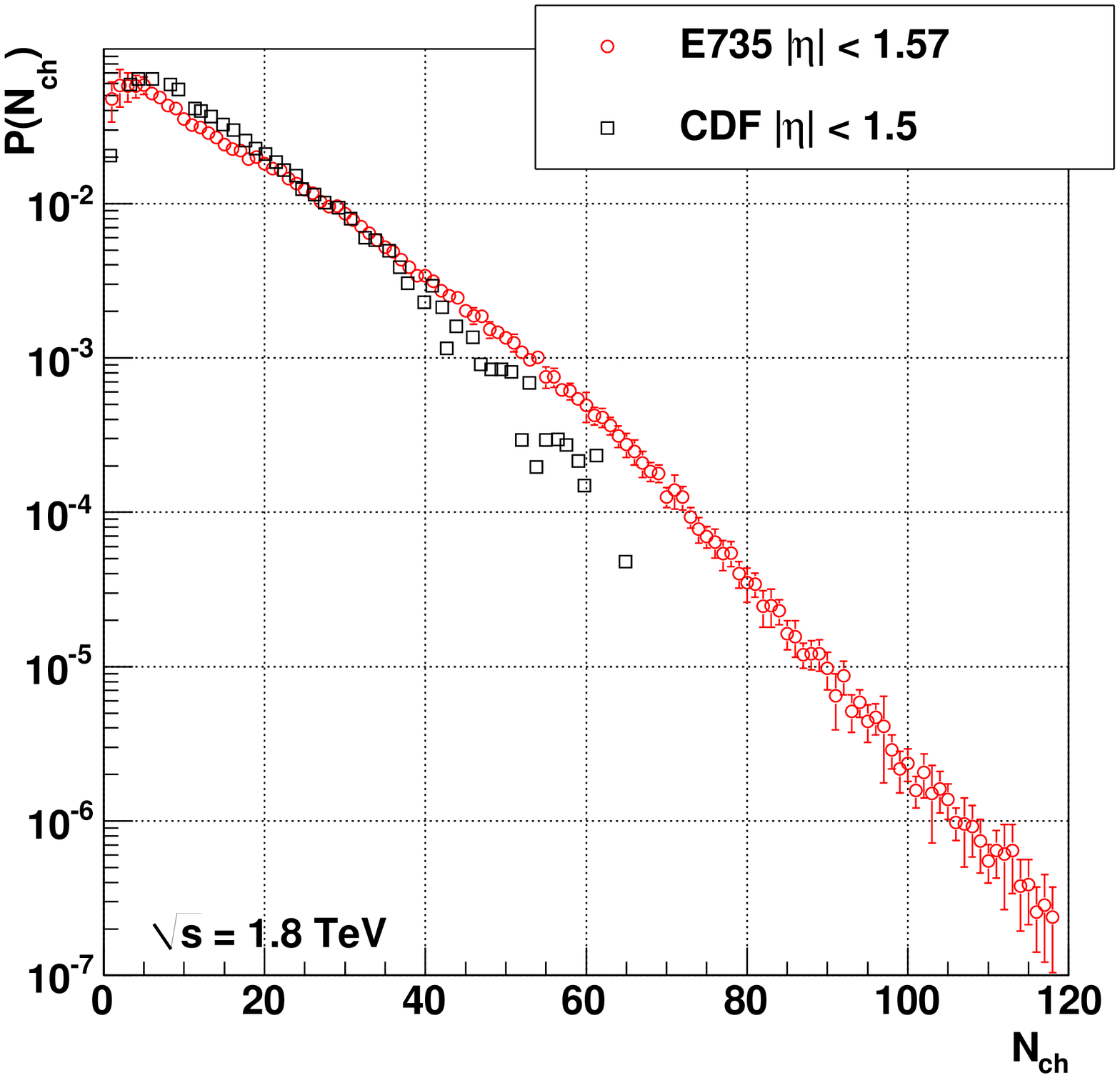}
			\hfill
			\includegraphics[width=0.48\linewidth]{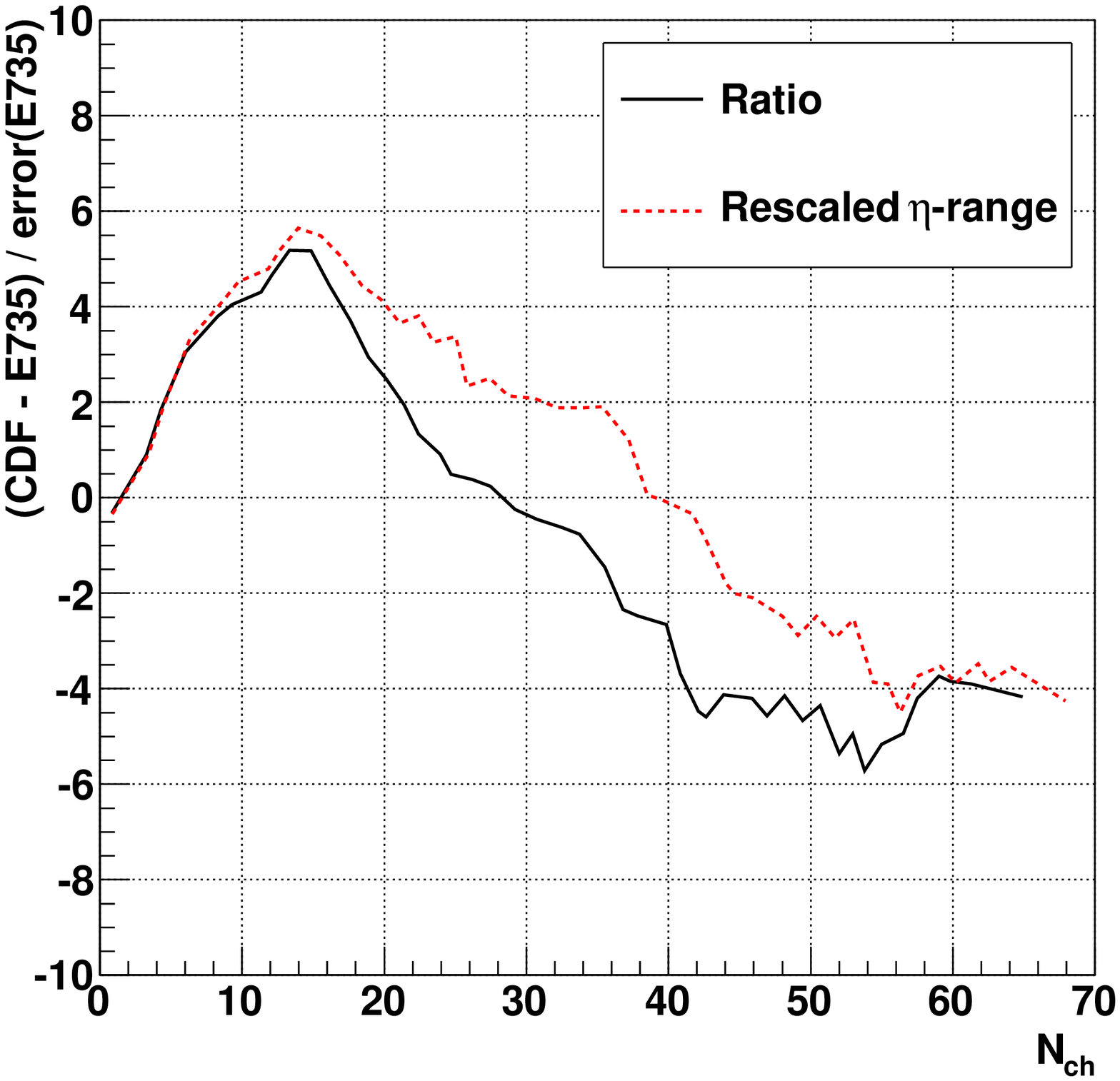}
			\caption{\label{fig_inconsistencies_cdf_e735} Data from CDF \cite{cdf_multiplicity1} and E735 \cite{wang_thesis} at \cmsofT{1.8} are compared. The right panel shows normalized residuals using the error of the E735 data (these have been smoothed over 5 data points to reduce fluctuations): the black curve compares the experiments directly; the dashed red curve takes the different $\eta$-ranges into account by simply scaling the CDF $N_\mathrm{ch}$ axis by a factor ($1.57/1.5$).}
			%CDFVsE735.C
		\efig
		
		In summary, there are various experimental inconsistencies, especially in the tail of the distributions which has a significant influence on, e.g., the calculation of moments of higher rank. It will be interesting to compare data taken at the LHC with the existing distributions.

\section{Predictions}
\label{sec_predictions}
The measurement of the charged-particle multiplicity at the LHC has the potential to improve our understanding of multi-particle productions mechanisms by rejecting models based on incorrect assumptions. In \figref{fig_predictions_dndeta_and_nch} predictions for $\dndetaZero$ and $\langle N_\mathrm{ch} \rangle$ in full phase space are summarized. Predictions for $\dndetaZero$ range between about $4.3-6.4$ at $\sqrt{s} = \unit[7]{TeV}$ and $4.5-7.8$ at $\sqrt{s} = \unit[14]{TeV}$. For the charged multiplicity in full phase space the range is about $55-75$ at $\sqrt{s} = \unit[7]{TeV}$ and $65-90$ at $\sqrt{s} = \unit[14]{TeV}$. Measured values outside these ranges would come as a surprise.

The predictions can be classified into several classes. First there are the simple extrapolations of trends observed at lower $\sqrt{s}$ (CDF \cite{Abe:1989td}, Busza \cite{Abreu:2007kv, Armesto:2009ug}). 
Those logarithmic extrapolations from Section~\ref{section_meas_energydependence} that fit the data reasonably well are shown. The $\ln^2 \sqrt{s}$ extrapolation performed in this review is conceptually identical to the extrapolation performed by CDF; however, additional data points from UA1 and P238 have been used in this review which leads to about a 5\% lower extrapolated value at larger energies than found by CDF.
The predictions based on the $p+p$/$e^+e^-$ universality discussed in Section~\ref{section_universality} also belong to this class. Other model predictions are based on the assumption of gluon saturation (Armesto, Salgado, Wiedemann \cite{Armesto:2004ud} and Kharzeev, Levin, Nardi \cite{Kharzeev:2004if}). 
QGSM is a representative of a class of models for soft scattering based on Regge theory and the parton structure of hadrons \cite{Kaidalov:2009hn}. In these models proton--proton interactions are described in terms of the exchange of colour-neutral objects called Pomerons. The multiple-particle production is governed by the fragmentation of strings that occur in the cut Feynman diagrams of these processes.

In many cases it is more practical to implement theoretical ideas in terms of Monte Carlo event generators. Phojet  \cite{Engel:1999xb} is such a generator based on the Dual Parton Model \cite{Capella:1992yb} whose concepts are similar to the concepts used in QGSM. Based on the Pomeron picture Phojet accounts both for soft and hard interactions. Epos is another event generator that aims at consistently treating soft and hard interactions \cite{Werner:2008zza, Abreu:2007kv}. This model has been compared and tested with data from high-energy cosmic rays. Epos can be run in a mode which allows the formation of a quark--gluon plasma in $p+p$ collisions. In the Pythia event generator \cite{Sjostrand:2006za} the picture of individual parton--parton scatterings, which successfully describes high-$p_T$ phenomena, is extrapolated to low $p_T$. Pythia has many parameters and several Pythia tunes exist which, e.g., describe Tevatron data well. The shown predictions are based on the default Pythia settings and three frequently-used tunes: A \cite{Albrow:2006rt}, D6T \cite{Albrow:2006rt}, and ATLAS MC09 which are the main tunes used by the CDF, CMS, and ATLAS experiments, respectively. Pythia 6.4.14 has been used with the structure functions CTEQ6L \cite{Pumplin:2002vw}.
The large increase in $\sqrt{s}$ from the Tevatron to the LHC will unveil whether certain Pythia tunes really capture the underlying physics or whether they are just ad hoc descriptions at specific energies.      

\bfig
\includegraphics[width=0.48\linewidth]{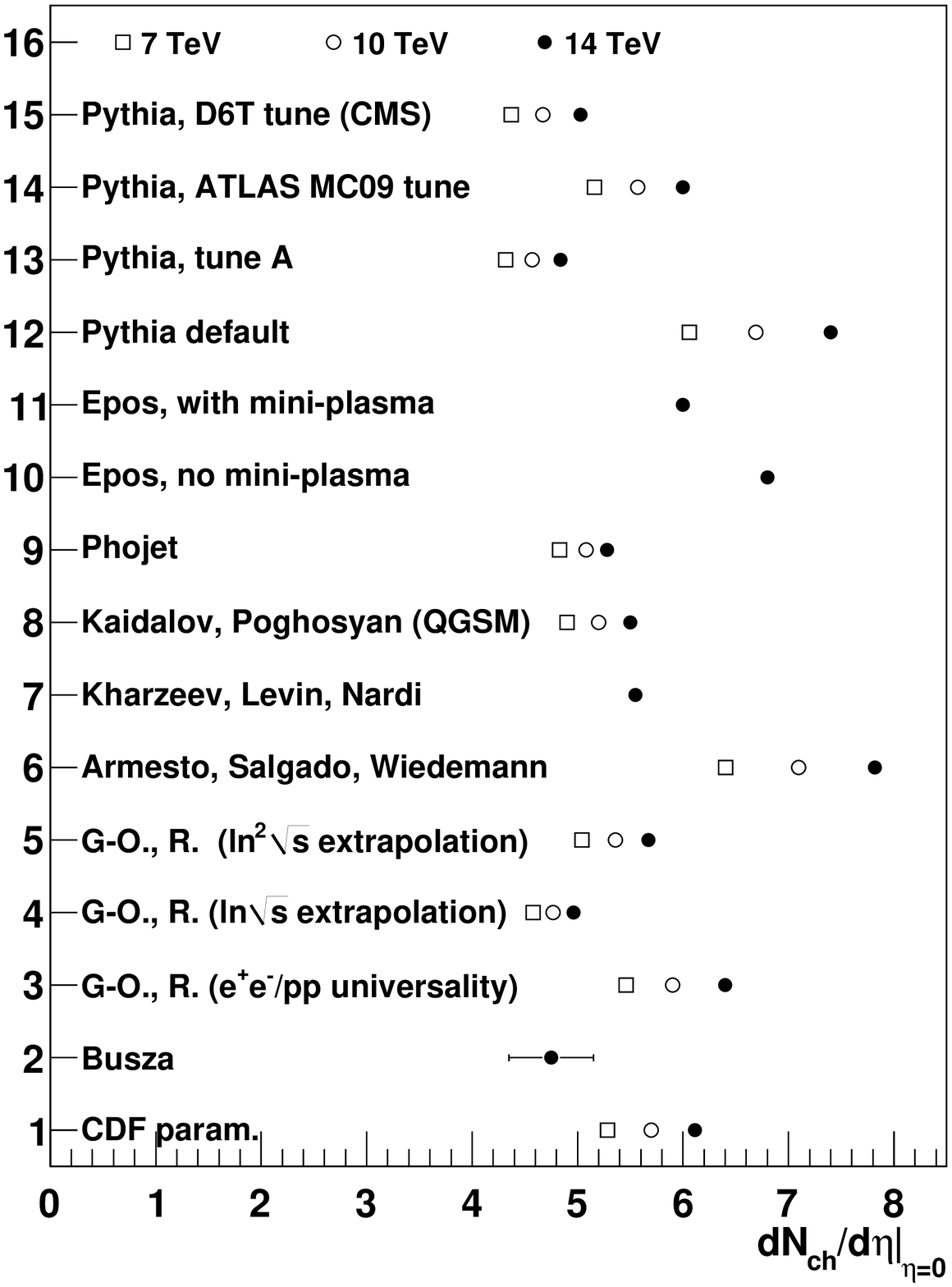}
\hfill
\includegraphics[width=0.48\linewidth]{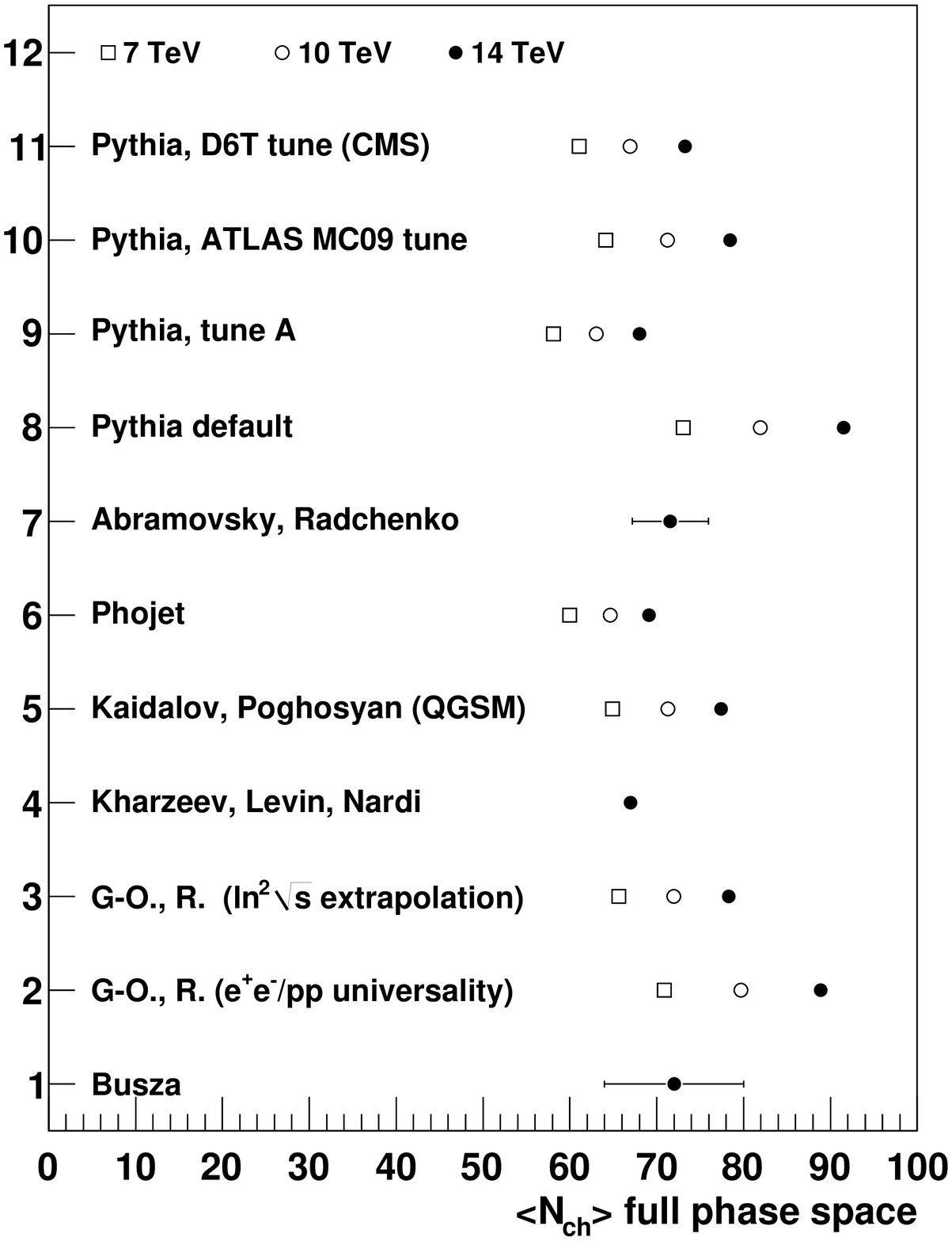}
\caption{\label{fig_predictions_dndeta_and_nch} Predictions for $\dndetaZero$ and $\langle N_\mathrm{ch} \rangle$ in full phase space in $p+p$ collisions at $\sqrt{s} = 7$, $10$, and $\unit[14]{TeV}$. For $\dndetaZero$ predictions 6 and 7 \cite{Armesto:2004ud, Kharzeev:2004if} it is not explicitly stated whether the predictions are for inelastic, NSD, or non-diffractive collisions; all other predictions \cite{Abe:1989td, Abreu:2007kv, Armesto:2009ug, Kaidalov:2009hn, Engel:1999xb, Werner:2008zza, Abramovsky:2008uh, Sjostrand:2006za} are for NSD events.}
%Predict.C
\efig

      \bfig
        \includegraphics[width=0.48\linewidth]{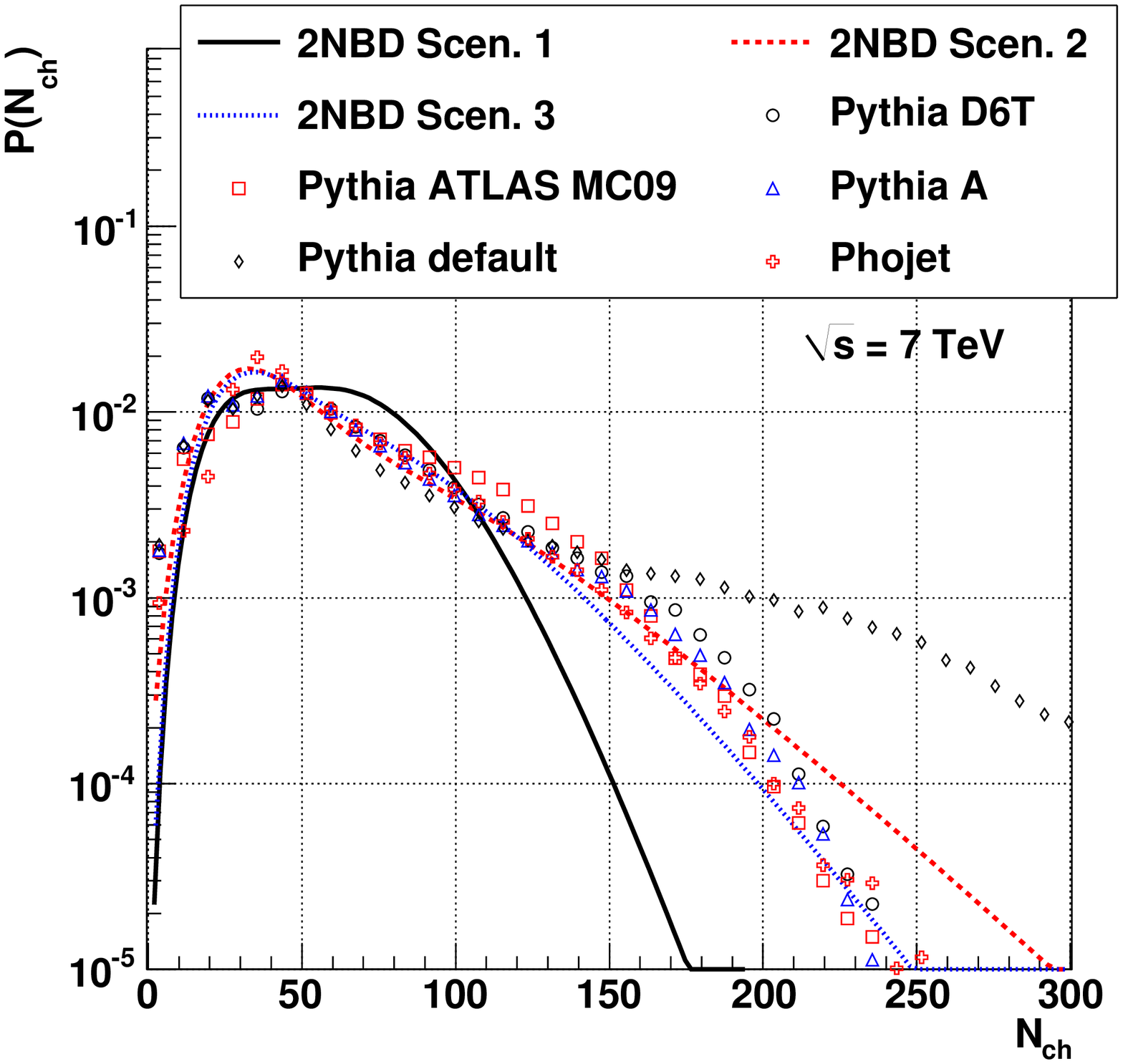}
        \hfill
        \includegraphics[width=0.48\linewidth]{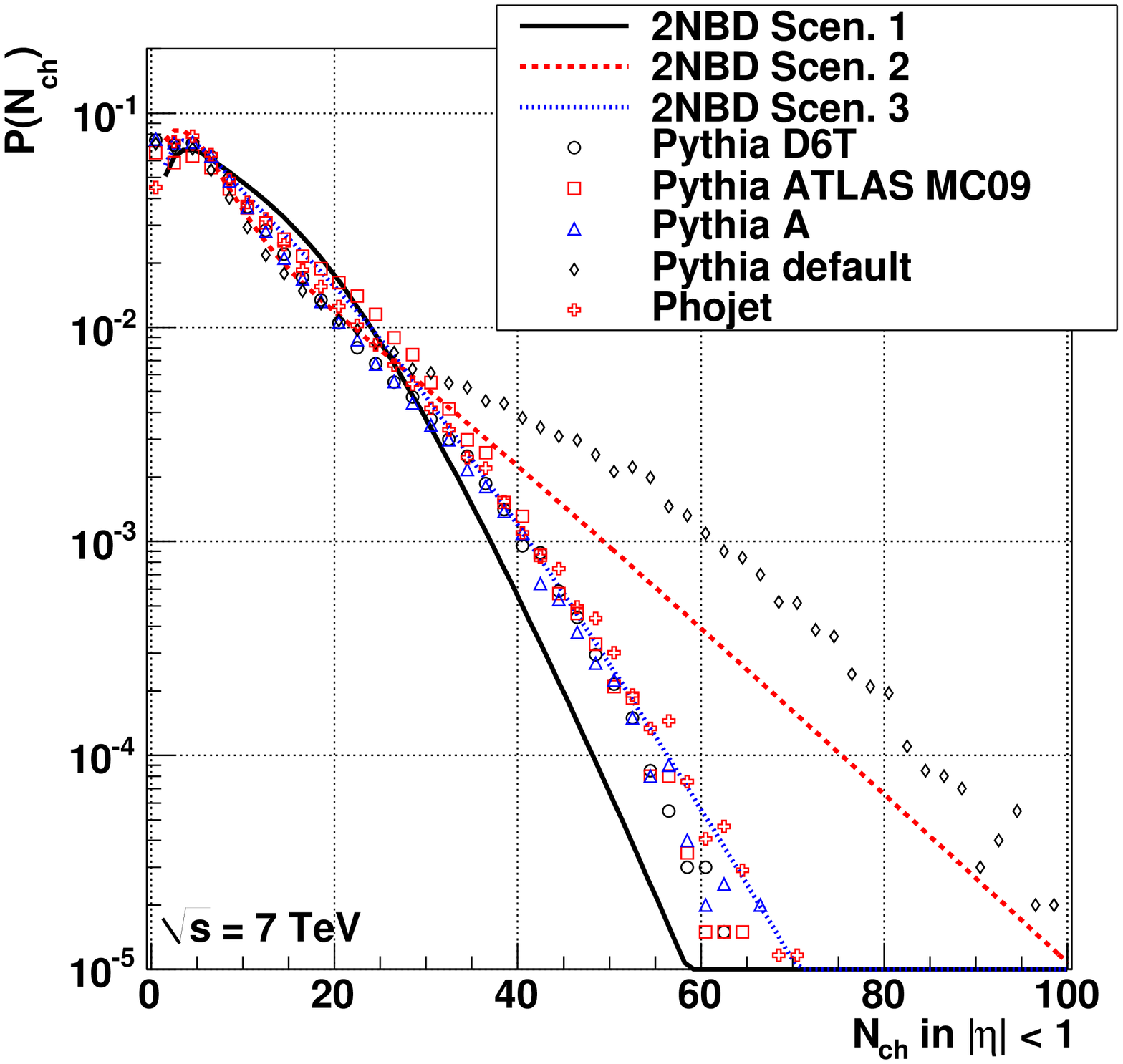}
        \caption{\label{fig_predictions_mult_7} Predictions for the multiplicity distribution of NSD events at \unit[7]{TeV} are shown in full phase space (left panel) and \etain{1} (right panel). The 2NBD predictions are from \cite{Giovannini:1998zb, Giovannini:1999tw}, for details see text. }
        %Prediction.C
      \efig

      \bfig
        \includegraphics[width=0.48\linewidth]{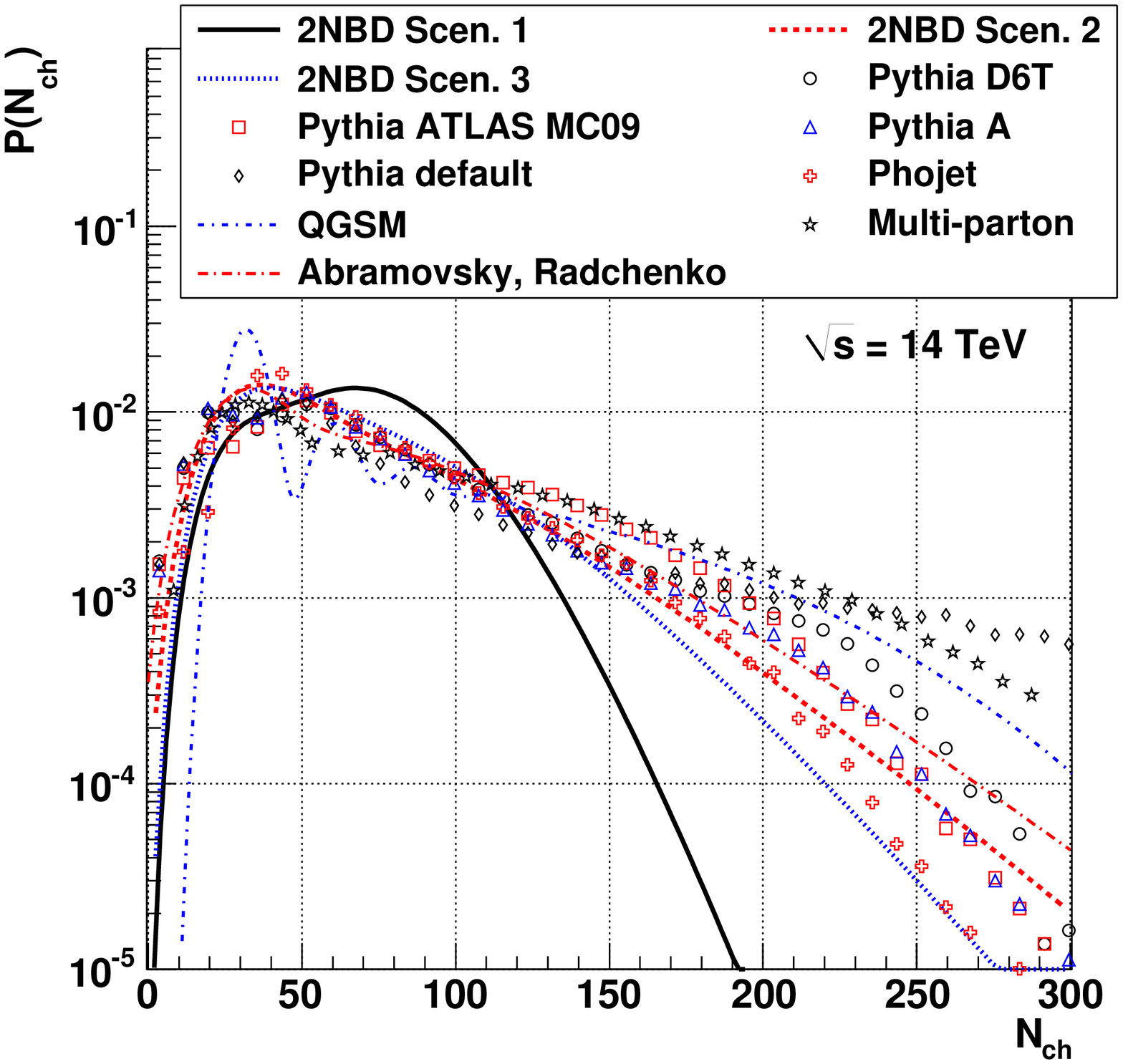}
        \hfill
        \includegraphics[width=0.48\linewidth]{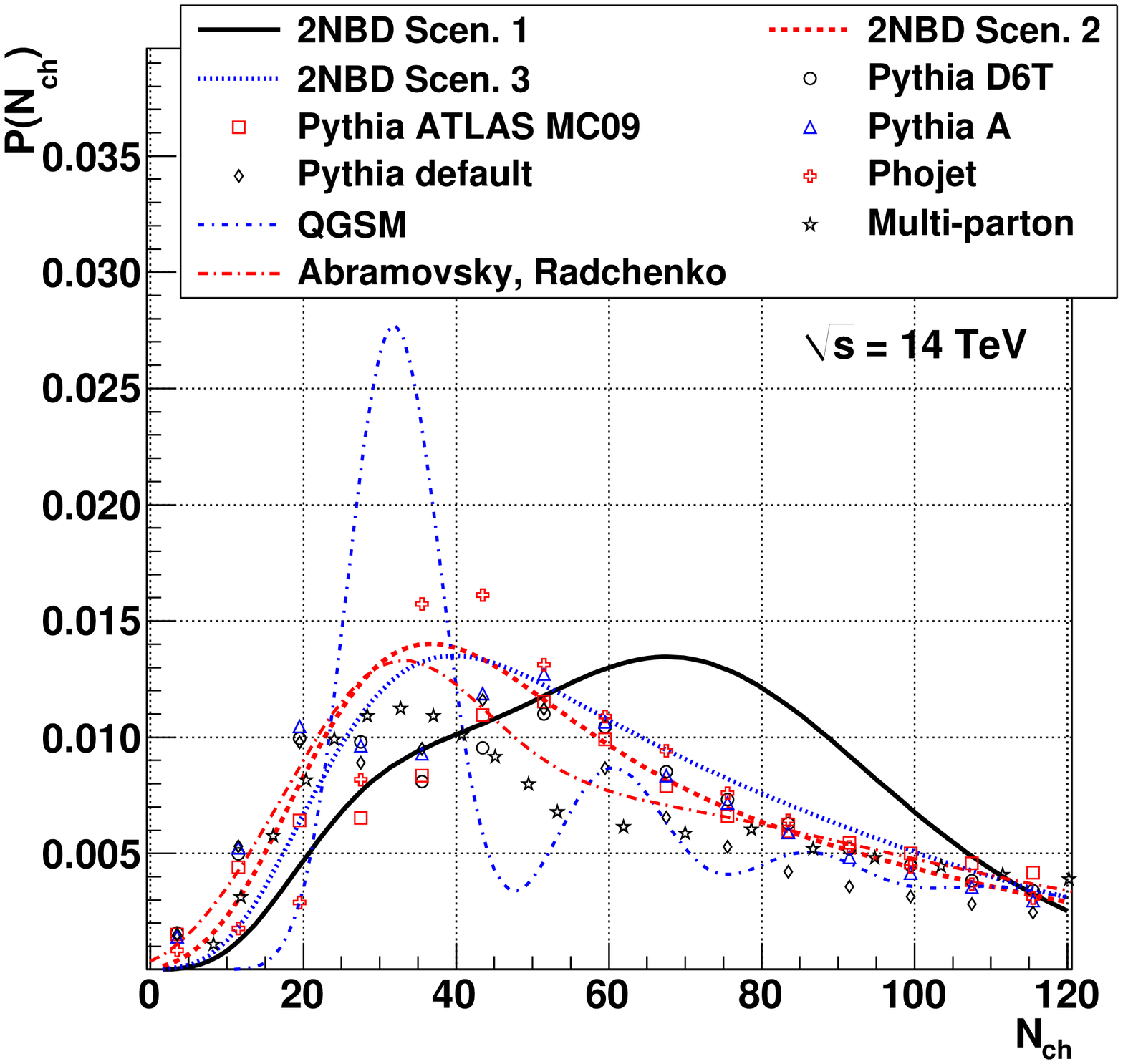}
        \caption{\label{fig_predictions_mult_14_fps} Predictions for the multiplicity distribution of NSD events at \unit[14]{TeV} are shown in full phase space. In addition to the logarithmic view, the right panel shows a linear scale and a zoom into the low-multiplicity region. The predictions are from \cite{Giovannini:1998zb, Giovannini:1999tw, Kaidalov:2009hn, Walker:2004tx, Abramovsky:2008uh}, for details see text. }
        %Prediction.C
      \efig
      
      \bfig
        \includegraphics[width=0.48\linewidth]{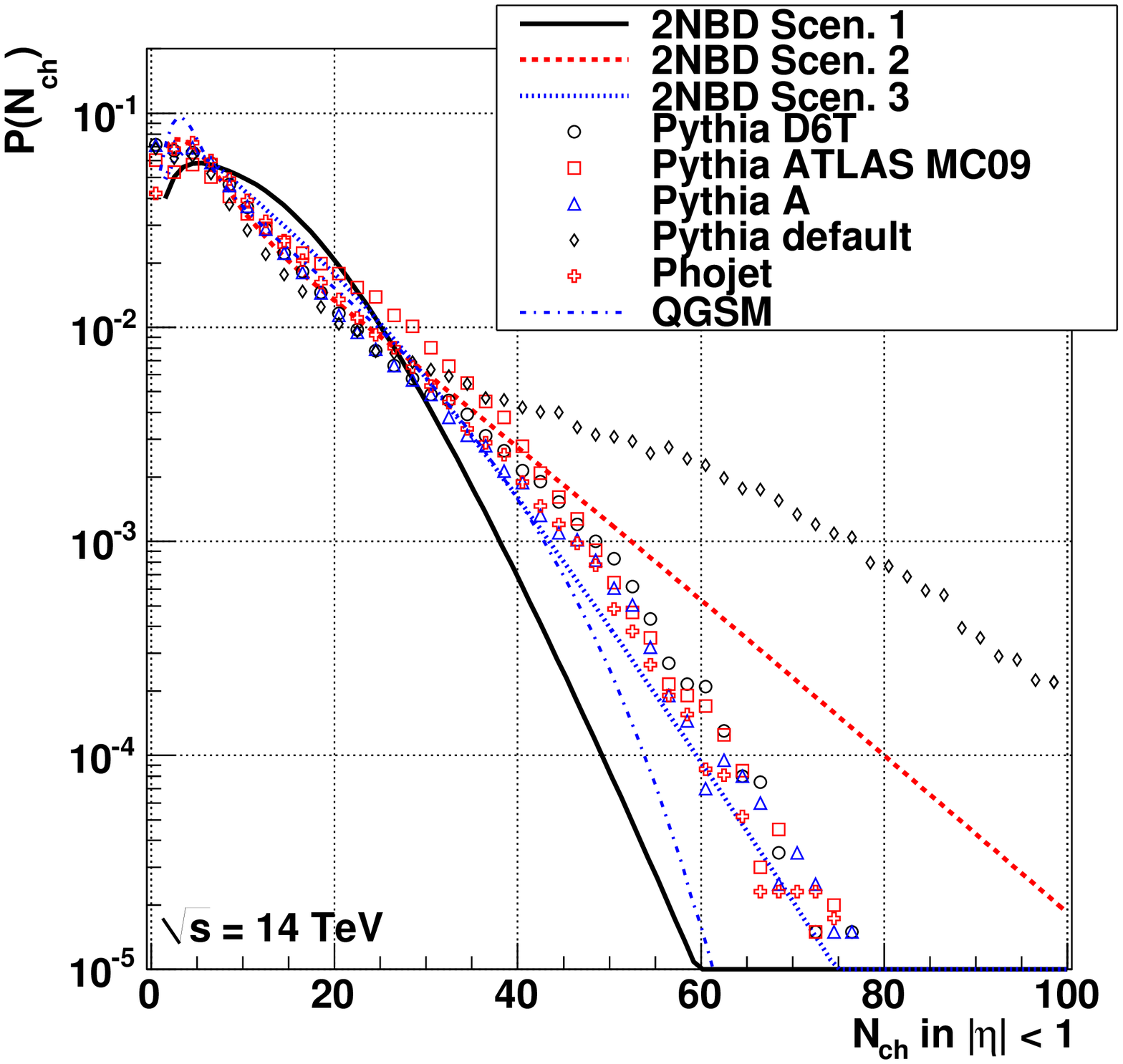}
        \hfill
        \includegraphics[width=0.48\linewidth]{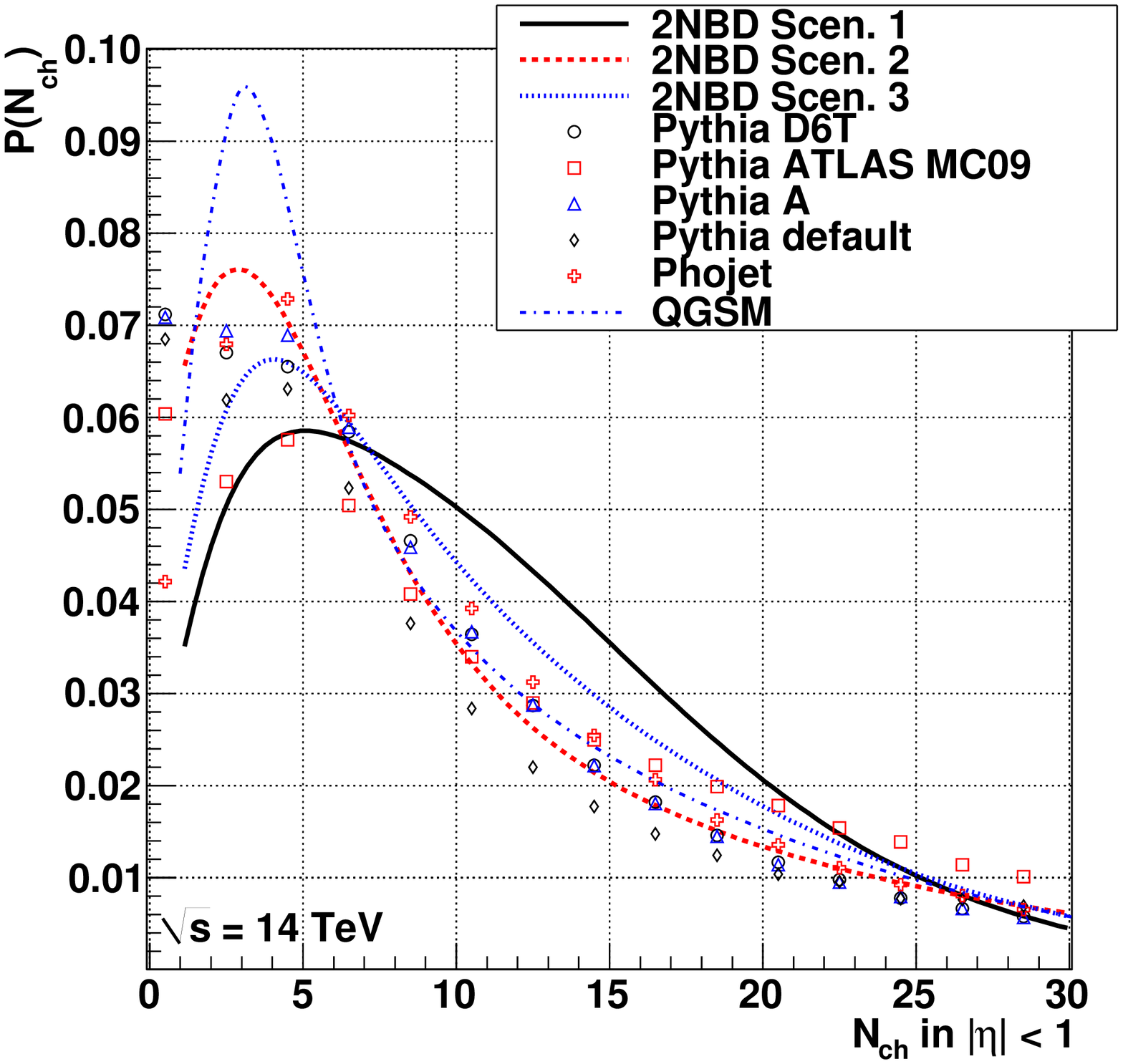}
        \caption{\label{fig_predictions_mult_14_eta1} Same as \figref{fig_predictions_mult_14_fps} but in \etain{1}.}
        %Prediction.C
      \efig
      
      Predictions for the multiplicity distribution in full phase space and in a limited range of \etain{1} are shown in \figref{fig_predictions_mult_7} for \cmsofT{7} and in Figures~\ref{fig_predictions_mult_14_fps} and \ref{fig_predictions_mult_14_eta1} for \unit[14]{TeV}. Shown are the extrapolations of the two-component model with NBDs for the three scenarios (from \cite{Giovannini:1998zb, Giovannini:1999tw}), Pythia with the four aforementioned tunes, and Phojet. Furthermore, for \unit[14]{TeV}, a prediction from QGSM \cite{Kaidalov:2009hn} and a prediction found in the framework of a multiple-parton interpretation of the collision \cite{Walker:2004tx} (only for full phase space, see Section~\ref{section_twocomponent_doubleparton}) are shown.
      
      The predictions differ significantly, which is most pronounced in the tails of the distributions where the deviation is more than an order of magnitude. This applies in full phase space as well as in limited $\eta$-ranges.
      However, also in the low-multiplicity region there are clear differences (see the right panels of \figref{fig_predictions_mult_14_fps} and \ref{fig_predictions_mult_14_eta1}). This difference is less pronounced in \etain{1}.
      From \cmsofT{7} to \unit[14]{TeV} the predicted differences increase further.

      Multiplicity distribution measurements at larger $\cms$ will allow one to decide which models best describe the data. For the specific case of Pythia it is clear that the parameter space is very large and several combinations of parameters may describe the data equally well. Nevertheless the measurement of the multiplicity distribution (together with the $p_T$ spectrum and the correlation of $\expval{p_T}$ and the multiplicity) will allow, e.g., one to learn about the colour correlations in the final-state \cite{Skands:2007zz}.
      
\section{Summary}
	
	This review summarizes measurements of charged-particle multiplicity distributions and pseudorapidity densities in high-energy $p+p(\bar{p})$ collisions. Moreover, related theoretical concepts have been briefly presented. The multiplicity cannot easily be described within QCD because it is related to soft interactions for which the strong coupling constant is large and perturbative methods are difficult to apply. The validity of the available theoretical descriptions has been assessed using data from collider experiments at centre-of-mass energies over about two orders of magnitude, from \unit[23.6]{GeV} to \unit[1.8]{TeV}. The energy dependence of $\langle N_\mathrm{ch} \rangle$ and $\dndetaZero$ shows that Feynman scaling is not satisfied at currently available energies; the moment analysis shows that KNO scaling does not hold except for very central and small regions of phase space where it is not ruled out by the present data. For high energies a single NBD does not fit the data anymore; a combination of two NBDs is more successful; however, additional assumptions are needed to identify general trends as a function of $\cms$. Although rapidity and multiplicity distributions differ between $p+p(\bar{p})$ and $e^+e^-$ collisions, their average multiplicities as function of $\cms$ show similar trends that can be unified using the concepts of effective energy and inelasticity. Without a correction for the multiplicity related to the leading protons the inelasticity $K$ in $p+p(\bar{p})$ collisions as defined by comparing to the multiplicity in $e^+e^-$ collisions decreases from about 0.6 at $\sqrt{s} = \unit[23.6]{GeV}$ to about 0.4 at $\sqrt{s} = \unit[1.8]{TeV}$. Taking this correction into account yields an energy-independent inelasticity of $K \approx 0.35$. Open experimental issues have been discussed and predictions for the LHC energy regime have been enumerated and briefly described. Interestingly, models that all more or less describe average multiplicities and multiplicity distributions up to Tevatron energies make significantly different predictions for the LHC. 

  At the LHC multiplicity measurements together with other global event properties will provide input to distinguish between the wealth of different model predictions including those from popular Monte Carlo event generators.  This will allow the amount of possible interpretations of the underlying physics to be reduced. In particular it will deepen the understanding of multiple-parton interactions and hadronisation as the LHC will allow for the first time to probe $p+p$ collisions in an energy regime where multiple hard parton interactions are present in most of the events. Understanding the underlying dynamics of multi-particle production is not only an interesting research topic in itself. Equally important is the characterization of the underlying event as prerequisite for more specialized studies of exotic and rare channels which LHC is aiming at.

\appendix
			
\section{Feynman Scaling}
  \label{section_derivation_feynmanscaling}

  In his paper \cite{Feynman:1969ej}, Feynman concluded that the mean number of particles rises logarithmically, but does not give a mathematic proof. However, one can assess the asymptotic behaviour by rewriting Eq.~\eqref{eq_feynman_d3sigma} in the form of the invariant cross section\footnote{The definition of the Feynman function is different in some publications (e.g. \cite{feynman_withoutsigma}), not considering the $1/\sigma$ term in Eq.~\eqref{eq_feynman_invxsection}. This approach, however, results in conclusions that are not confirmed by experiment. In detail compared to the results of the calculation presented in the following, the left sides of Eqs.~\eqref{eq_feynman_finalscaling} and \eqref{eq_feynman_finalscaling2} have to be multiplied by $\sigma$.}:
  \bq
    \frac{1}{\sigma} E \frac{\dd^3\sigma}{\dd p_z \dd^2p_T} = f_i(p_T, x_F). \label{eq_feynman_invxsection}
  \eq
  $f_i$ factorizes approximately (found experimentally) and a normalization of $g_i$ is chosen such that
  \bq
    \int f_i(p_T, x_F) \dd^2p_T = f_i(x_F) \underbrace{\int g_i(p_T) \dd^2p_T}_{= 1} = f_i(x_F). \label{eq_feynman_factorizing}
  \eq
  Integration of Eq.~\eqref{eq_feynman_invxsection} and application of Eq.~\eqref{eq_feynman_factorizing} yields:
  \bqq
    \int \frac{1}{\sigma} E \frac{\dd^3\sigma}{\dd p_z \dd^2p_T} \frac{\dd^3p}{E} = \N = \int f_i(p_T, x_F) \frac{\dd^3p}{E} \nonumber \\
    = \int f_i(x_F) \frac{\dd p_z}{\sqrt{W^2x^2 + m_T^2}},
  \eqq
  where on the left side the definition of the invariant cross section is used with the average particle multiplicity $\N$, and for $m_T$ an effective average-$p_T$ is used.

  Rewriting in $x_F$ yields the expression used to prove Feynman's hypothesis:
  \bq
    \N = \int_{-1}^1 f_i(x_F) \frac{\dd x_F}{\sqrt{x_F^2 + \frac{m_T^2}{W^2}}}. \label{eq_feynman_avgn}
  \eq
  The integral is symmetric because $f_i(x_F)$ is symmetric for collisions of identical particles. For other collision systems the integration can be performed separately for negative and positive $x_F$ and yields the same result.
  $f_i(x_F) \leq B$ is finite and bounded due to energy conservation. Furthermore, Feynman assumes that for $x_F=0$ a finite limit is reached. Therefore:
  \bqq
      2 \int_{0}^1 f_i(x_F) \frac{\dd x_F}{\sqrt{x_F^2 + \frac{m_T^2}{W^2}}} \leq 2 \int_{0}^1 B \frac{\dd x_F}{\sqrt{x_F^2 + \frac{m_T^2}{W^2}}} \\
      = 2 B \ln \left( x_F + \sqrt{x_F^2 + \frac{m_T^2}{W^2}} \right) \Biggr\vert_{0}^1 \\
      = 2 B \ln \left( 1 + \sqrt{1 + \frac{m_T^2}{W^2}} \right) - 2 B \ln \frac{m_T}{W}.
 \eqq
 The first term can be shown to be constant for $W \rightarrow \infty$ and the second is proportional to $\ln W$.

  In consequence, Feynman scaling implies that the average total multiplicity scales as
  \bq
    \N \propto \ln W \propto \ln \cms. \label{eq_feynman_finalscaling}
  \eq
  Considering that the maximum reachable rapidity in a collisions increases also with $\ln \cms$, and under the further assumption that the particles are evenly distributed in rapidity, it follows that $\dd N/\dd y$ is independent of $\cms$:
  \bq
    \frac{\dd N}{\dd y} = \mathrm{constant}. \label{eq_feynman_finalscaling2}
  \eq

\section{Uncertainties on Moments}
  \label{section_moments_uncertainty}
		
		Given a distribution $P(n)$ which is normalized to 1 with an uncertainty $e_n$, and assuming that the errors on the individual bins are uncorrelated (which may not be the case after an unfolding procedure is applied, see also Section~\ref{section_analysis_unfolding}) their errors can be calculated using the partial derivatives:
    \bq
      \frac{\partial C_q}{\partial P(n)} = \frac{n^q \n - \expval{n^q} q n}{\n^{q+1}},
    \eq
		\bq
      \frac{\partial F_q}{\partial P(n)} = \frac{n(n-1)...(n-q+1) \n - \expval{n(n-1)...(n-q+1)} q n}{\n^{q+1}},
		\eq
		\bq
  		\frac{\partial D_q}{\partial P(n)} =
  		  \frac{\expval{(n - \n)^q}^{\frac{1}{q}-1} \left[ \expval{-nq(n-\n)^{q-1}} + (n - \n)^q \right]}{q}.
		\eq
		The total is then 		
		\bq
      E_q^2 = \sum_n \left( \frac{\partial X_q}{\partial P(n)} e_n \right)^2,
    \eq
    where $X_q$ is $C_q$, $F_q$, or $D_q$.

\section{Relation of NBD and BD}
  \label{relation_nbd_bd}

  This section shows that a NBD becomes binomial when $k$ is a negative integer. To start with the NBD and the binomial distribution (BD) are recalled. The NBD is:
  \bq
    P^{\rm NBD}_{\n, k}(n) = \left ( \begin{array}{c} n+k-1 \\ n \\ \end{array} \right )
      \left ( \frac{ \n / k }{ 1 + \n/k } \right)^n \frac{1}{(1 + \n/k)^k}
  \eq
  with $n$ failures and $k$ successes.
  The BD is
  \bq
    P^{\rm BD}_{p, m}(n) = \left ( \begin{array}{c} m \\ n \\ \end{array} \right ) p^n (1-p)^{m-n}
  \eq
  with $m$ trials, $n$ successes, and success probability $p$. Important is that for both, the NBD and the BD, the running variable is $n$.
  
  Using Eq.~\eqref{eq_nbd_gamma}, the NBD is rewritten as
  \bqq
    P^{\rm NBD}_{\n, k}(n) =&& \frac{(n+k-1)\cdot(n+k-2)\cdot ... \cdot k}{n!} \cdot \nonumber \\
      && \left (  \n / k \right)^n (1 + \n/k)^{-k-n}.
  \eqq
  If we identify $m$ with $-k$ and $p$ with $-\n / k$ we find:
  \bq
    P^{\rm NBD}_{\n, k}(n) = \frac{(n-m-1)\cdot(n-m-2)\cdot ... \cdot (-m)}{n!} 
      \left (  -p \right)^n (1 - p)^{m-n}. \label{eq_nbd_eq1}
  \eq
  Assuming that $n - m - 1 < 0$ and $-m < 0$, all terms in the product are negative and the following relation holds:
  \bqq
    (n-m-1)&&\cdot(n-m-2)\cdot ... \cdot (-m) = \nonumber \\
    && (-1)^n \cdot (-n+m+1)\cdot(-n+m+2)\cdot ... \cdot m. \label{eq_nbd_product}
  \eqq
  Eq.~\eqref{eq_nbd_eq1} is then:
  \bq
    P^{\rm NBD}_{\n, k}(n) = \frac{(m-n+1)\cdot(m-n+2)\cdot ... \cdot m}{n!} p^n (1 - p)^{m-n}. \label{eq_nbd_eq2}
  \eq
  Applying Eq.~\eqref{eq_nbd_gamma} the first term can be identified as the binomial term of the BD:
  \bq
    \left ( \begin{array}{c} m \\ n \\ \end{array} \right ) = \frac{(m-n+1)\cdot(m-n+2)\cdot ... \cdot m}{n!}.
  \eq
  Thus the NBD with negative integer $k$ is a BD with the Bernoulli probability $p = -\n / k$ and the number of trials $m = -k$. For such a BD the assumptions made above are indeed fulfilled: $-m < k < 0$ ($m > 0$) follows trivially; the number of successes $n$ is smaller or equal than the number of trials $m$ and therefore also $n - m - 1 < 0$. It is required that $0 < p < 1$, thus $0 < \n < -k$.

\section*{Acknowledgements}
  We acknowledge extensive and fruitful discussions with Igor Dremin and Karel \v{S}afa\v{r}\'{i}k.
  
  Michelle Connor, Alberto Giovannini, Jochen Klein, Christian Klein-B\"osing, Andreas Morsch, Martin Poghosyan, Paul W. Stankus, Peter Steinberg, Michael J. Tannenbaum, and Johannes P. Wessels are thanked for comments and suggestions about the manuscript.

  We would like to thank Albert Erwin for providing some additional references of E735 data as well as Sandor Hegyi for sharing an electronic copy of E735 data sets that were published, but not available in electronic form. 

\section*{References}
\bibliographystyle{main3}
\bibliography{main3}

\end{document}